\documentclass[10pt,twoside,notitlepage,a4paper,fleqn,openbib]{report}

\usepackage{ifthen}
\newboolean{isARXIV}
\setboolean{isARXIV}{true}

\usepackage{epsf}
\usepackage{graphicx,amsmath,amssymb,pifont,dsfont,bbm}
\usepackage{leftidx}
\usepackage{psfrag}   % psfrag
\usepackage{tensind}  % for tensor notation
\usepackage{mathptmx} % use Times fonts if available on your TeX system
\usepackage[colorlinks]{hyperref}
\usepackage{fancyhdr}
\usepackage{palatino}
\usepackage{textcomp}
\usepackage{units}
\usepackage{color}
\usepackage{url}
\usepackage[rightcaption]{sidecap}

\usepackage[T1]{fontenc}
\newcommand{\verbfont}{\fontsize{7}{9}\selectfont}

\newcommand{\clearemptydoublepage}{\newpage{\pagestyle{empty}\cleardoublepage}}

\def\dst{\displaystyle}

\def\pdiff#1#2{\dst\frac{\partial#1}{\partial #2}}
\def\p2diff#1#2{\dst\frac{\partial^{2}#1}{\partial #2^{2}}}
\def\symml{\left(\right.}
\def\symmr{\left.\right)}
\def\antil{\left[\right.}
\def\antir{\left.\right]}

\def\sign{\textrm{sign}}
\def\sech{\textrm{sech}}
\def\arsinh{\textrm{arsinh}}

\def\defequal{:=}

\def\uLength{\textrm{length}}
\def\uMass{\textrm{mass}}
\def\uTime{\textrm{time}}

\def\setR{\mathbbm{R}}

\newcommand{\evalUnits}[1]{\left[#1\right]_{\text{\tiny U}}}

\newcommand{\SecMetric}{\vspace*{0.1cm}\noindent {\bf Metric-Tensor:}}
\newcommand{\SecChristoffel}{\vspace*{0.1cm}\noindent {\bf Christoffel symbols:}}
\newcommand{\SecRiemann}{\vspace*{0.1cm}\noindent {\bf Riemann-Tensor:}}
\newcommand{\SecRiemannLT}{\vspace*{0.1cm}\noindent {\bf Riemann-Tensor with respect to local tetrad:}}
\newcommand{\SecWeyl}{\vspace*{0.1cm}\noindent {\bf Weyl-Tensor:}}
\newcommand{\SecRicci}{\vspace*{0.1cm}\noindent {\bf Ricci-Tensor:}}
\newcommand{\SecRicciLT}{\vspace*{0.1cm}\noindent {\bf Ricci-Tensor with respect to local tetrad:}}
\newcommand{\SecWeylLT}{\vspace*{0.1cm}\noindent {\bf Weyl-Tensor with respect to local tetrad:}}
\newcommand{\SecKretsch}{\vspace*{0.1cm}\noindent {\bf Kretschmann scalar:}}
\newcommand{\SecLocal}{\vspace*{0.1cm}\noindent {\bf Local tetrad:}}
\newcommand{\SecStatLocal}{\vspace*{0.1cm}\noindent {\bf Static local tetrad:}}
\newcommand{\SecComLocal}{\vspace*{0.1cm}\noindent {\bf Comoving local tetrad:}}
\newcommand{\SecFreeLocal}{\vspace*{0.1cm}\noindent {\bf Freely falling local tetrad:}}

\newcommand{\SecLieCoef}{\vspace*{0.1cm}\noindent {\bf Structure coefficients:}}
\newcommand{\SecRicRotCoef}{\vspace*{0.1cm}\noindent {\bf Ricci rotation coefficients:}}
\newcommand{\SecKilling}{\vspace*{0.1cm}\noindent {\bf Killing vectors:}}
\newcommand{\SecEmbedding}{\vspace*{0.1cm}\noindent {\bf Embedding:}}
\newcommand{\SecEffPoti}{\vspace*{0.1cm}\noindent {\bf Effective potential:}}
\newcommand{\SecEulLag}{\vspace*{0.1cm}\noindent {\bf Euler-Lagrange:}}
\newcommand{\FurtherReading}{\vspace*{0.1cm}\noindent {\bf Further reading:}}

\newcommand{\ESCstr}{Case}

\newcommand{\Cchris}[2]{\Gamma_{#1}^{#2}}
\newcommand{\Clt}[1]{\mathbf{e}_{(#1)}}
\newcommand{\Cdlt}[1]{\boldsymbol{\theta}^{(#1)}}

\newcommand{\metricEq}[2]{
 \setlength{\fboxrule}{0.04cm}
 \setlength{\fboxsep}{0.2cm}
 \begin{equation}
  \fbox{$\dst #1 $}
  \label{eqM:#2}
 \end{equation} 
}

\newcommand{\metricEqAlign}[2]{
 \setlength{\fboxrule}{0.04cm}
 \setlength{\fboxsep}{0.2cm}
 \label{eqM:#2}
 \begin{align}
  \fbox{$\begin{array}{rcl} #1 \end{array}$}  
 \end{align} 
}
\definecolor{darkred}{rgb}{0.5,0.0,0.0}
\definecolor{darkgreen}{rgb}{0.0,0.5,0.0}
\hypersetup{
  linkcolor=darkred,
  citecolor=blue
}

\begin{document}
\tensordelimiter{?}
%% --------------------------------------------------------------------
%%                        T i t e l s e i t e
%% --------------------------------------------------------------------
\begin{titlepage}
  \vspace*{2cm}
  \begin{center}
    {\fontsize{32}{35}\selectfont \bf Catalogue of Spacetimes}\\[4em]
    %{\Large v2 rev16}\\[3em]
\ifthenelse{\boolean{isARXIV}}{
    {\includegraphics[scale=1.0]{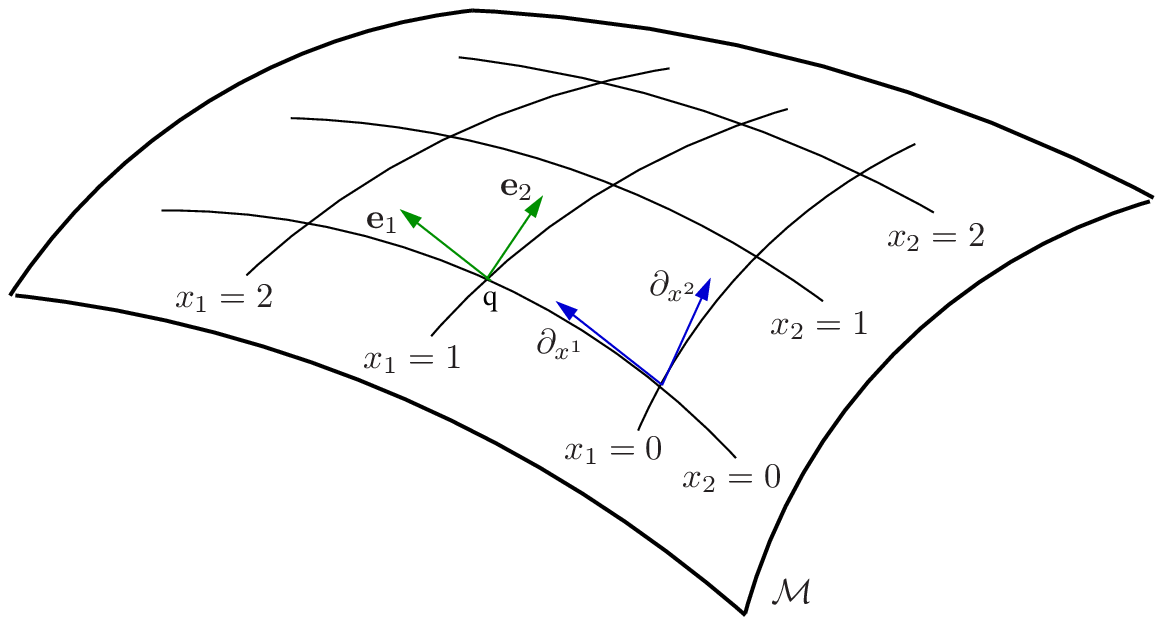}}
}{
    {\includegraphics[scale=1.0]{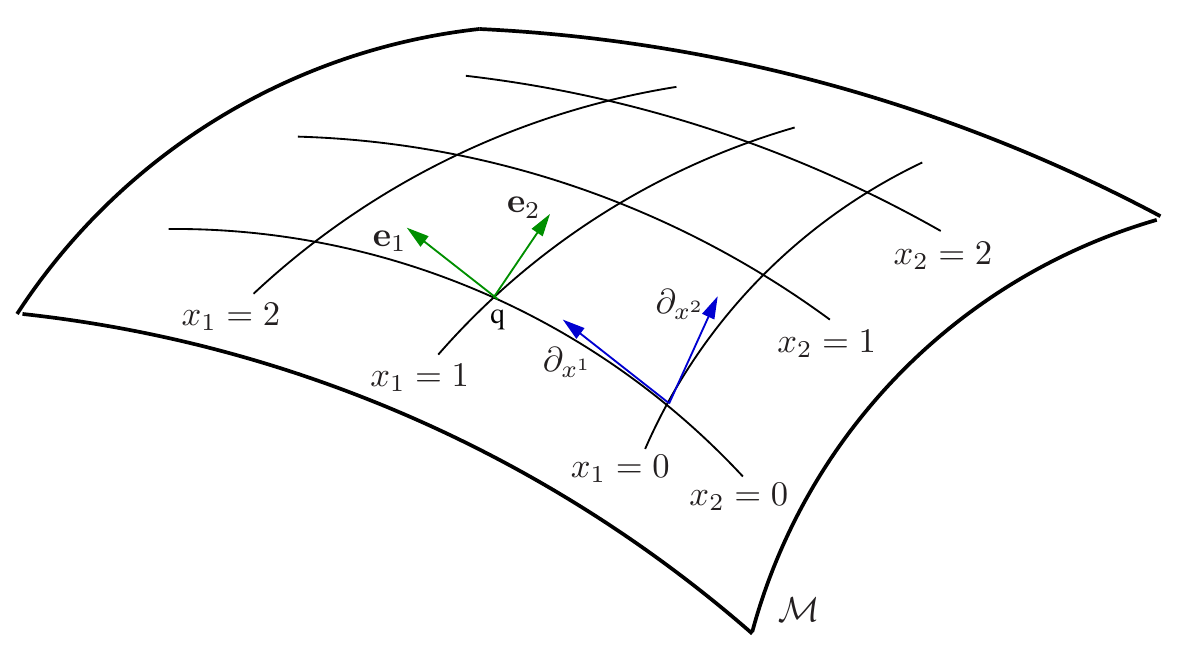}}
}    
  \end{center}
  \vspace*{4cm}
  \begin{tabbing}\hspace*{1.6cm}\=\\
    Authors:\>{\bf Thomas M{\"u}ller}\\
    \>Visualisierungsinstitut der Universit\"at Stuttgart (VISUS)\\
    \>Allmandring 19, 70569 Stuttgart, Germany\\[0.25em]
    \>Thomas.Mueller@vis.uni-stuttgart.de\\[1em]
    \>{\bf Frank Grave}\\
    \>formerly, Universit\"at Stuttgart, Institut f{\"u}r Theoretische Physik 1 (ITP1)\\
    \>Pfaffenwaldring 57 /\!\!/ IV, 70550 Stuttgart, Germany\\[0.25em]
    \>Frank.Grave@vis.uni-stuttgart.de\\[1em]
    URL:\>http://www.vis.uni-stuttgart.de/\texttildelow muelleta/CoS\\[1em]
    Date:\>04. Nov 2010
  \end{tabbing}
\end{titlepage}

% -----------------------------------------------------------------
%         Coauthors
% -----------------------------------------------------------------
\thispagestyle{empty}
\vspace*{17cm}
{\bf Co-authors}
\begin{tabbing}\hspace*{0.5cm}\=\\
Andreas Lemmer, formerly, Institut f{\"u}r Theoretische Physik 1 (ITP1), Universit\"at Stuttgart\\
\>{\it Alcubierre Warp}\\[0.5em]
Sebastian Boblest, Institut f{\"u}r Theoretische Physik 1 (ITP1), Universit\"at Stuttgart\\
\>{\it deSitter, Friedmann-Robertson-Walker}\\[0.5em]
Felix Beslmeisl, Institut f{\"u}r Theoretische Physik 1 (ITP1), Universit\"at Stuttgart\\
\>{\it Petrov-Type D}\\[0.5em]
Heiko Munz, Institut f{\"u}r Theoretische Physik 1 (ITP1), Universit\"at Stuttgart\\
\>{\it Bessel and plane wave}
\end{tabbing}

% -----------------------------------------------------------------
%         Contents
% -----------------------------------------------------------------
\clearemptydoublepage
\pagenumbering{roman}
\tableofcontents

% -----------------------------------------------------------------
%    chapter:  Introduction and Notation
% -----------------------------------------------------------------
\clearemptydoublepage
\setcounter{page}{1}
\pagenumbering{arabic}
\chapter{Introduction and Notation}
The {\sl Catalogue of Spacetimes} is a collection of four-dimensional Lorentzian spacetimes in the context of the General Theory of Relativity (GR). The aim of the catalogue is to give a quick reference for students who need some basic facts of the most well-known spacetimes in GR. For a detailed discussion of a metric, the reader is referred to the standard literature or the original articles. Important resources for exact solutions are the book by Stephani et al\cite{exact2003} and the book by Griffiths and Podolsk{\'y}\cite{griffiths2009}.\\[0.5em]
Most of the metrics in this catalogue are implemented in the Motion4D-library\cite{mueller2009b} and can be visualized using the GeodesicViewer\cite{mueller2010a}. Except for the Minkowski and Schwarzschild spacetimes, the metrics are sorted by their names.

% -----------------------------------------------------------------
%    sec:  Notation
% -----------------------------------------------------------------
\section{Notation}
\setcounter{equation}{0}
The notation we use in this catalogue is as follows:\\[0.3em]
{\bf Indices:} Coordinate indices are represented either by Greek letters or by coordinate names. Tetrad indices are indicated by Latin letters or coordinate names in brackets.\\[0.3em]
{\bf Einstein sum convention:} When an index appears twice in a single term, once as lower index and once as upper index, we build the sum over all indices:
\begin{equation}
   \zeta_{\mu}\zeta^{\mu} \equiv \sum\limits_{\mu=0}^{3}\zeta_{\mu}\zeta^{\mu}.
\end{equation}
{\bf Vectors:} A coordinate vector in $x^{\mu}$ direction is represented as $\partial_{x^{\mu}}\equiv\partial_{\mu}$. For arbitrary vectors, we use boldface symbols. Hence, a vector $\mathbf{a}$ in coordinate representation reads $\mathbf{a}=a^{\mu}\partial_{\mu}$.\\[0.3em]
{\bf Derivatives:} Partial derivatives are indicated by a comma, $\partial\psi/\partial x^{\mu}\equiv \partial_{\mu}\psi\equiv \psi_{,\mu}$, whereas covariant derivatives are indicated by a semicolon, $\nabla\psi = \psi_{;\mu}$.\\[0.3em]
{\bf Symmetrization and Antisymmetrization brackets:}
\begin{equation}
  a_{\symml\mu}b_{\nu\symmr} = \frac{1}{2}\left(a_{\mu}b_{\nu}+a_{\nu}b_{\mu}\right),\qquad  a_{\antil\mu}b_{\nu\antir} = \frac{1}{2}\left(a_{\mu}b_{\nu}-a_{\nu}b_{\mu}\right)
\end{equation}

% -----------------------------------------------------------------
%    sec:  General remarks
% -----------------------------------------------------------------
\section{General remarks}
\setcounter{equation}{0}
The Einstein field equation in the most general form reads\cite{mtw}
\begin{equation}
  \label{eq:EinsteinFieldEquations}
  G_{\mu\nu} = \varkappa T_{\mu\nu}-\Lambda g_{\mu\nu},\qquad \varkappa = \frac{8\pi G}{c^4},
\end{equation}
with the symmetric and divergence-free Einstein tensor $G_{\mu\nu}=R_{\mu\nu}-\frac{1}{2}Rg_{\mu\nu}$, the Ricci tensor $R_{\mu\nu}$, the Ricci scalar $R$, the metric tensor $g_{\mu\nu}$, the energy-momentum tensor $T_{\mu\nu}$, the cosmological constant $\Lambda$, Newton's gravitational constant $G$, and the speed of light $c$. Because the Einstein tensor is divergence-free, the conservation equation $?T^{\mu\nu}_{;\nu}?=0$ is automatically fulfilled.\\[0.5em]
A solution to the field equation is given by the line element
\begin{equation}
 ds^2 = g_{\mu\nu}dx^{\mu}dx^{\nu}
\end{equation}
with the symmetric, covariant metric tensor $g_{\mu\nu}$. The contravariant metric tensor $g^{\mu\nu}$ is related to the covariant tensor via $g_{\mu\nu}g^{\nu\lambda}=\delta_{\mu}^{\lambda}$ with the Kronecker-$\delta$. Even though $g_{\mu\nu}$ is only a component of the metric tensor $\mathbf{g}=g_{\mu\nu}dx^{\mu}\otimes dx^{\nu}$, we will also call $g_{\mu\nu}$ the metric tensor.\\[0.5em]
Note that, in this catalogue, we mostly use the convention that the signature of the metric is $+2$. In general, we will also keep the physical constants $c$ and $G$ within the metrics.

% -----------------------------------------------------------------
%    sec:  Basic objects of a metric
% -----------------------------------------------------------------
\section{Basic objects of a metric}
\setcounter{equation}{0}
The basic objects of a metric are the Christoffel symbols, the Riemann and Ricci tensors as well as the Ricci and Kretschmann scalars which are defined as follows:\\[0.5em]
{\bf Christoffel symbols of the first kind:}\footnote{The notation of the Christoffel symbols of the first kind differs from the one used by Rindler\cite{rindler}, $\Gamma_{\mu\nu\lambda}^{\text{Rindler}}=\Gamma_{\nu\lambda\mu}^{\text{CoS}}$.}
\begin{equation}
  \label{eq:ChristoffelSymbolsFK}
  \Gamma_{\nu\lambda\mu} = \frac{1}{2}\left(g_{\mu\nu,\lambda}+g_{\mu\lambda,\nu}-g_{\nu\lambda,\mu}\right)
\end{equation}
with the relation
\begin{equation}
  g_{\nu\lambda,\mu} = \Gamma_{\mu\nu\lambda}+\Gamma_{\mu\lambda\nu}
\end{equation}

{\bf Christoffel symbols of the second kind:}
\begin{equation}
  \label{eq:ChristoffelSymbols}
  \Gamma_{\nu\lambda}^{\mu} = \frac{1}{2}g^{\mu\rho}\left(g_{\rho\nu,\lambda}+g_{\rho\lambda,\nu}-g_{\nu\lambda,\rho}\right)
\end{equation}
which are related to the Christoffel symbols of the first kind via
\begin{equation}
  \Gamma_{\nu\lambda}^{\mu} = g^{\mu\rho}\Gamma_{\nu\lambda\rho}
\end{equation}

{\bf Riemann tensor:}
\begin{equation}
   \label{eq:RiemannTensor}
   ?R^{\mu}_{\nu\rho\sigma}? = \Gamma_{\nu\sigma,\rho}^{\mu}-\Gamma_{\nu\rho,\sigma}^{\mu}+\Gamma_{\rho\lambda}^{\mu}\Gamma_{\nu\sigma}^{\lambda}-\Gamma_{\sigma\lambda}^{\mu}\Gamma_{\nu\rho}^{\lambda}
\end{equation}
or
\begin{equation}
  R_{\mu\nu\rho\sigma} = g_{\mu\lambda}?R^{\lambda}_{\nu\rho\sigma}? = \Gamma_{\nu\sigma\mu,\rho}-\Gamma_{\nu\rho\mu,\sigma}+\Gamma_{\nu\rho}^{\lambda}\Gamma_{\mu\sigma\lambda}-\Gamma_{\nu\sigma}^{\lambda}\Gamma_{\mu\sigma\lambda}
\end{equation}
with symmetries
\begin{equation}
  R_{\mu\nu\rho\sigma} = -R_{\mu\nu\sigma\rho},\qquad R_{\mu\nu\rho\sigma} = -R_{\nu\mu\rho\sigma},\qquad R_{\mu\nu\rho\sigma} = R_{\rho\sigma\mu\nu}
\end{equation}
and
\begin{equation}
  R_{\mu\nu\rho\sigma} + R_{\mu\rho\sigma\nu} + R_{\mu\sigma\nu\rho} = 0
\end{equation}

{\bf Ricci tensor:}
\begin{equation}
  R_{\mu\nu} = g^{\rho\sigma}R_{\rho\mu\sigma\nu} = ?R^{\rho}_{\mu\rho\nu}?
\end{equation}

{\bf Ricci and Kretschmann scalar:}
\begin{equation}
  \mathcal{R} = g^{\mu\nu}R_{\mu\nu} = ?R^{\mu}_{\mu}?, \qquad \mathcal{K} = R_{\alpha\beta\gamma\delta}R^{\alpha\beta\gamma\delta} = ?R^{\gamma\delta}_{\alpha\beta}??R^{\alpha\beta}_{\gamma\delta}?
\end{equation}

{\bf Weyl tensor:}
\begin{equation}
  C_{\mu\nu\rho\sigma} = R_{\mu\nu\rho\sigma}-\frac{1}{2}\left(g_{\mu\antil\rho}R_{\sigma\antir\nu}-g_{\nu\antil\rho}R_{\sigma\antir\mu}\right)+\frac{1}{3}R\,g_{\mu\antil\rho}g_{\sigma\antir\nu}
\end{equation}

If we change the signature of a metric, these basic objects transform as follows:
\begin{subequations}
\begin{alignat}{5}
  \Gamma_{\nu\lambda}^{\mu}&\mapsto \Gamma_{\nu\lambda}^{\mu}, &\qquad R_{\mu\nu\rho\sigma}&\mapsto -R_{\mu\nu\rho\sigma}, &\qquad C_{\mu\nu\rho\sigma}&\mapsto -C_{\mu\nu\rho\sigma},\\
  R_{\mu\nu} &\mapsto R_{\mu\nu}, & \mathcal{R} &\mapsto -\mathcal{R}, & \mathcal{K} &\mapsto \mathcal{K}.
\end{alignat}
\end{subequations}

{\bf Covariant derivative}
\begin{equation}
 \nabla_{\lambda}g_{\mu\nu} = g_{\mu\nu;\lambda} = 0.
\end{equation}

{\bf Covariant derivative} of the vector field $\psi^{\mu}$:
\begin{equation}
  \nabla_{\nu}\psi^{\mu} = \psi^{\mu}_{;\nu} = \partial_{\nu}\psi^{\mu}+\Gamma_{\nu\lambda}^{\mu}\psi^{\lambda}
\end{equation}

{\bf Covariant derivative} of a r-s-tensor field:
\begin{equation}
\begin{aligned}
 \nabla_c?T^{a_1\ldots a_r}_{b_1\ldots b_s}? &= \partial_c?T^{a_1\ldots a_r}_{b_1\ldots b_s}?+\Gamma_{dc}^{a_1}?T^{d\ldots a_r}_{b_1\ldots b_s}?+\ldots +\Gamma_{dc}^{a_r}?T^{a_1\ldots a_{r-1}d}_{b_1\ldots b_s}?\\
  &\qquad - \Gamma_{b_1c}^d?T^{a_1\ldots a_r}_{d\ldots b_s}?-\ldots -\Gamma_{b_sc}^d?T^{a_1\ldots a_r}_{b_1\ldots b_{s-1}d}?
\end{aligned}
\end{equation}

{\bf Killing equation:}
\begin{equation}
 \label{eq:killing}
 \xi_{\mu;\nu}+\xi_{\nu;\mu}=0.
\end{equation}

% -----------------------------------------------------------------
%    sec:  Natural local tetrad and initial conditions
% -----------------------------------------------------------------
\section{Natural local tetrad and initial conditions for geodesics}\label{sec:natLocTetrad}
\setcounter{equation}{0}
We will call a local tetrad natural if it is adapted to the symmetries or the coordinates of the spacetime. The four base vectors $\mathbf{e}_{(i)}=e_{(i)}^{\mu}\partial_{\mu}$ are given with respect to coordinate directions $\partial/\partial x^{\mu}=\partial_{\mu}$, compare Nakahara\cite{nakahara} or Chandrasekhar\cite{chandrasekhar2006} for an introduction to the tetrad formalism. The inverse or dual tetrad is given by $\boldsymbol{\theta}^{(i)}=\theta^{(i)}_{\mu}dx^{\mu}$ with
\begin{equation}
  \boldsymbol{\theta}^{(i)}_{\mu}e_{(j)}^{\mu} = \delta_{(j)}^{(i)}\qquad\text{and}\qquad \theta^{(i)}_{\mu}e_{(i)}^{\nu} = \delta_{\mu}^{\nu}.
\end{equation}
Note that we us Latin indices in brackets for tetrads and Greek indices for coordinates.
% -----------------------------------------------------------------
%           subsec:  Orthonormality condition
% -----------------------------------------------------------------
\subsection{Orthonormality condition}
To be applicable as a local reference frame (Minkowski frame), a local tetrad $\mathbf{e}_{(i)}$ has to fulfill the orthonormality condition
\begin{equation}
  \left<\mathbf{e}_{(i)},\mathbf{e}_{(j)}\right>_{\mathbf{g}} = \mathbf{g}\left(\mathbf{e}_{(i)},\mathbf{e}_{(j)}\right) = g_{\mu\nu}e_{(i)}^{\mu}e_{(j)}^{\nu} \stackrel{!}{=} \eta_{(i)(j)},
\end{equation}
where $\eta_{(i)(j)}=\mathrm{diag}(\mp 1,\pm 1,\pm 1,\pm 1)$ depending on the signature $\sign(\mathbf{g})=\pm 2$ of the metric. Thus, the line element of a metric can be written as
\begin{equation}
  \label{eq:lineElemWithTheta}
  ds^2 = \eta_{(i)(j)}\boldsymbol{\theta}^{(i)}\boldsymbol{\theta}^{(j)} = \eta_{(i)(j)}\theta^{(i)}_{\mu}\theta^{(j)}_{\nu}dx^{\mu}dx^{\nu}.
\end{equation}
To obtain a local tetrad $\mathbf{e}_{(i)}$, we could first determine the dual tetrad $\boldsymbol{\theta}^{(i)}$ via Eq. (\ref{eq:lineElemWithTheta}). If we combine all four dual tetrad vectors into one matrix $\Theta$, we only have to determine its inverse $\Theta^{-1}$ to find the tetrad vectors,
\begin{equation}
 \Theta = \begin{pmatrix} \theta_0^{(0)} & \theta_1^{(0)} & \theta_2^{(0)} & \theta_3^{(0)} \\ \theta_0^{(1)} & \theta_1^{(1)} & \theta_2^{(1)} & \theta_3^{(1)} \\ \theta_0^{(2)} & \theta_1^{(2)} & \theta_2^{(2)} & \theta_3^{(2)} \\ \theta_0^{(3)} & \theta_1^{(3)} & \theta_2^{(3)} & \theta_3^{(3)}  \end{pmatrix} \quad\Rightarrow\quad \Theta^{-1} = \begin{pmatrix} e_{(0)}^0 & e_{(1)}^0 & e_{(2)}^0 & e_{(3)}^0 \\ e_{(0)}^1 & e_{(1)}^1 & e_{(2)}^1 & e_{(3)}^1 \\ e_{(0)}^2 & e_{(1)}^2 & e_{(2)}^2 & e_{(3)}^2 \\ e_{(0)}^3 & e_{(1)}^3 & e_{(2)}^3 & e_{(3)}^3  \end{pmatrix}.
\end{equation}
There are also several useful relations:
\begin{subequations}
\begin{alignat}{5}
  e_{(a)\mu} &= g_{\mu\nu}e_{(a)}^{\nu}, &\qquad \eta_{(a)(b)} &= e_{(a)}^{\mu}e_{(b)\mu}, &\qquad e_{(b)\mu} &= \eta_{(a)(b)}\theta_{\mu}^{(a)},\\
  \theta_{\mu}^{(b)} &= \eta^{(a)(b)}e_{(a)\mu}, & g_{\mu\nu} &= e_{(a)\mu}\theta_{\nu}^{(a)}, & \eta^{(a)(b)} &= \theta_{\mu}^{(a)}\theta_{\nu}^{(b)}g^{\mu\nu}.
\end{alignat}
\end{subequations}

% -----------------------------------------------------------------
%           subsec:  Tetrad transformations
% -----------------------------------------------------------------
\subsection{Tetrad transformations}
Instead of the above found local tetrad that was directly constructed from the spacetime metric, we can also use any other local tetrad
\begin{equation}
  \mathbf{\hat{e}}_{(i)} = A_i^k\mathbf{e}_{(k)},
\end{equation}
where $\mathbf{A}$ is an element of the Lorentz group $O(1,3)$. Hence $\mathbf{A}^T\boldsymbol{\eta}\mathbf{A}=\boldsymbol{\eta}$ and $(\det{\mathbf{A}})^2=1$.

Lorentz-transformation in the direction $n^a=(\sin\chi\cos\xi,\sin\chi\sin\xi,\cos\xi)^T=n_a$ with $\gamma=1/\sqrt{1-\beta^2}$,
\begin{equation}
 \Lambda_0^0 = \gamma,\qquad \Lambda_a^0 = -\beta\gamma n_a,\qquad \Lambda_0^a = -\beta\gamma n^a,\qquad \Lambda_b^a = (\gamma-1)n^an_b+\delta^a_b.
\end{equation}

% -----------------------------------------------------------------
%           subsec:  Connection coefficients with respect to the local tetrad
% -----------------------------------------------------------------
\subsection{Ricci rotation-, connection-, and structure coefficients}
The Ricci rotation coefficients $\gamma_{(i)(j)(k)}$ with respect to the local tetrad $\mathbf{e}_{(i)}$ are defined by 
\begin{equation}
  \label{eq:ricciRotCoeffs}
 \gamma_{(i)(j)(k)} \defequal g_{\mu\lambda}e_{(i)}^{\mu}\nabla_{\mathbf{e}_{(k)}}e_{(j)}^{\lambda} = g_{\mu\lambda} e_{(i)}^{\mu}e_{(k)}^{\nu}\nabla_{\nu}e_{(j)}^{\lambda} = g_{\mu\lambda} e_{(i)}^{\mu}e_{(k)}^{\nu}\left(\partial_{\nu}e_{(j)}^{\lambda}+\Gamma_{\nu\beta}^{\lambda}e_{(j)}^{\beta}\right).
\end{equation}
They are antisymmetric in the first two indices, $\gamma_{(i)(j)(k)} = -\gamma_{(j)(i)(k)}$, which follows from the definition, Eq. (\ref{eq:ricciRotCoeffs}), and the relation
\begin{equation}
 0 = \partial_{\mu}\eta_{(i)(j)} = \nabla_{\mu}\left(g_{\beta\nu}e_{(i)}^{\beta}e_{(j)}^{\nu}\right),
\end{equation}
where $\nabla_{\mu}g_{\beta\nu}=0$, compare \cite{chandrasekhar2006}. Otherwise, we have
\begin{equation}
 ?\gamma^{(i)}_{(j)(k)}? = \theta_{\lambda}^{(i)}e_{(k)}^{\nu}\nabla_{\nu}e_{(j)}^{\lambda} = -e_{(j)}^{\lambda}e_{(k)}^{\nu}\nabla_{\nu}\theta_{\lambda}^{(i)}.
\end{equation}
The contraction of the first and the last index is given by
\begin{equation}
 \gamma_{(j)} = ?\gamma^{(k)}_{(j)(k)}? = \eta^{(k)(i)}\gamma_{(i)(j)(k)} = -\gamma_{(0)(j)(0)}+\gamma_{(1)(j)(1)}+\gamma_{(2)(j)(2)}+\gamma_{(3)(j)(3)} = \nabla_{\nu}e_{(j)}^{\nu}.
\end{equation}

The connection coefficients $\boldsymbol{\omega}_{(j)(n)}^{(m)}$ with respect to the local tetrad $\mathbf{e}_{(i)}$ are defined by
\begin{equation}
 \boldsymbol{\omega}_{(j)(n)}^{(m)}\defequal\theta^{(m)}_{\mu}\nabla_{\mathbf{e}_{(j)}}e_{(n)}^{\mu}= \theta^{(m)}_{\mu}e_{(j)}^{\alpha}\nabla_{\alpha}e_{(n)}^{\mu}=\theta^{(m)}_{\mu}e_{(j)}^{\alpha}\left(\partial_{\alpha}e_{(n)}^{\mu}+\Gamma_{\alpha\beta}^{\mu}e_{(n)}^{\beta}\right),
\end{equation}
compare Nakahara\cite{nakahara}. They are related to the Ricci rotation coefficients via
\begin{equation}
 \gamma_{(i)(j)(k)} = \boldsymbol{\eta}_{(i)(m)}\boldsymbol{\omega}_{(k)(j)}^{(m)}.
\end{equation}

Furthermore, the local tetrad has a non-vanishing Lie-bracket $[X,Y]^{\nu}=X^{\mu}\partial_{\mu}Y^{\nu}-Y^{\mu}\partial_{\mu}X^{\nu}$. Thus,
\begin{equation}
 \left[\mathbf{e}_{(i)},\mathbf{e}_{(j)}\right]=c_{(i)(j)}^{(k)}\mathbf{e}_{(k)}\qquad\text{or}\qquad c_{(i)(j)}^{(k)} = \boldsymbol{\theta}^{(k)}\left[\mathbf{e}_{(i)},\mathbf{e}_{(j)}\right].
\end{equation}
The structure coefficients $c_{(i)(j)}^{(k)}$ are related to the connection coefficients or the Ricci rotation coefficients via
\begin{equation}
 c_{(i)(j)}^{(k)} = \boldsymbol{\omega}_{(i)(j)}^{(k)}-\boldsymbol{\omega}_{(j)(i)}^{(k)} = \boldsymbol{\eta}^{(k)(m)}\left(\gamma_{(m)(j)(i)}-\gamma_{(m)(i)(j)}\right) = ?\gamma^{(k)}_{(j)(i)}?-?\gamma^{(k)}_{(i)(j)}?.
\end{equation}

% -----------------------------------------------------------------
%           subsec:  Riemann tensor with respect to the local tetrad
% -----------------------------------------------------------------
\subsection{Riemann-, Ricci-, and Weyl-tensor with respect to a local tetrad}
The transformations between the coordinate representations of the Riemann-, Ricci-, and Weyl-tensors and their representation with respect to a local tetrad $\mathbf{e}_{(i)}$ are given by
\begin{subequations}
 \begin{align}
   R_{(a)(b)(c)(d)} &= R_{\mu\nu\rho\sigma}e_{(a)}^{\mu}e_{(b)}^{\nu}e_{(c)}^{\rho}e_{(d)}^{\sigma},\\
   R_{(a)(b)} &= R_{\mu\nu}e_{(a)}^{\mu}e_{(b)}^{\nu},\\
   \nonumber C_{(a)(b)(c)(d)} &= C_{\mu\nu\rho\sigma}e_{(a)}^{\mu}e_{(b)}^{\nu}e_{(c)}^{\rho}e_{(d)}^{\sigma}\\
   &= R_{(a)(b)(c)(d)} - \frac{1}{2}\left(\eta_{(a)\antil(c)}R_{(d)\antir(b)}-\eta_{(b)\antil(c)}R_{(d)\antir(a)}\right) + \frac{R}{3}\eta_{(a)\antil(c)}\eta_{(d)\antir(b)}.
 \end{align}
\end{subequations}

% -----------------------------------------------------------------
%           subsec:  Null or timelike directions
% -----------------------------------------------------------------
\subsection{Null or timelike directions}\label{subsec:initDir}
A null or timelike direction $\boldsymbol{\upsilon}=\upsilon^{(i)}\mathbf{e}_{(i)}$ with respect to a local tetrad $\mathbf{e}_{(i)}$ can be written as
\begin{equation}
  \boldsymbol{\upsilon} = \upsilon^{(0)}\mathbf{e}_{(0)} + \psi\left(\sin\chi\cos\xi\,\mathbf{e}_{(1)}+\sin\chi\sin\xi\,\mathbf{e}_{(2)} + \cos\chi\,\mathbf{e}_{(3)}\right) = \upsilon^{(0)}\mathbf{e}_{(0)} + \psi\mathbf{n}.
\end{equation}
In the case of a null direction we have $\psi=1$ and $\upsilon^{(0)}=\pm 1$. A timelike direction can be identified with an initial four-velocity $\mathbf{u}=c\gamma\left(\mathbf{e}_0+\beta\mathbf{n}\right)$, where
\begin{equation}
  \mathbf{u}^2 = \left<\mathbf{u},\mathbf{u}\right>_{\mathbf{g}} = c^2\gamma^2\left<\mathbf{e}_{(0)}+\beta\mathbf{n},\mathbf{e}_{(0)}+\beta\mathbf{n}\right> = c^2\gamma^2\left(-1+\beta^2\right) = \mp c^2,\qquad \sign(\mathbf{g})=\pm 2.
\end{equation}
Thus, $\psi=c\beta\gamma$ and $\upsilon^0=\pm c\gamma$. The sign of $\upsilon^{(0)}$ determines the time direction.
\begin{SCfigure}[1][ht]  
\ifthenelse{\boolean{isARXIV}}{
  \includegraphics[scale=1.0]{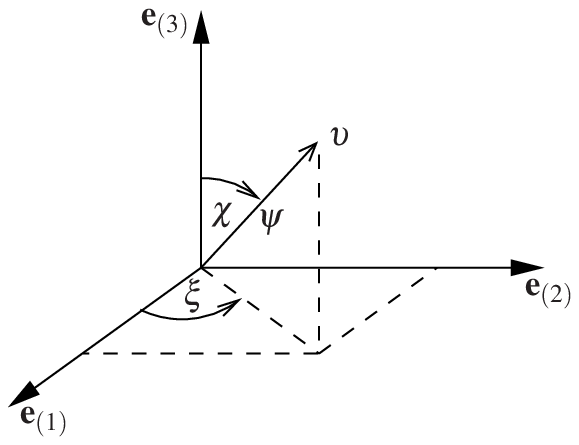}
}{
  \includegraphics[scale=1.0]{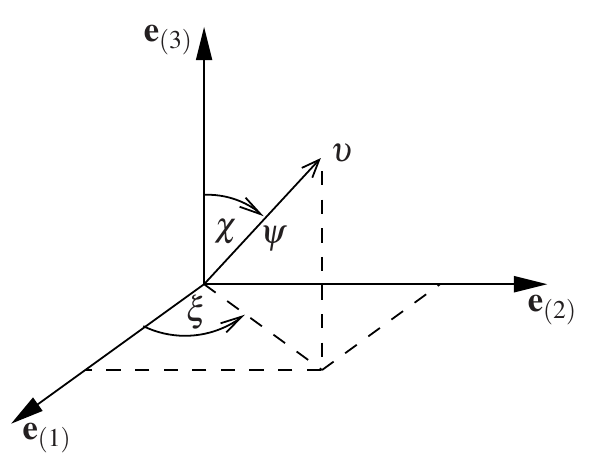}
}
  \caption{Null or timelike direction $\boldsymbol{\upsilon}$ with respect to the local tetrad $\mathbf{e}_{(i)}$.}
\end{SCfigure}

The transformations between a local direction $\upsilon^{(i)}$ and its coordinate representation $\upsilon^{\mu}$ read
\begin{equation}
  \upsilon^{\mu} = \upsilon^{(i)}e_{(i)}^{\mu} \qquad\text{and}\qquad \upsilon^{(i)}=\theta^{(i)}_{\mu}\upsilon^{\mu}.
\end{equation}
% -----------------------------------------------------------------
%           subsec:  Local tetrad for diagonal metrics
% -----------------------------------------------------------------
\subsection{Local tetrad for diagonal metrics}\label{subsec:LTdiagMetrics}
If a spacetime is represented by a diagonal metric
\begin{equation}
  ds^2 = g_{00}(dx^0)^2 + g_{11}(dx^1)^2 + g_{22}(dx^2)^2 + g_{33}(dx^3)^2,
\end{equation}
the natural local tetrad reads
\begin{equation}
 \mathbf{e}_{(0)} = \frac{1}{\sqrt{g_{00}}}\partial_0,\quad \mathbf{e}_{(1)} = \frac{1}{\sqrt{g_{11}}}\partial_1,\quad \mathbf{e}_{(2)} = \frac{1}{\sqrt{g_{22}}}\partial_2,\quad \mathbf{e}_{(3)} = \frac{1}{\sqrt{g_{33}}}\partial_3,
\end{equation}
given that the metric coefficients are well behaved. Analogously, the dual tetrad reads
\begin{equation}
 \boldsymbol{\theta}^{(0)} = \sqrt{g_{00}}\,dx^0,\quad \boldsymbol{\theta}^{(1)} = \sqrt{g_{11}}\,dx^1,\quad \boldsymbol{\theta}^{(2)} = \sqrt{g_{22}}\,dx^2,\quad \boldsymbol{\theta}^{(3)} = \sqrt{g_{33}}\,dx^3.
\end{equation}

% -----------------------------------------------------------------
%           subsec:  stationary axisymmetric spacetimes
% -----------------------------------------------------------------
\subsection{Local tetrad for stationary axisymmetric spacetimes}\label{subsec:axisymST}
The line element of a stationary axisymmetric spacetime is given by
\begin{equation}
  ds^2 = g_{tt}dt^2 + 2g_{t\varphi}\,dt\,d\varphi+g_{\varphi\varphi}d\varphi^2+g_{rr}dr^2+g_{\vartheta\vartheta}d\vartheta^2,
\end{equation}
where the metric components are functions of $r$ and $\vartheta$ only.\\
The local tetrad for an observer on a stationary circular orbit, $(r=\text{const},\vartheta=\text{const})$, with four velocity $\mathbf{u}=c\Gamma\left(\partial_t+\zeta\partial_{\varphi}\right)$ can be defined as, compare Bini\cite{bini2000},
\begin{subequations}
\begin{align}
  \mathbf{e}_{(0)} &=\Gamma\left(\partial_t+\zeta\partial_{\varphi}\right),\qquad \mathbf{e}_{(1)} = \frac{1}{\sqrt{g_{rr}}}\partial_r,\qquad \mathbf{e}_{(2)} = \frac{1}{\sqrt{g_{\vartheta\vartheta}}}\partial_{\vartheta},\\
  \mathbf{e}_{(3)} &= \Delta\Gamma\left[\pm(g_{t\varphi}+\zeta g_{\varphi\varphi})\partial_t\mp(g_{tt}+\zeta g_{t\varphi})\partial_{\varphi}\right],
\end{align}
\end{subequations}
where
\begin{equation}
  \Gamma = \frac{1}{\sqrt{-\left(g_{tt}+2\zeta g_{t\varphi}+\zeta^2g_{\varphi\varphi}\right)}}\qquad\text{and}\qquad\Delta=\frac{1}{\sqrt{g_{t\varphi}^2-g_{tt}g_{\varphi\varphi}}}.
\end{equation}%
The angular velocity $\zeta$ is limited due to $g_{tt}+2\zeta g_{t\varphi}+\zeta^2g_{\varphi\varphi}<0$
\begin{equation}
  \zeta_{\text{min}} = \omega-\sqrt{\omega^2-\frac{g_{tt}}{g_{\varphi\varphi}}}\qquad\text{and}\qquad \zeta_{\text{max}} = \omega+\sqrt{\omega^2-\frac{g_{tt}}{g_{\varphi\varphi}}}
\end{equation}
with $\omega = -g_{t\varphi}/g_{\varphi\varphi}$.%

For $\zeta=0$, the observer is static with respect to spatial infinity. The locally non-rotating frame (LNRF) has angular velocity $\zeta=\omega$, see also MTW\cite{mtw}, exercise 33.3.

Static limit: $\zeta_{\text{min}}=0\quad\Rightarrow\quad g_{tt}=0$.

The transformation between the local direction $\upsilon^{(i)}$ and the coordinate direction $\upsilon^{\mu}$ reads
\begin{equation}
  \upsilon^0 = \Gamma\left(\upsilon^{(0)}\pm \upsilon^{(3)}\Delta w_1\right),\qquad \upsilon^1 = \frac{\upsilon^{(1)}}{\sqrt{g_{rr}}},\qquad \upsilon^2 = \frac{\upsilon^{(2)}}{\sqrt{g_{\vartheta\vartheta}}},\qquad \upsilon^3 = \Gamma\left(\upsilon^{(0)}\zeta\mp \upsilon^{(3)}\Delta w_2\right),
\end{equation}
with
\begin{equation}
  w_1 = g_{t\varphi}+\zeta g_{\varphi\varphi}\qquad\text{and}\qquad w_2 = g_{tt}+\zeta g_{t\varphi}.
\end{equation}
The back transformation reads
\begin{equation}
  \upsilon^{(0)} = \frac{1}{\Gamma}\frac{\upsilon^0w_2+\upsilon^3w_1}{\zeta w_1+w_2},\qquad \upsilon^{(1)} = \sqrt{g_{rr}}\,\upsilon^1,\quad \upsilon^{(2)} = \sqrt{g_{\vartheta\vartheta}}\,\upsilon^2,\qquad \upsilon^{(3)} = \pm\frac{1}{\Delta\Gamma}\frac{\zeta \upsilon^0-\upsilon^3}{\zeta w_1+w_2}.
\end{equation}

Note, to obtain a right-handed local tetrad, $\det\left(e_{(i)}^{\mu}\right)>0$, the upper sign has to be used.

% -----------------------------------------------------------------
%    sec:  Newman-Penrose tetrad
% -----------------------------------------------------------------
\section{Newman-Penrose tetrad and spin-coefficients}\label{sec:newmanPenrose}
\setcounter{equation}{0}
The Newman-Penrose tetrad consists of four null vectors $\mathbf{e}^{\star}_{(i)}=\left\{\mathbf{l}, \mathbf{n}, \mathbf{m}, \mathbf{\bar{m}}\right\}$, where $\mathbf{l}$ and $\mathbf{n}$ are real and $\mathbf{m}$ and $\mathbf{\bar{m}}$ are complex conjugates; see Penrose and Rindler\cite{penroseRindler} or Chandrasekhar\cite{chandrasekhar2006} for a thorough discussion. The Newman-Penrose (NP) tetrad has to fulfill the orthonormality relation
\begin{equation}
  \left<\mathbf{e}^{\star}_{(i)},\mathbf{e}^{\star}_{(j)}\right> = \boldsymbol{\eta}^{\star}_{(i)(j)}\qquad\text{with}\qquad \boldsymbol{\eta}^{\star}_{(i)(j)} = \left(\begin{array}{cccc} 0 & 1 & 0 & 0 \\ 1 & 0 & 0 & 0 \\ 0 & 0 & 0 & -1 \\ 0 & 0 & -1 & 0\end{array}\right). 
\end{equation}
A straightforward relation between the NP tetrad and the natural local tetrad, as discussed in  Sec.~\ref{sec:natLocTetrad}, is given by
\begin{equation}
  \mathbf{l}=\mp\frac{1}{\sqrt{2}}\left(\mathbf{e}_{(0)}+\mathbf{e}_{(1)}\right),\quad \mathbf{n}=\mp\frac{1}{\sqrt{2}}\left(\mathbf{e}_{(0)}-\mathbf{e}_{(1)}\right),\quad \mathbf{m}=\mp\frac{1}{\sqrt{2}}\left(\mathbf{e}_{(2)}+i\mathbf{e}_{(3)}\right),
\end{equation}
where the upper/lower sign has to be used for metrics with positive/negative signature. The Ricci rotation-coefficients of a NP tetrad are now called {\it spin coefficients} and are designated by specific symbols:
\begin{subequations}
\begin{alignat}{5}
 \kappa &= \gamma_{(2)(1)(1)}, &\qquad \rho &= \gamma_{(2)(0)(3)}, &\qquad \epsilon &= \frac{1}{2}\left(\gamma_{(1)(0)(0)}+\gamma_{(2)(3)(0)}\right),\\
 \sigma &= \gamma_{(2)(0)(2)}, &\qquad \mu &= \gamma_{(1)(3)(2)}, &\qquad \gamma &= \frac{1}{2}\left(\gamma_{(1)(0)(1)}+\gamma_{(2)(3)(1)}\right),\\
 \lambda &= \gamma_{(1)(3)(3)}, &\qquad \tau &= \gamma_{(2)(0)(1)}, &\qquad \alpha &= \frac{1}{2}\left(\gamma_{(1)(0)(3)}+\gamma_{(2)(3)(3)}\right),\\
 \nu &= \gamma_{(1)(3)(1)}, &\qquad \pi &= \gamma_{(1)(3)(0)}, &\qquad \beta &= \frac{1}{2}\left(\gamma_{(1)(0)(2)}+\gamma_{(2)(3)(2)}\right).
\end{alignat}
\end{subequations} 

% -----------------------------------------------------------------
%    sec:   coordinate relations
% -----------------------------------------------------------------
\section{Coordinate relations}\label{sec:coordRels}
\setcounter{equation}{0}
% -----------------------------------------------------------------
%    subsec:         subRelation between spherical and Cartesian coordinates
% -----------------------------------------------------------------
\subsection{Spherical and Cartesian coordinates}\label{subsec:sphCartRel}
\begin{SCfigure}[1][ht]
\ifthenelse{\boolean{isARXIV}}{
  \includegraphics[scale=1.0]{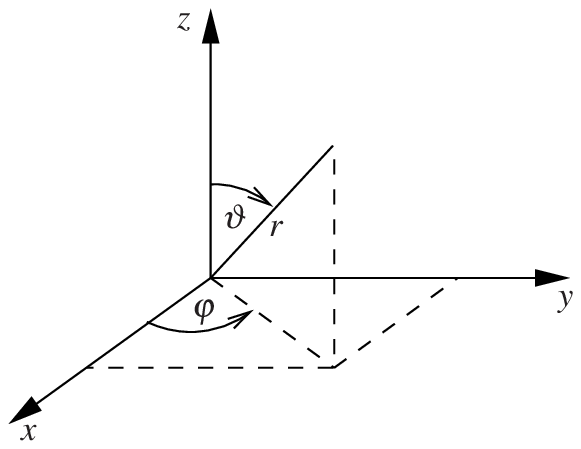}
}{
  \includegraphics[scale=1.0]{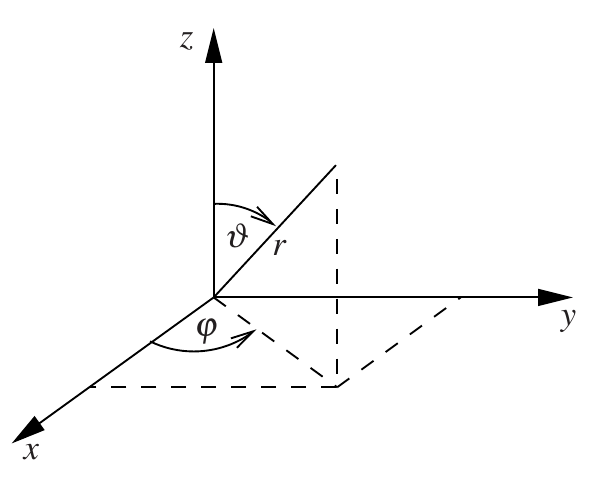}
}
  \caption{\label{fig:relSphCart}Relation between spherical and Cartesian coordinates.}
\end{SCfigure}

The well-known relation between the spherical coordinates $(r,\vartheta,\varphi)$ and the Cartesian coordinates $(x,y,z)$, compare Fig. \ref{fig:relSphCart}, are
\begin{equation}
    x = r\sin\vartheta\cos\varphi, \qquad y = r\sin\vartheta\sin\varphi, \qquad  z = r\cos\vartheta,
\end{equation}
and
\begin{equation}
    r = \sqrt{x^2+y^2+z^2}, \qquad \vartheta = \arctan2(\sqrt{x^2+y^2},z), \qquad   \varphi = \arctan2(y,x),
\end{equation}
where $\arctan2()$ ensures that $\varphi\in [0,2\pi)$ and $\vartheta\in (0,\pi)$.

The total differentials of the spherical coordinates read
\begin{equation}
  dr         = \frac{x\,dx+y\,dy + z\,dz}{r}, \qquad
  d\vartheta = \frac{xz\,dx+yz\,dy-(x^2+y^2)dz}{r^2\sqrt{x^2+y^2}},\qquad
  d\varphi   = \frac{-y\,dx+x\,dy}{x^2+y^2},
\end{equation}
whereas the coordinate derivatives read
\begin{subequations}
\begin{alignat}{3}
  \partial_r &= \pdiff{x}{r}\partial_x+\pdiff{y}{r}\partial_y+\pdiff{z}{r}\partial_z  &\, &= \sin\vartheta\cos\varphi\,\partial_x+\sin\vartheta\sin\varphi\,\partial_y+\cos\vartheta\,\partial_z,\\[0.5em]
  \partial_{\vartheta} &= \pdiff{x}{\vartheta}\partial_x+\pdiff{y}{\vartheta}\partial_y+\pdiff{z}{\vartheta}\partial_z &\, &= r\cos\vartheta\cos\varphi\,\partial_x+r\cos\vartheta\sin\varphi\,\partial_y-r\sin\vartheta\,\partial_z,\\[0.5em]
  \partial_{\varphi} &= \pdiff{x}{\varphi}\partial_x+\pdiff{y}{\varphi}\partial_y+\pdiff{z}{\varphi}\partial_z &\, &= -r\sin\vartheta\sin\varphi\,\partial_x+r\sin\vartheta\cos\varphi\,\partial_y,
\end{alignat} 
\end{subequations}
and
\begin{subequations}
\begin{alignat}{3}
  \partial_x &= \pdiff{r}{x}\partial_r+\pdiff{\vartheta}{x}\partial_{\vartheta}+\pdiff{\varphi}{x}\partial_{\varphi} &\, &= \sin\vartheta\cos\varphi\,\partial_r+\frac{\cos\vartheta\cos\varphi}{r}\,\partial_{\vartheta}-\frac{\sin\varphi}{r\sin\vartheta}\partial_{\varphi},\\[0.5em]
  \partial_y &= \pdiff{r}{y}\partial_r+\pdiff{\vartheta}{y}\partial_{\vartheta}+\pdiff{\varphi}{y}\partial_{\varphi} &\, &= \sin\vartheta\sin\varphi\,\partial_r+\frac{\cos\vartheta\sin\varphi}{r}\,\partial_{\vartheta}+\frac{\cos\varphi}{r\sin\vartheta}\partial_{\varphi},\\[0.5em]
  \partial_z &= \pdiff{r}{z}\partial_r+\pdiff{\vartheta}{z}\partial_{\vartheta}+\pdiff{\varphi}{z}\partial_{\varphi} &\, &= \cos\vartheta\,\partial_r-\frac{\sin\vartheta}{r}\,\partial_{\vartheta}.
\end{alignat}
\end{subequations}

% -----------------------------------------------------------------
%           subsec:  Relation between cylindrical and Cartesian coordinates
% -----------------------------------------------------------------
\subsection{Cylindrical and Cartesian coordinates}\label{subsec:cylCartRel}
The relation between cylindrical coordinates $(r,\varphi,z)$ and Cartesian coordinates $(x,y,z)$ is given by%
\begin{equation}
   x = r\cos\varphi, \quad y = r\sin\varphi,\qquad\text{and}\qquad r = \sqrt{x^2+y^2},\quad \varphi = \arctan2(y,x),
\end{equation}
where $\arctan2()$ again ensures that the angle $\varphi\in [0,2\pi)$.%

\begin{SCfigure}[1][ht]
\ifthenelse{\boolean{isARXIV}}{
  \includegraphics[scale=1.0]{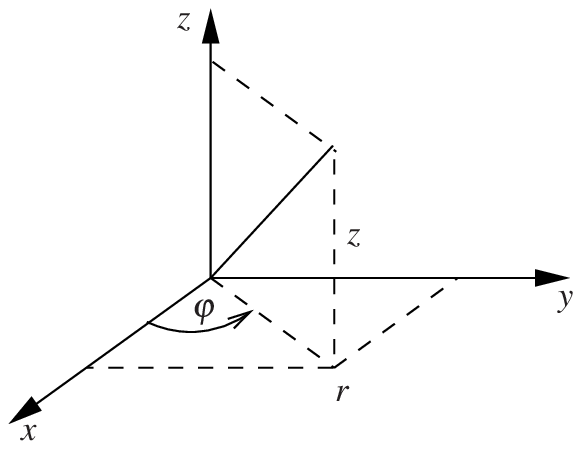}
}{
  \includegraphics[scale=1.0]{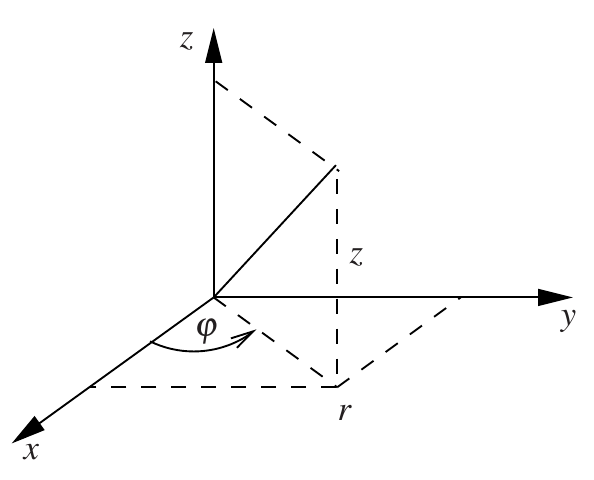}
}
  \caption{Relation between cylindrical and Cartesian coordinates.}
\end{SCfigure}

The total differentials of the spherical coordinates are given by%
\begin{equation}
  dr  = \frac{x\,dx+y\,dy}{r}, \qquad d\varphi = \frac{-y\,dx+x\,dy}{r^2},
\end{equation}
and
\begin{equation}
  dx  = \cos\varphi\, dr - r\sin\varphi\, d\varphi,\qquad  dy  = \sin\varphi\, dr + r\cos\varphi\, d\varphi.
\end{equation}
The coordinate derivatives are
\begin{subequations}
\begin{alignat}{3}
  \partial_r &= \pdiff{x}{r}\partial_x+\pdiff{y}{r}\partial_y &\, &= \cos\varphi\,\partial_x+\sin\varphi\,\partial_y,\\[0.5em]
  \partial_{\varphi} &= \pdiff{x}{\varphi}\partial_x+\pdiff{y}{\varphi}\partial_y&\, &= -r\sin\varphi\,\partial_x+r\cos\varphi\,\partial_ym
\end{alignat} 
\end{subequations}
and
\begin{subequations}
\begin{alignat}{3}
  \partial_x &= \pdiff{r}{x}\partial_r+\pdiff{\varphi}{x}\partial_{\varphi} &\, &= \cos\varphi\,\partial_r-\frac{\sin\varphi}{r}\,\partial_y,\\[0.5em]
  \partial_y &= \pdiff{r}{y}\partial_r+\pdiff{\varphi}{y}\partial_{\varphi} &\, &= \sin\varphi\,\partial_r+\frac{\cos\varphi}{r}\,\partial_y.
\end{alignat} 
\end{subequations}

% -----------------------------------------------------------------
%    sec:   Embedding diagram
% -----------------------------------------------------------------
\section{Embedding diagram}\label{sec:embedding}
\setcounter{equation}{0}
A two-dimensional hypersurface with line segment
\begin{equation}
 d\sigma^2 = g_{rr}(r)dr^2+g_{\varphi\varphi}(r)d\varphi^2
\end{equation}
can be embedded in a three-dimensional Euclidean space with cylindrical coordinates,
\begin{equation}
 d\sigma^2 = \left[1+\left(\frac{dz}{d\rho}\right)^2\right]d\rho^2+\rho^2 d\varphi^2.
\end{equation}
With $\rho(r)^2=g_{\varphi\varphi}(r)$ and $dr=(dr/d\rho)d\rho$, we obtain for the embedding function $z=z(r)$,
\begin{equation}
 \frac{dz}{dr} = \pm\sqrt{g_{rr}-\left(\frac{d\sqrt{g_{\varphi\varphi}}}{dr}\right)^2}.
\end{equation}
If $g_{\varphi\varphi}(r)=r^2$, then $d\sqrt{g_{\varphi\varphi}}/dr = 1$.

% -----------------------------------------------------------------
%    sec:   Equations of motion
% -----------------------------------------------------------------
\section{Equations of motion and transport equations}\label{sec:eqmotion}
\setcounter{equation}{0}

% -----------------------------------------------------------------
%           subsec:  Geodesic equation
% -----------------------------------------------------------------
\subsection{Geodesic equation}
The geodesic equation reads
\begin{equation}
  \label{eq:geodEq}
  \frac{D^2x^{\mu}}{d\lambda^2} = \frac{d^2x^{\mu}}{d\lambda^2} + \Gamma_{\rho\sigma}^{\mu}\frac{dx^{\rho}}{d\lambda}\frac{dx^{\sigma}}{d\lambda} = 0
\end{equation}
with the affine parameter $\lambda$. For timelike geodesics, however, we replace the affine parameter by the proper time $\tau$.

The geodesic equation (\ref{eq:geodEq}) is a system of ordinary differential equations of second order. Hence, to solve these differential equations, we need an initial position $x^{\mu}(\lambda=0)$ as well as an initial direction $(dx^{\mu}/d\lambda)(\lambda=0)$. This initial direction has to fulfill the constraint equation
\begin{equation}
  \label{eq:constrEq}
  g_{\mu\nu}\frac{dx^{\mu}}{d\lambda}\frac{dx^{\nu}}{d\lambda} = \kappa c^2,
\end{equation}
where $\kappa=0$ for lightlike and $\kappa=\mp 1$, $(\sign(\mathbf{g})=\pm 2)$, for timelike geodesics.

The initial direction can also be determined by means of a local reference frame, compare sec. \ref{subsec:initDir}, that automatically fulfills the constraint equation (\ref{eq:constrEq}). If we use the natural local tetrad as local reference frame, we have
\begin{equation}
 \frac{dx^{\mu}}{d\lambda}\bigg|_{\lambda=0}=\upsilon^{\mu}=\upsilon^{(i)}e_{(i)}^{\mu}.
\end{equation}

% -----------------------------------------------------------------
%           subsec:  Fermi-Walker transport
% -----------------------------------------------------------------
\subsection{Fermi-Walker transport}
The Fermi-Walker transport, see e.g. Stephani\cite{stephani1990}, of a vector $\mathbf{X}=X^{\mu}\partial_{\mu}$ along the worldline $x^{\mu}(\tau)$ with four-velocity $\mathbf{u}=u^{\mu}(\tau)\partial_{\mu}$ is given by $\mathbb{F}_{\mathbf{u}}X^{\mu}=0$ with
\begin{equation}
  \mathbb{F}_{\mathbf{u}}X^{\mu}\defequal\frac{dX^{\mu}}{d\tau} + \Gamma_{\rho\sigma}^{\mu}u^{\rho}X^{\sigma} + \frac{1}{c^2}\left(u^{\sigma}a^{\mu}-a^{\sigma}u^{\mu}\right)g_{\rho\sigma}X^{\rho}.
\end{equation}
The four-acceleration follows from the four-velocity via
\begin{equation}
  a^{\mu} = \frac{D^2x^{\mu}}{d\tau^2}=\frac{Du^{\mu}}{d\tau} = \frac{du^{\mu}}{d\tau} + \Gamma_{\rho\sigma}^{\mu}u^{\rho}u^{\sigma}.
\end{equation}

% -----------------------------------------------------------------
%           subsec:  Parallel transport
% -----------------------------------------------------------------
\subsection{Parallel transport}
If the four-acceleration vanishes, the Fermi-Walker transport simplifies to the parallel transport $\mathbb{P}_{\mathbf{u}}X^{\mu}=0$ with
\begin{equation}
  \mathbb{P}_{\mathbf{u}}X^{\mu}\defequal\frac{DX^{\mu}}{d\tau} = \frac{dX^{\mu}}{d\tau} + \Gamma_{\rho\sigma}^{\mu}u^{\rho}X^{\sigma}.
\end{equation}

% -----------------------------------------------------------------
%           subsec:  Euler Lagrange
% -----------------------------------------------------------------
\subsection{Euler-Lagrange formalism}\label{subsec:EL}
A detailed discussion of the Euler-Lagrange formalism can be found, e.g., in Rindler\cite{rindler}. The Lagrangian $\mathcal{L}$ is defined as
\begin{equation}
 \mathcal{L} \defequal g_{\mu\nu}\dot{x}^{\mu}\dot{x}^{\nu},\qquad \mathcal{L}\stackrel{!}{=}\kappa c^2,
\end{equation}
where $x^{\mu}$ are the coordinates of the metric, and the dot means differentiation with respect to the affine parameter $\lambda$. For timelike geodesics, $\kappa=\mp 1$ depending on the signature of the metric, $\sign(\mathbf{g})=\pm 2$. For lightlike geodesics, $\kappa=0$.\\[0.3em]
The Euler-Lagrange equations read
\begin{equation}
 \frac{d}{d\lambda}\frac{\partial\mathcal{L}}{\partial\dot{x}^{\mu}}-\frac{\partial\mathcal{L}}{\partial x^{\mu}} = 0.
\end{equation}
If $\mathcal{L}$ is independent of $x^{\rho}$, then $x^{\rho}$ is a cyclic variable and
\begin{equation}
  p_{\rho} = g_{\rho\nu}\dot{x}^{\nu} = \text{const}.
\end{equation}
Note that $\evalUnits{\mathcal{L}} = \frac{\uLength^2}{\uTime^2}$ for timelike and $\evalUnits{\mathcal{L}} = 1$ for lightlike geodesics, see Sec.~\ref{sec:units}.

% -----------------------------------------------------------------
%           subsec:  Hamilton
% -----------------------------------------------------------------
\subsection{Hamilton formalism}\label{subsec:Hamilton}
The super-Hamiltonian $\mathcal{H}$ is defined as
\begin{equation}
  \mathcal{H} \defequal \frac{1}{2}g^{\mu\nu}p_{\mu}p_{\nu},\qquad \mathcal{H}\stackrel{!}{=}\frac{1}{2}\kappa c^2,
\end{equation}
where $p_{\mu}=g_{\mu\nu}\dot{x}^{\nu}$ are the canonical momenta, see e.g. MTW\cite{mtw}, para.~21.1. As in classical mechanics, we have
\begin{equation}
   \frac{dx^{\mu}}{d\lambda}=\frac{\partial\mathcal{H}}{\partial p_{\mu}}\qquad\text{and}\qquad \frac{dp_{\mu}}{d\lambda}=-\frac{\partial\mathcal{H}}{\partial x^{\mu}}.
\end{equation}

% -----------------------------------------------------------------
%    sec:   Units
% -----------------------------------------------------------------
\section{Units}\label{sec:units}
\setcounter{equation}{0}
A first test in analyzing whether an equation is correct is to check the units. Newton's gravitational constant $G$, for example, has the following units
\begin{equation}
 \evalUnits{G} = \frac{\uLength^3}{\uMass\cdot\uTime^2},
\end{equation}
where $\evalUnits{\cdot}$ indicates that we evaluate the units of the enclosed expression. Further examples are
\begin{equation}
 \evalUnits{ds} = \uLength,\qquad \evalUnits{\mathbf{u}} = \frac{\uLength}{\uTime},\qquad\evalUnits{R_{trtr}^{\text{\tiny Schwarzschild}}} = \frac{1}{\uTime^2},\qquad\evalUnits{R_{\vartheta\varphi\vartheta\varphi}^{\text{\tiny Schwarzschild}}} = \uLength^2.
\end{equation}

% -----------------------------------------------------------------
%    sec:   Tools
% -----------------------------------------------------------------
\section{Tools}\label{sec:tools}
\setcounter{equation}{0}
% -----------------------------------------------------------------
%           subsec:  Maple/GRTensorII
% -----------------------------------------------------------------
\subsection{Maple/GRTensorII}
The Christoffel symbols, the Riemann- and Ricci-tensors as well as the Ricci and Kretschmann scalars in this catalogue were determined by means of the software Maple together with the GRTensorII package by Musgrave, Pollney, and Lake.\footnote{The commercial software Maple can be found here: {http://www.maplesoft.com}. The GRTensorII-package is free: {http://grtensor.phy.queensu.ca}.}\\[1em]
A typical worksheet to enter a new metric may look like this:
{\verbfont
\begin{verbatim}
  > grtw();
  > makeg(Schwarzschild);

  Makeg 2.0: GRTensor metric/basis entry utility
  To quit makeg, type 'exit' at any prompt.
  Do you wish to enter a 1) metric [g(dn,dn)],
                         2) line element [ds],
                         3) non-holonomic basis [e(1)...e(n)], or
                         4) NP tetrad [l,n,m,mbar]?
  > 2:

  Enter coordinates as a LIST (eg. [t,r,theta,phi]):
  > [t,r,theta,phi]:

  Enter the line element using d[coord] to indicate differentials.
  (for example,  r^2*(d[theta]^2 + sin(theta)^2*d[phi]^2)
  [Type 'exit' to quit makeg]
  ds^2 = 

  If there are any complex valued coordinates, constants or functions
  for this spacetime, please enter them as a SET ( eg. { z, psi } ).

  Complex quantities [default={}]: 
  > {}:

  You may choose to 0) Use the metric WITHOUT saving it,
                    1) Save the metric as it is,
                    2) Correct an element of the metric,
                    3) Re-enter the metric,
                    4) Add/change constraint equations, 
                    5) Add a text description, or
                    6) Abandon this metric and return to Maple.
  > 0: 
\end{verbatim}
}
The worksheets for some of the metrics in this catalogue can be found on the authors homepage.
To determine the objects that are defined with respect to a local tetrad, the metric must be given as non-holonomic basis.\\
The various basic objects can be determined via
\vspace*{-0.5cm}

\begin{tabbing}\hspace*{0.5cm}\=\hspace*{5cm}\=\hspace*{4.5cm}\=\\
 \>Christoffel symbols $\Gamma_{\nu\rho}^{\mu}$ \> \verb+grcalc(Chr2);+ \> \verb+grcalc(Chr(dn,dn,up));+\\
 \>partial derivatives $\Gamma_{\nu\rho,\sigma}^{\mu}$ \> \> \verb+grcalc(Chr(dn,dn,up,pdn));+\\
 \>Riemann tensor $R_{\mu\nu\rho\sigma}$ \> \verb+grcalc(Riemman);+ \> \verb+grcalc(R(dn,dn,dn,dn));+\\
 \>Ricci tensor $R_{\mu\nu}$ \> \verb+grcalc(Ricci);+ \> \verb+grcalc(R(dn,dn));+\\
 \>Ricci scalar $\mathcal{R}$ \> \verb+grcalc(Ricciscalar);+\\
 \>Kretschmann scalar $\mathcal{K}$ \> \verb+grcalc(RiemSq);+\\
\end{tabbing}

% -----------------------------------------------------------------
%           subsec:  Mathematica
% -----------------------------------------------------------------
\subsection{Mathematica}
The calculation of the Christoffel symbols, the Riemann- or Ricci-tensor within {\sl Mathematica} could read like this:
{\verbfont
\begin{verbatim}
  Clearing the values of symbols:
      In[1]:=  Clear[coord, metric, inversemetric, affine, 
                      t, r, Theta, Phi]

  Setting the dimension:
      In[2]:=  n := 4

  Defining a list of coordinates:
      In[3]:=  coord := {t, r, Theta, Phi}

  Defining the metric:
      In[4]:=  metric := {{-(1 - rs/r) c^2, 0, 0, 0}, 
                          {0, 1/(1 - rs/r), 0, 0}, 
                          {0, 0, r^2, 0}, 
                          {0, 0, 0, r ^2 Sin[Theta]^2}}
      In[5]:=  metric // MatrixForm

  Calculating the inverse metric:
      In[6]:=  inversemetric := Simplify[Inverse[metric]]
    
      In[7]:=  inversemetric // MatrixForm

  Calculating the Christoffel symbols of the second kind:
      In[8]:=  affine :=  affine = Simplify[
        Table[(1/2) Sum[inversemetric[[Mu, Rho]] (
           D[metric[[Rho, Nu]], coord[[Lambda]]] + 
           D[metric[[Rho, Lambda]], coord[[Nu]]] - 
           D[metric[[Nu, Lambda]], coord[[Rho]]]), 
        {Rho, 1, n}], {Nu, 1, n}, {Lambda, 1, n}, {Mu, 1, n}]]

  Displaying the Christoffel symbols of the second kind:
      In[9]:=  listaffine := 
        Table[If[UnsameQ[affine[[Nu, Lambda, Mu]], 0], 
          {Style[ Subsuperscript[\[CapitalGamma], 
             Row[{coord[[Nu]], coord[[Lambda]]}], coord[[Mu]]], 18],
             "=", 
             Style[affine[[Nu, Lambda, Mu]], 14]}], 
          {Lambda, 1, n}, {Nu, 1, Lambda}, {Mu, 1, n}]

     In[10]:=  TableForm[Partition[DeleteCases[Flatten[listaffine], 
                                               Null], 3], 
                         TableSpacing -> {1, 2}]

  Defining the Riemann tensor:
     In[11]:=  riemann :=  riemann = 
      Table[D[affine[[Nu, Sigma, Mu]], coord[[Rho]]] - 
            D[affine[[Nu, Rho, Mu]], coord[[Sigma]]] + 
            Sum[affine[[Rho, Lambda, Mu]]
                affine[[Nu, Sigma, Lambda]] - 
                affine[[Sigma, Lambda, Mu]] 
                affine[[Nu, Rho, Lambda]], 
              {Lambda, 1, n}], 
      {Mu, 1, n}, {Nu, 1, n}, {Rho, 1, n}, {Sigma, 1, n}]

  Defining the Riemann tensor with lower indices:
     In[12]:= riemannDn :=  riemannDn = 
       Table[Simplify[
          Sum[metric[[Mu, Kappa]] riemann[[Kappa, Nu, Rho, Sigma]], 
          {Kappa, 1, n}]], 
       {Mu, 1, n}, {Nu, 1, n}, {Rho, 1, n}, {Sigma, 1, n}]

    In[13]:= listRiemann := 
       Table[If[UnsameQ[riemannDn[[Mu, Nu, Rho, Sigma]], 0], 
      {Style[Subscript[R, Row[{coord[[Mu]], coord[[Nu]], coord[[Rho]], 
        coord[[Sigma]]}]], 16], "=", 
        riemannDn[[Mu, Nu, Rho, Sigma]]}],
      {Nu, 1, n}, {Mu, 1, Nu}, {Sigma, 1, n}, {Rho, 1, Sigma}]
    
    In[14]:= TableForm[Partition[DeleteCases[Flatten[listRiemann],
                                             Null], 3], 
                       TableSpacing -> {2, 2}]

  Defining the Ricci tensor:
    In[15]:= ricci := ricci = 
      Table[Simplify[
       Sum[riemann[[Rho, Mu, Rho, Nu]], {Rho, 1, n}]], 
      {Mu, 1, n}, {Nu, 1, n}]

    In[16]:= listRicci := 
     Table[If[UnsameQ[ricci[[Mu, Nu]], 0], 
        {Style[Subscript[R, Row[{coord[[Mu]], coord[[Nu]]}]], 16], 
        "=", 
       Style[ricci[[Mu, Nu]], 16]}], {Nu, 1, 4}, {Mu, 1, Nu}]

    In[17]:= TableForm[Partition[DeleteCases[Flatten[listRicci],
                                             Null], 3], 
                       TableSpacing -> {1, 2}]

  Defining the Ricci scalar:
    In[18]:= ricciscalar :=  ricciscalar = 
      Simplify[Sum[
        Sum[inversemetric[[Mu, Nu]] ricci[[Nu, Mu]], 
        {Mu, 1, n}], {Nu, 1, n}]]

  Defining the Kretschmann scalar:
    In[19]:= riemannUp :=  riemannUp = 
      Table[Simplify[
       Sum[inversemetric[[Nu, Kappa]] 
             riemann[[Mu, Kappa, Rho, Sigma]], {Kappa, 1, n}]], 
      {Mu, 1, n}, {Nu, 1, n}, {Rho, 1, n}, {Sigma, 1, n}]

    In[20]:= kretschmann := kretschmann = 
       Simplify[Sum[ Sum[Sum[Sum[
          riemannUp[[Mu, Nu, Rho, Sigma]] 
          riemannUp[[Rho, Sigma, Mu, Nu]], 
         {Mu, 1, n}], {Nu, 1, n}], {Rho, 1, n}], {Sigma, 1, n}]]
\end{verbatim}
}

Some example notebooks can be found on the authors homepage.

% -----------------------------------------------------------------
%           subsec:  Maxima
% -----------------------------------------------------------------
\subsection{Maxima}
Instead of using commercial software like {\sl Maple} or {\sl Mathematica}, Maxima also offers a tensor package that helps to calculate the Christoffel symbols etc. The above example for the Schwarz\-schild metric can be written as a maxima worksheet as follows:
{\verbfont
\begin{verbatim}
  /* load ctensor package */
  load(ctensor);

  /* define coordinates to use */
  ct_coords:[t,r,theta,phi];

  /* start with the identity metric */
  lg:ident(4);
  lg[1,1]:c^2*(1-rs/r);
  lg[2,2]:-1/(1-rs/r);
  lg[3,3]:-r^2;
  lg[4,4]:-r^2*sin(theta)^2;
  cmetric();

  /* calculate the christoffel symbols of the second kind */
  christof(mcs);

  /* calculate the riemann tensor */
  lriemann(mcs);

  /* calculate the ricci tensor */
  ricci(mcs);

  /* calculate the ricci scalar */
  scurvature();

  /* calculate the Kretschmann scalar */
  uriemann(mcs);
  rinvariant();
  ratsimp(%);
\end{verbatim}
}

As you may have noticed, the Schwarzschild metric must be given with negative signature.

% -----------------------------------------------------------------
%      Spacetimes
% -----------------------------------------------------------------
\chapter{Spacetimes}
% -----------------------------------------------------------------
%       M i n k o w s k i
% -----------------------------------------------------------------
\section{Minkowski}
\setcounter{equation}{0}
\ifthenelse{\boolean{isARXIV}}{
% ******** Start of file minkowski.tex *********
%
%  Copyright (c) 2009 Thomas Mueller,
%                     Universitaet Stuttgart, VISUS
%

% -------------------------------------------------------------------
%   cartesian coordinates
% -------------------------------------------------------------------
\subsection{Cartesian coordinates}
The Minkowski metric in Cartesian coordinates $\left\{t,x,y,z\in\setR\right\}$ reads
\metricEq{
  ds^2 = -c^2dt^2 + dx^2 + dy^2 + dz^2.
}{minkCart}
All Christoffel symbols as well as the Riemann- and Ricci-tensor vanish identically.
\noindent The natural local tetrad is trivial,
\begin{equation}
  \mathbf{e}_{(t)} = \frac{1}{c}\partial_t,\qquad \mathbf{e}_{(x)} = \partial_x,\qquad \mathbf{e}_{(y)} = \partial_y,\qquad \mathbf{e}_{(z)} = \partial_z,
\end{equation}
with dual
\begin{equation}
  \boldsymbol{\theta}^{(t)} = c\,dt,\qquad \boldsymbol{\theta}^{(x)} = dx,\qquad \boldsymbol{\theta}^{(y)} = dy,\qquad \boldsymbol{\theta}^{(z)} = dz.
\end{equation}

% -------------------------------------------------------------------
%   cylindrical coordinates
% -------------------------------------------------------------------
\subsection{Cylindrical coordinates}
The Minkowski metric in cylindrical coordinates $\left\{t\in\setR,r\in\setR^{+}, \varphi\in[0,2\pi),z\in\setR\right\}$,
\metricEq{
  ds^2 = -c^2dt^2 + dr^2 + r^2d\varphi^2 + dz^2,
}{minkCyl}
has the natural local tetrad
\begin{equation}
  \mathbf{e}_{(t)} = \frac{1}{c}\partial_t,\qquad \mathbf{e}_{(r)} = \partial_r,\qquad \mathbf{e}_{(\varphi)} = \frac{1}{r}\partial_{\varphi},\qquad \mathbf{e}_{(z)} = \partial_z.
\end{equation}
%% -------------------- Christoffel symbols --------------------
\SecChristoffel
\begin{equation}
  \Gamma_{\varphi\varphi}^r = -r,\qquad \Gamma_{r\varphi}^{\varphi} = \frac{1}{r}.
\end{equation}

Partial derivatives
\begin{equation}
  \Gamma_{r\varphi,r}^{\varphi} = -\frac{1}{r^2},\qquad \Gamma_{\varphi\varphi,r}^r = -1.
\end{equation}

%% -------------------- Ricci rotation coefficients --------------------
\SecRicRotCoef
\begin{equation}
  \gamma_{(\varphi)(r)(\varphi)} = \frac{1}{r}\qquad\text{and}\qquad \gamma_{(r)} = \frac{1}{r}.
\end{equation}

% -------------------------------------------------------------------
%   spherical coordinates
% -------------------------------------------------------------------
\subsection{Spherical coordinates}
In spherical coordinates $\left\{t\in\setR,r\in\setR^{+},\vartheta\in(0,\pi),\varphi\in[0,2\pi)\right\}$, the Minkowski metric reads
\metricEq{
  ds^2 = -c^2dt^2 + dr^2 + r^2\left(d\vartheta^2+\sin^2\vartheta d\varphi^2\right).
}{minkSph}

%% -------------------- Christoffel symbols --------------------
\SecChristoffel
\begin{subequations}
\begin{alignat}{5}
  \Gamma_{\vartheta\vartheta}^r &= -r, &\qquad \Gamma_{\varphi\varphi}^r &= -r\sin^2\vartheta, &\qquad \Gamma_{r\vartheta}^{\vartheta} &= \frac{1}{r},\\
  \Gamma_{\varphi\varphi}^{\vartheta} &= -\sin\vartheta\cos\vartheta, & \Gamma_{r\varphi}^{\varphi} &= \frac{1}{r}, & \Gamma_{\vartheta\varphi}^{\varphi} &= \cot\vartheta.
\end{alignat}
\end{subequations}

Partial derivatives
\begin{subequations}
 \begin{alignat}{5}
    \Gamma_{r\vartheta,r}^{\vartheta} &= -\frac{1}{r^2}, &\qquad \Gamma_{r\varphi,r}^{\varphi} &= -\frac{1}{r^2}, &\qquad \Gamma_{\vartheta\vartheta,r}^r &= -1,\\    
    \Gamma_{\vartheta\varphi,\vartheta}^{\varphi} &= -\frac{1}{\sin^2\vartheta}, & \Gamma_{\varphi\varphi,r}^r &= -\sin^2\vartheta, & \Gamma_{\varphi\varphi,\vartheta}^{\vartheta} &= -\cos(2\vartheta),\\
    \Gamma_{\varphi\varphi,\vartheta}^r &= -\sin(2\vartheta).
 \end{alignat}
\end{subequations}

%% -------------------- Local tetrad --------------------
\SecLocal
\begin{equation}
  \mathbf{e}_{(t)} = \frac{1}{c}\partial_t,\qquad \mathbf{e}_{(r)} = \partial_r,\qquad \mathbf{e}_{(\vartheta)} = \frac{1}{r}\partial_{\vartheta},\qquad \mathbf{e}_{(\varphi)} = \frac{1}{r\sin\vartheta}\partial_{\varphi}.
\end{equation}

%% -------------------- Ricci rotation coefficients --------------------
\SecRicRotCoef
\begin{equation}
  \gamma_{(\vartheta)(r)(\vartheta)} = \gamma_{(\varphi)(r)(\varphi)} = \frac{1}{r},\qquad \gamma_{(\varphi)(\vartheta)(\varphi)} = \frac{\cot\vartheta}{r}.
\end{equation}
The contractions of the Ricci rotation coefficients read
\begin{equation}
 \gamma_{(r)} = \frac{2}{r},\qquad \gamma_{(\vartheta)} = \frac{\cot\vartheta}{r}.
\end{equation}

% -------------------------------------------------------------------
%   conformal coordinates
% -------------------------------------------------------------------
\subsection{Conform-compactified coordinates}
The Minkowski metric in conform-compactified coordinates $\left\{\psi\in[-\pi,\pi],\xi\in(0,\pi),\vartheta\in(0,\pi),\varphi\in[0,2\pi)\right\}$ reads\cite{hawking1999}
\metricEq{
  ds^2 = -d\psi^2+d\xi^2+\sin^2\xi\left(d\vartheta^2+\sin^2\vartheta d\varphi^2\right).
}{minkConf}
This form follows from the spherical Minkowski metric (\ref{eqM:minkSph}) by means of the coordinate transformation
\begin{equation}
  ct+r=\tan\frac{\psi+\xi}{2},\qquad ct-r = \tan\frac{\psi-\xi}{2},
\end{equation}
resulting in the metric
\begin{equation}
  d\tilde{s}^2 = \frac{-d\psi^2+d\xi^2}{4\cos^2\frac{\psi+\xi}{2}\cos^2\frac{\psi-\xi}{2}} + \frac{\sin^2\xi}{4\cos^2\frac{\psi+\xi}{2}\cos^2\frac{\psi-\xi}{2}}\left(d\vartheta^2+\sin^2\vartheta d\varphi^2\right),
\end{equation}
and by the conformal transformation $ds^2=\Omega^2d\tilde{s}^2$ with $\Omega^2=4\cos^2\frac{\psi+\xi}{2}\cos^2\frac{\psi-\xi}{2}$.

%% -------------------- Christoffel symbols --------------------
\SecChristoffel
\begin{subequations}
\begin{alignat}{5}
  \Gamma_{\xi\vartheta}^{\vartheta} &= \cot\xi, &\qquad \Gamma_{\xi\varphi}^{\varphi} &= \cot\xi, &\qquad \Gamma_{\vartheta\vartheta}^{\xi} &= -\sin\xi\cos\xi,\\
  \Gamma_{\vartheta\varphi}^{\varphi} &= \cot\vartheta, & \Gamma_{\varphi\varphi}^{\xi} &= -\sin\xi\cos\xi\sin^2\vartheta, & \Gamma_{\varphi\varphi}^{\vartheta} &= -\sin\vartheta\cos\vartheta.
\end{alignat}
\end{subequations}

Partial derivatives
\begin{subequations}
 \begin{alignat}{5}
    \Gamma_{\xi\vartheta,\xi}^{\vartheta} &= -\frac{1}{\sin^2\xi}, &\quad \Gamma_{\xi\varphi,\xi}^{\varphi} &= -\frac{1}{\sin^2\xi}, &\quad \Gamma_{\vartheta\vartheta,\xi}^{\xi} &= -\cos(2\xi),\\    
    \Gamma_{\vartheta\varphi,\vartheta}^{\varphi} &= -\frac{1}{\sin^2\vartheta}, & \Gamma_{\varphi\varphi,\xi}^{\xi} &= -\cos(2\xi)\sin^2\vartheta, & \Gamma_{\varphi\varphi,\vartheta}^{\vartheta} &= -\cos(2\vartheta),\\
    \Gamma_{\varphi\varphi,\vartheta}^{\xi} &= -\frac{1}{2}\sin(2\xi)\sin(2\vartheta).
 \end{alignat}
\end{subequations}

%% -------------------- Riemann tensor --------------------
\SecRiemann
\begin{equation}
  R_{\xi\vartheta\xi\vartheta} = \sin^2\xi,\qquad R_{\xi\varphi\xi\varphi} = \sin^2\xi\sin^2\vartheta,\qquad R_{\vartheta\varphi\vartheta\varphi} = \sin^4\xi\sin^2\vartheta.
\end{equation}

%% -------------------- Ricci tensor --------------------
\SecRicci
\begin{equation}
  R_{\xi\xi} = 2,\qquad R_{\vartheta\vartheta} = 2\sin^2\xi,\qquad R_{\varphi\varphi} = 2\sin^2\xi\sin^2\vartheta.
\end{equation}

\noindent {\bf Ricci and Kretschmann scalars:}
\begin{equation}
  \mathcal{R} = 6,\qquad \mathcal{K} = 12.
\end{equation}

\noindent The Weyl tensor vanishs identically.

%% -------------------- Local tetrad --------------------
\SecLocal
\begin{equation}
  \mathbf{e}_{(\psi)} = \partial_{\psi},\qquad \mathbf{e}_{(\xi)} = \partial_{\xi},\qquad \mathbf{e}_{(\vartheta)} = \frac{1}{\sin\xi}\partial_{\vartheta},\qquad \mathbf{e}_{(\varphi)} = \frac{1}{\sin\xi\sin\vartheta}\partial_{\varphi}.
\end{equation}

%% -------------------- Ricci rotation coefficients --------------------
\SecRicRotCoef
\begin{equation}
  \gamma_{(\vartheta)(\xi)(\vartheta)} = \gamma_{(\varphi)(\xi)(\varphi)} = \cot\xi,\qquad \gamma_{(\varphi)(\vartheta)(\varphi)} = \frac{\cot\vartheta}{\sin\xi}.
\end{equation}
The contractions of the Ricci rotation coefficients read
\begin{equation}
  \gamma_{(\xi)} = 2\cot\xi,\qquad \gamma_{(\vartheta)} = \frac{\cot\vartheta}{\sin\xi}.
\end{equation}

%% -------------------- Riemann tensor LT--------------------
\SecRiemannLT
\begin{equation} 
  R_{(\xi)(\vartheta)(\xi)(\vartheta)} = R_{(\xi)(\varphi)(\xi)(\varphi)} = R_{(\vartheta)(\varphi)(\vartheta)(\varphi)} = 1.
\end{equation}

%% -------------------- Ricci tensor LT--------------------
\SecRicciLT
\begin{equation} 
  R_{(\xi)(\xi)} = R_{(\vartheta)(\vartheta)} = R_{(\varphi)(\varphi)} = 2.
\end{equation}

% -------------------------------------------------------------------
%   rotating coordinates
% -------------------------------------------------------------------
\subsection{Rotating coordinates}
The transformation $d\varphi\mapsto d\varphi+\omega\,dt$ brings the Minkowski metric (\ref{eqM:minkCyl}) into the rotating form\cite{rindler} with coordinates $\left\{t\in\setR,r\in\setR^{+},\varphi\in[0,2\pi),z\in\setR\right\}$,
\metricEq{
  ds^2 = -\left(1-\frac{\omega^2r^2}{c^2}\right)\left[c\,dt-\Omega(r)d\varphi\right]^2+dr^2+\frac{r^2}{1-\omega^2r^2/c^2}d\varphi^2 +dz^2
}{minkRot}
with $\Omega(r)=(r^2\omega/c)/(1-\omega^2r^2/c^2)$.

%% -------------------- Metric tensor --------------------
\SecMetric
\begin{equation}
 g_{tt} = -c^2+\omega^2r^2,\qquad g_{t\varphi} = \omega r^2,\qquad g_{rr} = g_{zz} = 1,\qquad g_{\varphi\varphi} = r^2.
\end{equation}

%% -------------------- Christoffel symbols --------------------
\SecChristoffel
\begin{equation}
  \Gamma_{tt}^r = -\omega^2r, \qquad \Gamma_{tr}^{\varphi} = \frac{\omega}{r}, \qquad \Gamma_{t\varphi}^r = -\omega r,\qquad \Gamma_{r\varphi}^{\varphi} = \frac{1}{r},\qquad \Gamma_{\varphi\varphi}^r = -r.
\end{equation}

Partial derivatives
\begin{equation}
  \Gamma_{tt,r}^r = -\omega^2,\quad \Gamma_{tr,r}^{\varphi} = -\frac{\omega}{r^2},\quad \Gamma_{t\varphi,r}^r = -\omega,\quad \Gamma_{r\varphi,r}^{\varphi} = -\frac{1}{r^2},\quad \Gamma_{\varphi\varphi,r}^r = -1.
\end{equation}

The local tetrad of the comoving observer is
\begin{equation}
  \mathbf{e}_{(t)} = \frac{1}{c}\partial_t-\frac{\omega}{c}\partial_{\varphi},\qquad \mathbf{e}_{(r)} = \partial_r,\qquad \mathbf{e}_{(\varphi)} = \frac{1}{r}\partial_{\varphi},\qquad \mathbf{e}_{(z)} = \partial_z,
\end{equation}
whereas the static observer has the local tetrad
\begin{subequations}
\begin{align}
  \mathbf{e}_{(t)} &= \frac{1}{c\sqrt{1-\omega^2r^2/c^2}}\partial_t,\qquad \mathbf{e}_{(r)} = \partial_r, \qquad \mathbf{e}_{(z)} = \partial_z,\\
  \mathbf{e}_{(\varphi)} &= \frac{\omega r}{c^2\sqrt{1-\omega^2r^2/c^2}}\partial_t + \frac{\sqrt{1-\omega^2r^2/c^2}}{r}\partial_{\varphi}.
\end{align}
\end{subequations}

% -------------------------------------------------------------------
%   Rindler coordinates
% -------------------------------------------------------------------
\subsection{Rindler coordinates}
The worldline of an observer in the Minkowski spacetime who moves with constant proper acceleration $\alpha$ along the $x$ direction reads
\begin{equation}
  x = \frac{c^2}{\alpha}\cosh\frac{\alpha t'}{c},\qquad ct=\frac{c^2}{\alpha}\sinh\frac{\alpha t'}{c},
\end{equation}
where $t'$ is the observer's proper time. The observer starts at $x=1$ with zero velocity.\\
However, such an observer could also be described with Rindler coordinates. With the coordinate transformation 
\begin{equation}
  (ct,x)\mapsto (\tau,\rho):\qquad ct=\frac{1}{\rho}\sinh\tau,\qquad x = \frac{1}{\rho}\cosh\tau,
\end{equation}
where $\rho=\alpha/c^2$, the Rindler metric reads
\metricEq{
  ds^2 = -\frac{1}{\rho^2}d\tau^2+\frac{1}{\rho^4}d\rho^2+dy^2+dz^2.
}{minkRindler}

%% -------------------- Christoffel symbols --------------------
\SecChristoffel
\begin{equation}
  \Gamma_{\tau\tau}^{\rho} = -\rho,\qquad \Gamma_{\tau\rho}^{\tau} = -\frac{1}{\rho},\qquad \Gamma_{\rho\rho}^{\rho} = -\frac{2}{\rho}.
\end{equation}

Partial derivatives
\begin{equation}
  \Gamma_{\tau\tau,\rho}^{\rho} = -1,\qquad \Gamma_{\tau\rho,\rho}^{\tau} = \frac{1}{\rho^2},\qquad \Gamma_{\rho\rho,\rho}^{\rho} = \frac{2}{\rho^2}.
\end{equation}

The Riemann and Ricci tensors as well as the Ricci and Kretschmann scalar vanish identically. 

%% -------------------- Local tetrad --------------------
\SecLocal
\begin{equation}
  \mathbf{e}_{(\tau)} = \rho\partial_{\tau},\qquad \mathbf{e}_{(\rho)} = \rho^2\partial_{\rho},\qquad \mathbf{e}_{(y)} = \partial_y,\qquad \mathbf{e}_{(z)} = \partial_z.
\end{equation}
 
%% -------------------- Ricci rotation coefficients --------------------
\SecRicRotCoef
\begin{equation}
  \gamma_{(\tau)(\rho)(\tau)} = \rho,\qquad\text{and}\qquad \gamma_{(\rho)} = -\rho.
\end{equation}

}{

}

%% ------------------------------------------------------------------------
%%       S c h w a r z s c h i l d
%% ------------------------------------------------------------------------
\clearpage
\section{Schwarzschild spacetime}
\setcounter{equation}{0}
\ifthenelse{\boolean{isARXIV}}{
% ******** Start of file schwarzschild.tex *********
%
%  Copyright (c) 2009 Thomas Mueller,
%                     Universitaet Stuttgart, VISUS
%

\subsection{Schwarzschild coordinates}
In Schwarzschild coordinates $\left\{t\in\setR,r\in\setR^{+},\vartheta\in(0,\pi),\varphi\in[0,2\pi)\right\}$, the Schwarzschild metric reads
\metricEq{
  ds^2 = -\left(1-\frac{r_s}{r}\right)c^2dt^2+\frac{1}{1-r_s/r}dr^2 + r^2\left(d\vartheta^2+\sin^2\vartheta d\varphi^2\right),
}{schwarzschildSpherical}
where $r_s=2GM/c^2$ is the Schwarzschild radius, $G$ is Newton's constant, $c$ is the speed of light, and $M$ is the mass of the black hole. The critical point $r=0$ is a real curvature singularity while the event horizon, $r=r_s$, is only a coordinate singularity, see e.g. the Kretschmann scalar.

%% -------------------- Christoffel symbols --------------------
\SecChristoffel
\begin{subequations}
\begin{alignat}{5}
  \Cchris{tt}{r} &= \frac{c^2r_s(r-r_s)}{2r^3}, &\qquad\Cchris{tr}{t} &= \frac{r_s}{2r(r-r_s)}, &\qquad\Cchris{rr}{r} &= -\frac{r_s}{2r(r-r_s)},\\
  \Cchris{r\vartheta}{\vartheta} &= \frac{1}{r}, &\qquad \Cchris{r\varphi}{\varphi} &= \frac{1}{r}, &\qquad \Cchris{\vartheta\vartheta}{r}&=-(r-r_s),\\
   \Cchris{\vartheta\varphi}{\varphi} &= \cot\vartheta, &\qquad \Cchris{\varphi\varphi}{r}&=-(r-r_s)\sin^2\vartheta,&\quad\Cchris{\varphi\varphi}{\vartheta} &=-\sin\vartheta\cos\vartheta.
\end{alignat}
\end{subequations}

Partial derivatives
\begin{subequations}
 \begin{alignat}{5}
    \Cchris{tt,r}{r} &= -\frac{(2r-3r_s)c^2r_s}{2r^4}, &\qquad \Cchris{tr,r}{t} &= -\frac{(2r-r_s)r_s}{2r^2(r-r_s)^2}, &\qquad \Cchris{rr,r}{r} &= \frac{(2r-r_s)r_s}{2r^2(r-r_s)^2},\\
    \Cchris{r\vartheta,r}{\vartheta} &= -\frac{1}{r^2}, & \Cchris{r\varphi,r}{\varphi} &= -\frac{1}{r^2}, & \Cchris{\vartheta\vartheta,r}{r} &= -1,\\
    \Cchris{\vartheta\varphi,\vartheta}{\varphi} &= -\frac{1}{\sin^2\vartheta}, & \Cchris{\varphi\varphi,r}{r} &= -\sin^2\vartheta, & \Cchris{\varphi\varphi,\vartheta}{\vartheta} &= -\cos(2\vartheta),\\
    \Cchris{\varphi\varphi,\vartheta}{r} &= -(r-r_s)\sin(2\vartheta).
 \end{alignat}
\end{subequations}

%% -------------------- Riemann tensor --------------------
\SecRiemann
\begin{subequations}
\begin{alignat}{3}
  R_{trtr} &= -\frac{c^2r_s}{r^3}, &\qquad R_{t\vartheta t\vartheta} &= \frac{1}{2}\frac{c^2\left(r-r_s\right)r_s}{r^2}, &\qquad R_{t\varphi t\varphi} &= \frac{1}{2}\frac{c^2\left(r-r_s\right)r_s\sin^2\vartheta}{r^2},\\
  R_{r\vartheta r\vartheta} &= -\frac{1}{2}\frac{r_s}{r-r_s}, &\qquad R_{r\varphi r\varphi} &= -\frac{1}{2}\frac{r_s\sin^2\vartheta}{r-r_s}, & R_{\vartheta\varphi\vartheta\varphi} &= rr_s\sin^2\vartheta.
\end{alignat}
\end{subequations}

\noindent As aspected, the Ricci tensor as well as the Ricci scalar vanish identically because the Schwarz\-schild spacetime is a vacuum solution of the field equations. Hence, the Weyl tensor is identical to the Riemann tensor. The Kretschmann scalar reads
\begin{equation}
  \mathcal{K} = 12\frac{r_s^2}{r^6}.
\end{equation}
Here, it becomes clear that at $r=r_s$ there is no real singularity.

%% -------------------- Local tetrad --------------------
\SecLocal
\begin{equation}
  \Clt{t} = \frac{1}{c\sqrt{1-r_s/r}}\partial_t, \qquad \Clt{r} = \sqrt{1-\frac{r_s}{r}}\partial_r,\qquad \Clt{\vartheta} = \frac{1}{r}\partial_{\vartheta}, \qquad \Clt{\varphi} = \frac{1}{r\sin\vartheta}\partial_{\varphi}.
\end{equation}
Dual tetrad:
\begin{equation}
 \Cdlt{t} = c\sqrt{1-\frac{r_s}{r}}\,dt,\qquad \Cdlt{r} = \frac{dr}{\sqrt{1-r_s/r}},\qquad \Cdlt{\vartheta}=r\,d\vartheta,\qquad \Cdlt{\varphi} = r\sin\vartheta\,d\varphi.
\end{equation}

%% -------------------- Ricci rotation coefficients --------------------
\SecRicRotCoef
\begin{equation}
  \gamma_{(r)(t)(t)} = \frac{r_s}{2r^2\sqrt{1-r_s/r}},\quad \gamma_{(\vartheta)(r)(\vartheta)} = \gamma_{(\varphi)(r)(\varphi)} = \frac{1}{r}\sqrt{1-\frac{r_s}{r}},\quad \gamma_{(\varphi)(\vartheta)(\varphi)} = \frac{\cot\vartheta}{r}.
\end{equation}
The contractions of the Ricci rotation coefficients read
\begin{equation}
   \gamma_{(r)} = \frac{4r-3r_s}{2r^2\sqrt{1-r_s/r}},\qquad \gamma_{(\vartheta)} = \frac{\cot\vartheta}{r}.
\end{equation}

%% -------------------- Lie coefficients --------------------
\SecLieCoef
\begin{equation}
  c_{(t)(r)}^{(t)} = \frac{r_s}{2r^2\sqrt{1-r_s/r}},\qquad
  c_{(r)(\vartheta)}^{(\vartheta)} = c_{(r)(\varphi)}^{(\varphi)} = -\frac{1}{r}\sqrt{1-\frac{r_s}{r}},\qquad
  c_{(\vartheta)(\varphi)}^{(\varphi)} = \frac{\cot\vartheta}{r}.
\end{equation}

%% -------------------- Riemann tensor LT--------------------
\SecRiemannLT
\begin{subequations}
 \begin{align}
  R_{(t)(r)(t)(r)} &= -R_{(\vartheta)(\varphi)(\vartheta)(\varphi)} = -\frac{r_s}{r^3},\\
  R_{(t)(\vartheta)(t)(\vartheta)} &= R_{(t)(\varphi)(t)(\varphi)} = -R_{(r)(\vartheta)(r)(\vartheta)} = -R_{(r)(\varphi)(r)(\varphi)} = \frac{r_s}{2r^3}.
 \end{align}
\end{subequations}
The covariant derivatives of the Riemann tensor read
\begin{subequations}
 \begin{align}
   R_{(t)(r)(t)(r);(r)} &= -R_{(\vartheta)(\varphi)(\vartheta)(\varphi);(r)} = \frac{3r_s}{r^5}\sqrt{r(r-r_s)},\\
   \nonumber R_{(t)(r)(r)(\vartheta);(\vartheta)} &= R_{(t)(r)(t)(\varphi);(\varphi)} = R_{(t)(\vartheta)(t)(\vartheta);(r)} = R_{(t)(\varphi)(t)(\varphi);(r)} =\\
  &= R_{(r)(\varphi)(\vartheta)(\varphi);(\vartheta)} = -\frac{3r_s}{2r^5}\sqrt{r(r-r_s)},\\
   R_{(r)(\vartheta)(r)(\vartheta);(r)} &= R_{(r)(\vartheta)(\vartheta)(\varphi);(\varphi)} = R_{(r)(\varphi)(r)(\varphi);(r)} = \frac{3r_s}{2r^5}\sqrt{r(r-r_s)}.
 \end{align}
\end{subequations}

{\bf Newman-Penrose tetrad:}
\begin{equation}
 \mathbf{l} = \frac{1}{\sqrt{2}}\left(\mathbf{e}_{(t)}+\mathbf{e}_{(r)}\right),\qquad \mathbf{n} = \frac{1}{\sqrt{2}}\left(\mathbf{e}_{(t)}-\mathbf{e}_{(r)}\right),\qquad \mathbf{m} = \frac{1}{\sqrt{2}}\left(\mathbf{e}_{(\vartheta)}+i\mathbf{e}_{(\varphi)}\right).
\end{equation}
Non-vanishing spin coefficients:
\begin{equation}
 \rho = \mu = -\frac{1}{\sqrt{2}r}\sqrt{1-\frac{r_s}{r}},\quad \gamma = \epsilon = \frac{r_s}{4\sqrt{2}r^2\sqrt{1-r_s/r}},\quad \alpha = -\beta = -\frac{\cot\vartheta}{2\sqrt{2}r}.
\end{equation}

%% -------------------- Embedding --------------------
\SecEmbedding

The embedding function reads
\begin{equation}
  \label{eq:schwEmb}
  z = 2\sqrt{r_s}\sqrt{r-r_s}.
\end{equation}

%% -------------------- Euler Lagrange --------------------
\SecEulLag

The Euler-Lagrangian formalism, Sec.~\ref{subsec:EL}, for geodesics in the $\vartheta=\pi/2$ hyperplane yields 
\begin{equation}
  \frac{1}{2}\dot{r}^2+V_{\text{eff}} = \frac{1}{2}\frac{k^2}{c^2},\qquad V_{\text{eff}}=\frac{1}{2}\left(1-\frac{r_s}{r}\right)\left(\frac{h^2}{r^2}-\kappa c^2\right)
\end{equation}
with the constants of motion $k=(1-r_s/r)c^2\dot{t}$, $h=r^2\dot{\varphi}$, and $\kappa$ as in Eq. (\ref{eq:constrEq}). For timelike geodesics, the effective potential has the extremal points
\begin{equation}
  r_{\pm} = \frac{h^2\pm h\sqrt{h^2-3c^2r_s^2}}{c^2r_s},
\end{equation}
where $r_{+}$ is a maximum and $r_{-}$ is a minimum. The innermost timelike circular geodesic follows from $h^2=3c^2r_s^2$ and reads $r_{\text{itcg}}=3r_s$. Null geodesics, however, have only a maximum at $r_{\text{po}}=\frac{3}{2}r_s$. The corresponding circular orbit is called photon orbit.

%% -------------------- Further reading --------------------
\FurtherReading

Schwarzschild\cite{schwarzschild1916a,schwarzschild1916b}, MTW\cite{mtw}, Rindler\cite{rindler}, Wald\cite{wald}, Chandrasekhar\cite{chandrasekhar2006},\\ M{\"u}ller\cite{mueller2008grg,mueller2009grg}.

% ******** Start of file schwarzschildCart.tex *********
%
%  Copyright (c) 2009 Thomas Mueller,
%                     Universitaet Stuttgart, VISUS
%

\subsection{Schwarzschild in pseudo-Cartesian coordinates}
The Schwarzschild spacetime in pseudo-Cartesian coordinates $(t,x,y,z)$ reads
% \begin{subequations}
% \begin{align}
%   \label{eq:schwarzschildCart}
%   ds^2 &= -\left(1-\frac{r_s}{r}\right)c^2dt^2 + \left(\frac{x^2}{1-r_s/r}+y^2+z^2\right)\frac{dx^2}{r^2} + \left(x^2+\frac{y^2}{1-r_s/r}+z^2\right)\frac{dy^2}{r^2}\\
%   &\quad + \left(x^2+y^2+\frac{z^2}{1-r_s/r}\right)\frac{dz^2}{r^2} + \frac{2r_s}{r^2(r-r_s)}\left(xy\,dxdy + xz\,dxdz + yz\,dydz\right).
% \end{align}
% \end{subequations}
\metricEqAlign{ds^2 &= \dst -\left(1-\frac{r_s}{r}\right)c^2dt^2 + \left(\frac{x^2}{1-r_s/r}+y^2+z^2\right)\frac{dx^2}{r^2} + \left(x^2+\frac{y^2}{1-r_s/r}+z^2\right)\frac{dy^2}{r^2}\\
   &\quad  \dst + \left(x^2+y^2+\frac{z^2}{1-r_s/r}\right)\frac{dz^2}{r^2} + \frac{2r_s}{r^2(r-r_s)}\left(xy\,dxdy + xz\,dxdz + yz\,dydz\right),
}{schwarzschildCart}
where $r^2=x^2+y^2+z^2$.
For a natural local tetrad that is adapted to the x-axis, we make the following ansatz:
\begin{equation}
  \mathbf{e}_{(0)} = \frac{1}{c\sqrt{1-r_s/r}}\partial_t,\qquad \mathbf{e}_{(1)} = A\partial_x,\qquad \mathbf{e}_{(2)} = B\partial_x + C\partial_y,\qquad \mathbf{e}_{(3)} = D\partial_x+E\partial_y+F\partial_z.
\end{equation}

\begin{subequations}
\begin{alignat}{5}
  A &= \frac{1}{\sqrt{g_{xx}}}, &\qquad B &= \frac{-g_{xy}}{g_{xx}\sqrt{-g_{xy}^2/g_{xx}+g_{yy}}}, &\qquad C &= \frac{1}{\sqrt{-g_{xy}^2/g_{xx}+g_{yy}}},\\
  D &= \frac{g_{xy}g_{yz}-g_{xz}g_{yy}}{\sqrt{NW}}, &\qquad E &= \frac{g_{xz}g_{xy}-g_{xx}g_{yz}}{\sqrt{NW}}, &\qquad F &= \frac{\sqrt{N}}{\sqrt{W}},
\end{alignat}
\end{subequations}
with
\begin{subequations}
\begin{align}
  N &= g_{xx}g_{yy}-g_{xy}^2,\\
  W &= g_{xx}g_{yy}g_{zz} - g_{xz}^2g_{yy} + 2g_{xz}g_{xy}g_{yz} - g_{xy}^2g_{zz} - g_{xx}g_{yz}^2.
\end{align}
\end{subequations}

}{

}

%% ------------------------------------------------------------------------
%%       isotropic
%% ------------------------------------------------------------------------
\ifthenelse{\boolean{isARXIV}}{
% ******** Start of file isotropic.tex *********
%
%  Copyright (c) 2009 Thomas Mueller,
%                     Universitaet Stuttgart, VISUS
%

\subsection{Isotropic coordinates}

\subsubsection{Spherical isotropic coordinates}
The Schwarzschild metric (\ref{eqM:schwarzschildSpherical}) in spherical isotropic coordinates $(t,\rho,\vartheta,\varphi)$ reads
\metricEq{
  ds^2 = -\left(\frac{1-\rho_s/\rho}{1+\rho_s/\rho}\right)^2c^2dt^2 + \left(1+\frac{\rho_s}{\rho}\right)^4\left[d\rho^2+\rho^2\left(d\vartheta^2+\sin^2\vartheta d\varphi^2\right)\right],
}{isotropic}
where
\begin{equation}
  r = \rho\left(1+\frac{\rho_s}{\rho}\right)^2\qquad\text{or}\qquad \rho=\frac{1}{4}\left(2r-r_s\pm 2\sqrt{r(r-r_s)}\right)
\end{equation}
is the coordinate transformation between the Schwarzschild radial coordinate $r$ and the isotropic radial coordinate $\rho$, see e.g. MTW\cite{mtw} page 840. The event horizon is given by $\rho_s=r_s/4$. The photon orbit and the innermost timelike circular geodesic read
\begin{equation}
  \rho_{\text{po}} = \left(2+\sqrt{3}\right)\rho_s\qquad\text{and}\qquad \rho_{\text{itcg}} = \left(5+2\sqrt{6}\right)\rho_s. 
\end{equation}

%% -------------------- Christoffel symbols --------------------
\SecChristoffel
\begin{subequations}
\begin{alignat}{5}
  \Gamma_{tt}^{\rho} &= \frac{2(\rho-\rho_s)\rho^4\rho_sc^2}{(\rho+\rho_s)^7}, &\,\quad \Gamma_{t\rho}^t &= \frac{2\rho_s}{\rho^2-\rho_s^2}, &\, \Gamma_{\rho\rho}^{\rho} &= -\frac{2\rho_s}{(\rho+\rho_s)\rho},\\
  \Gamma_{\rho\vartheta}^{\vartheta} &= \frac{\rho-\rho_s}{(\rho+\rho_s)\rho}, & \Gamma_{\rho\varphi}^{\varphi} &= \frac{\rho-\rho_s}{(\rho+\rho_s)\rho}, & \Gamma_{\vartheta\vartheta}^{\rho} &= -\rho\frac{\rho-\rho_s}{\rho+\rho_s},\\
  \Gamma_{\vartheta\varphi}^{\varphi} &= \cot\vartheta, & \Gamma_{\varphi\varphi}^{\rho} &= -\frac{(\rho-\rho_s)\rho\sin^2\vartheta}{\rho+\rho_s}, &\quad \Gamma_{\varphi\varphi}^{\vartheta} &= -\sin\vartheta\cos\vartheta.
\end{alignat}
\end{subequations}

%% -------------------- Riemann tensor --------------------
\SecRiemann
\begin{subequations}
\begin{alignat}{3}
  R_{t\rho t\rho} &= -4\frac{(\rho-\rho_s)^2\rho_sc^2}{(\rho+\rho_s)^4\rho}, &\qquad R_{t\vartheta t\vartheta} &= 2\frac{(\rho-\rho_s)^2\rho \rho_sc^2}{(\rho+\rho_s)^4},\\
  R_{t\varphi t\varphi} &= 2\frac{(\rho-\rho_s)^2\rho c^2\rho_s\sin^2\vartheta}{(\rho+\rho_s)^4}, & R_{\rho\vartheta\rho\vartheta} &= -2\frac{(\rho+\rho_s)^2\rho_s}{\rho^3},\\
  R_{\rho\varphi\rho\varphi} &= -2\frac{(\rho+\rho_s)^2\rho_s\sin^2\vartheta}{\rho^3}, & R_{\vartheta\varphi\vartheta\varphi} &= \frac{4(\rho+\rho_s)^2\rho_s\sin^2\vartheta}{\rho}.
\end{alignat}
\end{subequations}

The Ricci tensor and the Ricci scalar vanish identically.

%% -------------------- Kretschmann scalar --------------------
\SecKretsch
\begin{equation}
  \mathcal{K}=192\frac{r_s^2}{\rho^6\left(1+\rho_s/\rho\right)^{12}} = 12\frac{r_s^2}{r(\rho)^6}.
\end{equation}

%% -------------------- Local tetrad --------------------
\SecLocal
\begin{subequations}
\begin{alignat}{3}
  \mathbf{e}_{(t)} &= \frac{1+\rho_s/\rho}{1-\rho_s/\rho}\frac{\partial_t}{c}, &\qquad \mathbf{e}_{(r)} &= \frac{1}{\left[1+\rho_s/\rho\right]^2}\partial_{\rho},\\
  \mathbf{e}_{(\vartheta)} &= \frac{1}{\rho\left[1+\rho_s/\rho\right]^2}\partial_{\vartheta}, & \mathbf{e}_{(\varphi)} &=\frac{1}{\rho\left[1+\rho_s/\rho\right]^2\sin^2\vartheta}\partial_{\varphi}.
\end{alignat}
\end{subequations}

%% -------------------- Ricci rotation coefficients --------------------
\SecRicRotCoef
\begin{subequations}
\begin{align} 
  \gamma_{(\rho)(t)(t)} &= \frac{2\rho_s\rho^2}{(\rho+\rho_s)^3(\rho-\rho_s)},\quad \gamma_{(\vartheta)(\rho)(\vartheta)} = \gamma_{(\varphi)(\rho)(\varphi)} = \frac{\rho(\rho-\rho_s)}{(\rho+\rho_s)^3},\\
  \gamma_{(\varphi)(\vartheta)(\varphi)} &= \frac{\rho\cot\vartheta}{(\rho+\rho_s)^2}.
\end{align}
\end{subequations}
The contractions of the Ricci rotation coefficients read
\begin{equation}
   \gamma_{(\rho)} = \frac{2\rho(\rho^2-\rho\rho_s+\rho_s^2)}{(\rho+\rho_s)^3(\rho-\rho_s)},\qquad \gamma_{(\vartheta)} = \frac{\rho\cot\vartheta}{(\rho+\rho_s)^2}.
\end{equation}

%% -------------------- Riemann tensor LT--------------------
\SecRiemannLT
\begin{subequations}
 \begin{align}
  R_{(t)(\rho)(t)(\rho)} &= -R_{(\vartheta)(\varphi)(\vartheta)(\varphi)} = -\frac{r_s}{r(\rho)^3},\\
  R_{(t)(\vartheta)(t)(\vartheta)} &= R_{(t)(\varphi)(t)(\varphi)} = -R_{(\rho)(\vartheta)(\rho)(\vartheta)} = -R_{(\rho)(\varphi)(\rho)(\varphi)} = \frac{r_s}{2r(\rho)^3}.
 \end{align}
\end{subequations}

%% -------------------- Further reading --------------------
\FurtherReading

Buchdahl\cite{buchdahl1985}.

\subsubsection{Cartesian isotropic coordinates}
The Schwarzschild metric (\ref{eqM:schwarzschildSpherical}) in Cartesian isotropic coordinates $(t,x,y,z)$ reads, 
\metricEq{
  ds^2 = -\left(\frac{1-\rho_s/\rho}{1+\rho_s/\rho}\right)^2c^2dt^2 + \left(1+\frac{\rho_s}{\rho}\right)^4\left[dx^2+dy^2+dz^2\right],
}{cartIsotropic}
where $\rho^2=x^2+y^2+z^2$ and, as before,
\begin{equation}
  r = \rho\left(1+\frac{\rho_s}{\rho}\right)^2.
\end{equation}

%% -------------------- Christoffel symbols --------------------
\SecChristoffel
\begin{subequations}
\begin{align}
   \Cchris{tt}{x} &= \frac{2c^2\rho^3\rho_s\left(\rho-\rho_s\right)x}{\left(\rho+\rho_s\right)^7},\qquad \Cchris{tt}{y} = \frac{2c^2\rho^3\rho_s\left(\rho-\rho_s\right)y}{\left(\rho+\rho_s\right)^7},\qquad \Cchris{tt}{z} = \frac{2c^2\rho^3\rho_s\left(\rho-\rho_s\right)z}{\left(\rho+\rho_s\right)^7},\\
   \Cchris{tx}{t} &= \frac{2\rho_s x}{\rho^3\left[1-\rho_s^2/\rho^2\right]}, \qquad \Cchris{ty}{t} = \frac{2\rho_s y}{\rho^3\left[1-\rho_s^2/\rho^2\right]}, \qquad \Cchris{tz}{t} = \frac{2\rho_s z}{\rho^3\left[1-\rho_s^2/\rho^2\right]},\\
   \Cchris{xx}{x} &= \Cchris{xy}{y} = \Cchris{xz}{z} = -\Cchris{yy}{x} = -\Cchris{zz}{x} = -\frac{2\rho_s}{\rho^3}\frac{x}{1+\rho_s/\rho},\\
   \Cchris{xx}{y} &= -\Cchris{xy}{x} = -\Cchris{yy}{y} = -\Cchris{yz}{z} = \Cchris{zz}{y} = \frac{2\rho_s}{\rho^3}\frac{y}{1+\rho_s/\rho},\\
   \Cchris{xx}{z} &= -\Cchris{xz}{x} = \Cchris{yy}{z} = -\Cchris{yz}{y} = -\Cchris{zz}{z} = \frac{2\rho_s}{\rho^3}\frac{z}{1+\rho_s/\rho}.
\end{align}
\end{subequations}

}{

}

%% ------------------------------------------------------------------------
%%       E d d i n g t o n  -  F i n k e l s t e i n
%% ------------------------------------------------------------------------
\ifthenelse{\boolean{isARXIV}}{
% ******** Start of file eddingtonfinkelstein.tex *********
%
%  Copyright (c) 2009 Thomas Mueller,
%                     Universitaet Stuttgart, VISUS
%

\subsection{Eddington-Finkelstein}
The transformation of the Schwarzschild metric (\ref{eqM:schwarzschildSpherical}) from the usual Schwarzschild time coordinate $t$ to the advanced null coordinate $v$ with
\begin{equation}
  cv = ct+r+r_s\ln(r-r_s)
\end{equation}
leads to the ingoing Eddington-Finkelstein\cite{eddington1924,finkelstein1958} metric with coordinates $(v,r,\vartheta,\varphi)$,
\metricEq{
  ds^2 = -\left(1-\frac{r_s}{r}\right)c^2dv^2 + 2c\,dv\,dr + r^2\left(d\vartheta^2+\sin^2\vartheta d\varphi^2\right).
}{eddFinkel}

%% -------------------- Metric tensor --------------------
\SecMetric
\begin{equation}
 g_{vv} = -c^2\left(1-\frac{r_s}{r}\right),\qquad g_{vr} = c,\qquad g_{\vartheta\vartheta} = r^2,\qquad g_{\varphi\varphi}=r^2\sin^2\vartheta.
\end{equation}

%% -------------------- Christoffel symbols --------------------
\SecChristoffel
\begin{subequations}
\begin{alignat}{7}
  \Gamma_{vv}^v &= \frac{cr_s}{2r^2}, &\quad \Gamma_{vv}^r &= \frac{c^2r_s(r-r_s)}{2r^3}, &\quad \Gamma_{vr}^r &= -\frac{cr_s}{2r^2}, &\quad \Gamma_{r\vartheta}^{\vartheta} &= \frac{1}{r},\\
  \Gamma_{r\varphi}^{\varphi} &= \frac{1}{r}, & \Gamma_{\vartheta\vartheta}^v &= -\frac{r}{c}, & \Gamma_{\vartheta\vartheta}^r &= -(r-r_s), & \Gamma_{\vartheta\varphi}^{\varphi} &= \cot\vartheta,\\
  \Gamma_{\varphi\varphi}^v &= -\frac{r\sin^2\vartheta}{c}, & \Gamma_{\varphi\varphi}^r &= -(r-r_s)\sin^2\vartheta, & \Gamma_{\varphi\varphi}^{\vartheta} &= -\sin\vartheta\cos\vartheta.
\end{alignat}
\end{subequations}

Partial derivatives
\begin{subequations}
 \begin{alignat}{5}
    \Gamma_{vv,r}^v &= -\frac{cr_s}{r^3}, &\qquad \Gamma_{vv,r}^r &= -\frac{(2r-3r_s)c^2r_s}{2r^4}, &\qquad \Gamma_{vr,r}^r &= \frac{cr_s}{r^3},\\
    \Gamma_{r\vartheta,r}^{\vartheta} &= -\frac{1}{r^2}, & \Gamma_{r\varphi,r}^{\varphi} &= -\frac{1}{r^2}, & \Gamma_{\vartheta\vartheta,r}^v &= -\frac{1}{c},\\
    \Gamma_{\vartheta\vartheta,r}^r &= -1, & \Gamma_{\vartheta\varphi,\vartheta}^{\varphi} &= -\frac{1}{\sin^2\vartheta}, & \Gamma_{\varphi\varphi,r}^v &= -\frac{\sin^2\vartheta}{c},\\
    \Gamma_{\varphi\varphi,\vartheta}^v &= -\frac{r\sin(2\vartheta)}{c}, & \Gamma_{\varphi\varphi,r}^r &= -\sin^2\vartheta, & \Gamma_{\varphi\varphi,\vartheta}^{\vartheta} &= -\cos(2\vartheta),\\
    \Gamma_{\varphi\varphi,\vartheta}^r &= -(r-r_s)\sin(2\vartheta).
 \end{alignat}
\end{subequations}

%% -------------------- Riemann tensor --------------------
\SecRiemann
\begin{subequations}
\begin{alignat}{5}
  R_{vrvr} &= -\frac{c^2r_s}{r^3}, &\qquad R_{v\vartheta v\vartheta} &= \frac{c^2r_s(r-r_s)}{2r^2}, &\qquad R_{v\vartheta r\vartheta} &= -\frac{cr_s}{2r},\\
  R_{v\varphi v\varphi} &= \frac{c^2r_s(r-r_s)\sin^2\vartheta}{2r^2}, & R_{v\varphi r\varphi} &= -\frac{cr_s\sin^2\vartheta}{2r}, & R_{\vartheta\varphi\vartheta\varphi} &= rr_s\sin^2\vartheta.
\end{alignat}
\end{subequations}

While the Ricci tensor and the Ricci scalar vanish identically, the Kretschmann scalar is $\mathcal{K}=12 r_s^2/r^6$. 

%% -------------------- Local tetrad --------------------
\SecStatLocal
\begin{equation}
  \mathbf{e}_{(v)} = \frac{1}{c\sqrt{1-r_s/r}}\partial_v,\quad \mathbf{e}_{(r)} = \frac{1}{c\sqrt{1-r_s/r}}\partial_v+\sqrt{1-\frac{r_s}{r}}\partial_r,\quad \mathbf{e}_{(\vartheta)} = \frac{1}{r}\partial_{\vartheta},\quad \mathbf{e}_{(\varphi)} = \frac{1}{r\sin\vartheta}\partial_{\varphi}.
\end{equation}
Dual tetrad:
\begin{equation}
 \boldsymbol{\theta}^{(v)}=c\sqrt{1-\frac{r_s}{r}}dv-\frac{dr}{\sqrt{1-r_s/r}},\quad \boldsymbol{\theta}^{(r)} = \frac{dr}{\sqrt{1-r_s/r}},\quad \boldsymbol{\theta}^{(\vartheta)} = r\,d\vartheta,\quad \boldsymbol{\theta}^{(\varphi)} = r\sin\vartheta d\varphi.
\end{equation}

%% -------------------- Ricci rotation coefficients --------------------
\SecRicRotCoef
\begin{equation}
  \gamma_{(r)(v)(v)} = \frac{r_s}{2r^2\sqrt{1-r_s/r}},\quad \gamma_{(\vartheta)(r)(\vartheta)} = \gamma_{(\varphi)(r)(\varphi)} = \frac{1}{r}\sqrt{1-\frac{r_s}{r}},\quad \gamma_{(\varphi)(\vartheta)(\varphi)} = \frac{\cot\vartheta}{r}.
\end{equation}
The contractions of the Ricci rotation coefficients read
\begin{equation}
   \gamma_{(r)} = \frac{4r-3r_s}{2r^2\sqrt{1-r_s/r}},\qquad \gamma_{(\vartheta)} = \frac{\cot\vartheta}{r}.
\end{equation}

%% -------------------- Riemann tensor LT--------------------
\SecRiemannLT
\begin{subequations}
 \begin{align}
  R_{(v)(r)(v)(r)} &= -R_{(\vartheta)(\varphi)(\vartheta)(\varphi)} = -\frac{r_s}{r^3},\\
  R_{(v)(\vartheta)(v)(\vartheta)} &= R_{(v)(\varphi)(v)(\varphi)} = -R_{(r)(\vartheta)(r)(\vartheta)} = -R_{(r)(\varphi)(r)(\varphi)} = \frac{r_s}{2r^3}.
 \end{align}
\end{subequations}

}{

}

%% ------------------------------------------------------------------------
%%       K r u s k a l  -  S z e k e r e s
%% ------------------------------------------------------------------------
\ifthenelse{\boolean{isARXIV}}{
% ******** Start of file kruskal.tex *********
%
%  Copyright (c) 2009 Thomas Mueller,
%                     Universitaet Stuttgart, VISUS
%

\subsection{Kruskal-Szekeres}
The Schwarzschild metric in Kruskal-Szekeres\cite{kruskal1960,wald} coordinates $(T,X,\vartheta,\varphi)$ reads
\metricEq{
  ds^2 = \frac{4r_s^3}{r}e^{-r/r_s}\left(-dT^2+dX^2\right) + r^2d\Omega^2,
}{kruskSzek}
where $r\in\mathds{R}_{+}\setminus\left\{0\right\}$ is given by means of the LambertW-function $\mathcal{W}$,
\begin{equation}
  \left(\frac{r}{r_s}-1\right)e^{r/r_s} = X^2-T^2\qquad\text{or}\qquad r=r_s\left[\mathcal{W}\left(\frac{X^2-T^2}{e}\right)+1\right].
\end{equation}
The Schwarzschild coordinate time $t$ in terms of the Kruskal coordinates $T$ and $X$ reads
\begin{subequations}
\begin{alignat}{3}
   t &= 2r_s\textrm{arctanh}\frac{T}{X}, &\qquad r &> r_s,\\
   t &= 2r_s\textrm{arctanh}\frac{X}{T}, &\qquad r &< r_s,\\
   t &= \infty, &\qquad r &= r_s.
\end{alignat}
\end{subequations}

The transformations between Kruskal- and Schwarzschild coordinates read
\begin{subequations}
\begin{alignat}{5}
  X &= \sqrt{1-\frac{r}{r_s}}\,e^{r/(2r_s)}\sinh\frac{ct}{2r_s}, &\quad T &=\sqrt{1-\frac{r}{r_s}}\,e^{r/(2r_s)}\cosh\frac{ct}{2r_s}, &\qquad 0&<r<r_2,\\
  X &= \sqrt{\frac{r}{r_s}-1}\,e^{r/(2r_s)}\cosh\frac{ct}{2r_s}, & T &= \sqrt{\frac{r}{r_s}-1}\,e^{r/(2r_s)}\sinh\frac{ct}{2r_s}, &\qquad r&\geq r_s.
\end{alignat}
\end{subequations}

%% -------------------- Christoffel symbols --------------------
\SecChristoffel
\begin{subequations}
\begin{alignat}{3}
  \Gamma_{TT}^T &= \Gamma_{TX}^X = \Gamma_{XX}^T = \frac{Tr_s(r+r_s)}{r^2}e^{-r/r_s},\\
  \Gamma_{TT}^X &= \Gamma_{TX}^T = \Gamma_{XX}^X = -\frac{Xr_s(r+r_s)}{r^2}e^{-r/r_s},\\
  \Gamma_{T\vartheta}^{\vartheta} &= -\frac{2r_s^2T}{r^2}e^{-r/r_s}, &\qquad \Gamma_{X\vartheta}^{\vartheta} &= \frac{2r_s^2X}{r^2}e^{-r/r_s},\\
  \Gamma_{\vartheta\vartheta}^T &= -\frac{r}{2r_s}T, & \Gamma_{\vartheta\vartheta}^X &= \frac{r}{2r_s}X,\\
  \Gamma_{\vartheta\vartheta}^T &= -\frac{r}{2r_s}T\sin^2\vartheta, & \Gamma_{\vartheta\vartheta}^X &= \frac{r}{2r_s}X\sin^2\vartheta,\\
  \Gamma_{\vartheta\varphi}^{\varphi} &= \cot\vartheta, & \Gamma_{\varphi\varphi}^{\vartheta} &= -\sin\vartheta\cos\vartheta.
\end{alignat}
\end{subequations}

%% -------------------- Riemann tensor --------------------
\SecRiemann
\begin{subequations}
\begin{alignat}{3}
  R_{TXTX} &= -16\frac{r_s^7}{r^5}e^{-2r/r_s}, &\qquad R_{T\vartheta T\vartheta} &= \frac{2r_s^4}{r^2}e^{-r/r_s},\\
  R_{T\varphi T\varphi} &= \frac{2r_s^4}{r^2}e^{-r/r_s}\sin^2\vartheta, & R_{X\vartheta X\vartheta} &= -\frac{2r_s^4}{r^2}e^{-r/r_s},\\
  R_{X\varphi X\varphi} &= -\frac{2r_s^4}{r^2}e^{-r/r_s}\sin^2\vartheta, & R_{\vartheta\varphi\vartheta\varphi} &= rr_s\sin^2\vartheta.
\end{alignat}
\end{subequations}

\noindent The {\sl Ricci-Tensor} as well as the {\sl Ricci-scalar} vanish identically.

%% -------------------- Kretschmann scalar --------------------
\SecKretsch
\begin{equation}
  \mathcal{K} = \frac{12r_s^2}{r^6}.
\end{equation}

%% -------------------- Local tetrad --------------------
\SecLocal
\begin{equation}
  \mathbf{e}_{(T)} = \frac{\sqrt{r}}{2r_s\sqrt{r_s}}e^{r/(2r_s)}\partial_T,\quad \mathbf{e}_{(X)} = \frac{\sqrt{r}}{2r_s\sqrt{r_s}}e^{r/(2r_s)}\partial_X,\quad\mathbf{e}_{(\vartheta)} = \frac{1}{r}\partial_{\vartheta},\quad\mathbf{e}_{(\varphi)} = \frac{1}{r\sin\vartheta}\partial_{\varphi}
\end{equation}

%% -------------------- Riemann tensor LT--------------------
\SecRiemannLT
\begin{subequations}
 \begin{align}
  R_{(T)(X)(T)(X)} &= R_{(X)(\vartheta)(X)(\vartheta)} = R_{(X)(\varphi)(X)(\varphi)} = -R_{(\vartheta)(\varphi)(\vartheta)(\varphi)} = -\frac{r_s}{r^3},\\
  R_{(T)(\vartheta)(T)(\vartheta)} &= R_{(T)(\varphi)(T)(\varphi)} = \frac{r_s}{2r^3}.
 \end{align}
\end{subequations}

}{

}

%% ------------------------------------------------------------------------
%%       tortoise
%% ------------------------------------------------------------------------
\ifthenelse{\boolean{isARXIV}}{
% ******** Start of file schwarzschildTortoise.tex *********
%
%  Copyright (c) 2009,2010 Thomas Mueller,
%                          Universitaet Stuttgart, VISUS
%

\subsection{Tortoise coordinates}
The Schwarzschild metric represented by tortoise coordinates $(t,\rho,\vartheta,\varphi)$ reads
\metricEq{
  ds^2 = -\left(1-\frac{r_s}{r(\rho)}\right)c^2dt^2+\left(1-\frac{r_s}{r(\rho)}\right)d\rho^2 + r(\rho)^2\left(d\vartheta^2+\sin^2\vartheta d\varphi^2\right),
}{schwarzschildTortoise}
where $r_s=2GM/c^2$ is the Schwarzschild radius, $G$ is Newton's constant, $c$ is the speed of light, and $M$ is the mass of the black hole. The tortoise radial coordinate $\rho$ and the Schwarzschild radial coordinate $r$ are related by
\begin{equation}
 \rho=r+r_s\ln\left(\frac{r}{r_s}-1\right)\qquad\text{or}\qquad r=r_s\left\{1+\mathcal{W}\left[\exp\left(\frac{\rho}{r_s}-1\right)\right]\right\}.
\end{equation}

%% -------------------- Christoffel symbols --------------------
\SecChristoffel
\begin{subequations}
\begin{alignat}{5}
  \Gamma_{tt}^{\rho} &= \frac{c^2r_s}{2r(\rho)^2}, &\qquad\Gamma_{t\rho}^t &= \frac{r_s}{2r(\rho)^2}, &\qquad\Gamma_{\rho\rho}^{\rho} &= \frac{r_s}{2r(\rho)^2},\\
  \Gamma_{\rho\vartheta}^{\vartheta} &= \frac{1}{r(\rho)}-\frac{1}{r_s}, &\qquad \Gamma_{\rho\varphi}^{\varphi} &= \frac{1}{r(\rho)}-\frac{1}{r_s}, &\qquad \Gamma_{\vartheta\vartheta}^{\rho}&=-r(\rho),\\
   \Gamma_{\vartheta\varphi}^{\varphi} &= \cot\vartheta, &\qquad \Gamma_{\varphi\varphi}^{\rho}&=-r(\rho)\sin^2\vartheta,&\quad\Gamma_{\varphi\varphi}^{\vartheta}&=-\sin\vartheta\cos\vartheta.
\end{alignat}
\end{subequations}

% Partial derivatives
% \begin{subequations}
%  \begin{alignat}{5}
%     \Gamma_{tt,r}^r &= -\frac{(2r-3r_s)c^2r_s}{2r^4}, &\qquad \Gamma_{tr,r}^t &= -\frac{(2r-r_s)r_s}{2r^2(r-r_s)^2}, &\qquad \Gamma_{rr,r}^r &= \frac{(2r-r_s)r_s}{2r^2(r-r_s)^2},\\
%     \Gamma_{r\vartheta,r}^{\vartheta} &= -\frac{1}{r^2}, & \Gamma_{r\varphi,r}^{\varphi} &= -\frac{1}{r^2}, & \Gamma_{\vartheta\vartheta,r}^r &= -1,\\
%     \Gamma_{\vartheta\varphi,\vartheta}^{\varphi} &= -\frac{1}{\sin^2\vartheta}, & \Gamma_{\varphi\varphi,r}^r &= -\sin^2\vartheta, & \Gamma_{\varphi\varphi,\vartheta}^{\vartheta} &= -\cos(2\vartheta),\\
%     \Gamma_{\varphi\varphi,\vartheta}^r &= -(r-r_s)\sin(2\vartheta).
%  \end{alignat}
% \end{subequations}

%% -------------------- Riemann tensor --------------------
\SecRiemann
\begin{subequations}
\begin{alignat}{2}
  R_{t\rho t\rho} &= -\frac{c^2r_s}{r(\rho)^3}\left(1-\frac{r_s}{r(\rho)}\right)^2, &\qquad R_{t\vartheta t\vartheta} &= \frac{c^2}{2}\left(1-\frac{r_s}{r(\rho)}\right)\frac{r_s}{r(\rho)},\\
  R_{t\varphi t\varphi} &= \frac{c^2\sin^2\vartheta}{2}\left(1-\frac{r_s}{r(\rho)}\right)\frac{r_s}{r(\rho)}, &\qquad
  R_{\rho\vartheta\rho\vartheta} &= -\frac{1}{2}\left(1-\frac{r_s}{r(\rho)}\right)\frac{r_s}{r(\rho)}\\
  R_{\rho\varphi\rho\varphi} &= -\frac{\sin^2\vartheta}{2}\left(1-\frac{r_s}{r(\rho)}\right)\frac{r_s}{r(\rho)}, & R_{\vartheta\varphi\vartheta\varphi} &= r(\rho)r_s\sin^2\vartheta.
\end{alignat}
\end{subequations}

\noindent The Ricci tensor as well as the Ricci scalar vanish identically because the Schwarz\-schild spacetime is a vacuum solution of the field equations. Hence, the Weyl tensor is identical to the Riemann tensor. The Kretschmann scalar reads
\begin{equation}
  \mathcal{K} = 12\frac{r_s^2}{r(\rho)^6}.
\end{equation}

%% -------------------- Local tetrad --------------------
\SecLocal
\begin{equation}
  \mathbf{e}_{(t)} = \frac{1}{c\sqrt{1-r_s/r(\rho)}}\partial_t, \quad \mathbf{e}_{(\rho)} = \frac{1}{\sqrt{1-r_s/r(\rho)}}\partial_{\rho},\quad \mathbf{e}_{(\vartheta)} = \frac{1}{r(\rho)}\partial_{\vartheta}, \quad \mathbf{e}_{(\varphi)} = \frac{1}{r(\rho)\sin\vartheta}\partial_{\varphi}.
\end{equation}
Dual tetrad:
\begin{equation}
 \boldsymbol{\theta}^{(t)} = c\sqrt{1-\frac{r_s}{r(\rho)}}\,dt,\quad \boldsymbol{\theta}^{(\rho)} = \sqrt{1-\frac{r_s}{r(\rho)}}\,d\rho,\quad \boldsymbol{\theta}^{(\vartheta)}=r(\rho)\,d\vartheta,\quad \boldsymbol{\theta}^{(\varphi)} = r(\rho)\sin\vartheta\,d\varphi.
\end{equation}

%% -------------------- Riemann tensor LT--------------------
\SecRiemannLT
\begin{subequations}
 \begin{align}
  R_{(t)(\rho)(t)(\rho)} &= -R_{(\vartheta)(\varphi)(\vartheta)(\varphi)} = -\frac{r_s}{r(\rho)^3},\\
  R_{(t)(\vartheta)(t)(\vartheta)} &= R_{(t)(\varphi)(t)(\varphi)} = -R_{(\rho)(\vartheta)(\rho)(\vartheta)} = -R_{(\rho)(\varphi)(\rho)(\varphi)} = \frac{r_s}{2r(\rho)^3}.
 \end{align}
\end{subequations}

%% -------------------- Further reading --------------------
\FurtherReading

MTW\cite{mtw}
}{

}

%% ------------------------------------------------------------------------
%%      P a i n l e v e  -  G u l l s t r a n d
%% ------------------------------------------------------------------------
\ifthenelse{\boolean{isARXIV}}{
% ******** Start of file painlevegullstrand.tex *********
%
%  Copyright (c) 2009 Thomas Mueller,
%                     Universitaet Stuttgart, VISUS
%

\subsection{Painlev{\'e}-Gullstrand}
The Schwarzschild metric expressed in Painlev{\'e}-Gullstrand coordinates\cite{martel2001} reads
\metricEq{
  ds^2 = -c^2dT^2+\left(dr+\sqrt{\frac{r_s}{r}}c\,dT\right)^2+r^2\left(d\vartheta^2+\sin^2\vartheta d\varphi^2\right),
}{painleve}
where the new time coordinate $T$ follows from the Schwarzschild time $t$ in the following way:
\begin{equation}
  cT = ct+2r_s\left(\sqrt{\frac{r}{r_s}}+\frac{1}{2}\ln\bigg|\frac{\sqrt{r/r_s}-1}{\sqrt{r/r_s}+1}\bigg|\right).
\end{equation}

%% -------------------- Metric tensor --------------------
\SecMetric
\begin{equation}
 g_{TT} = -c^2\left(1-\frac{r_s}{r}\right),\qquad g_{Tr} = c\sqrt{\frac{r_s}{r}},\qquad g_{rr} = 1,\qquad g_{\vartheta\vartheta} = r^2,\qquad g_{\varphi\varphi}=r^2\sin^2\vartheta.
\end{equation}

%% -------------------- Christoffel symbols --------------------
\SecChristoffel
\begin{subequations}
\begin{alignat}{5}
  \Gamma_{TT}^T &= \frac{cr_s}{2r^2}\sqrt{\frac{r_s}{r}}, &\qquad \Gamma_{TT}^r &= \frac{c^2r_s(r-r_s)}{2r^3}, &\qquad \Gamma_{Tr}^T &= \frac{r_s}{2r^2},\\
  \Gamma_{Tr}^r &= -\frac{cr_s}{2r^2}\sqrt{\frac{r_s}{r}}, & \Gamma_{rr}^T &= \frac{r_s}{2cr^2}\sqrt{\frac{r}{r_s}}, & \Gamma_{rr}^r &= -\frac{r_s}{2r^2},\\
  \Gamma_{r\vartheta}^{\vartheta} &= \frac{1}{r}, & \Gamma_{r\varphi}^{\varphi} &= \frac{1}{r}, & \Gamma_{\vartheta\vartheta}^T &= -\frac{r}{c}\sqrt{\frac{r_s}{r}},\\
  \Gamma_{\vartheta\vartheta}^r &= -(r-r_s), & \Gamma_{\vartheta\varphi}^{\varphi} &= \cot\vartheta, & \Gamma_{\varphi\varphi}^T &= -\frac{r}{c}\sqrt{\frac{r_s}{r}}\sin^2\vartheta,\\
  \Gamma_{\varphi\varphi}^r &= -(r-r_s)\sin^2\vartheta, & \Gamma_{\varphi\varphi}^{\vartheta} &= -\sin\vartheta\cos\vartheta.
\end{alignat}
\end{subequations}

%% -------------------- Riemann tensor --------------------
\SecRiemann
\begin{subequations}
\begin{alignat}{3}
  R_{TrTr} &= -\frac{c^2r_s}{r^3}, &\quad R_{T\vartheta T\vartheta} &= \frac{c^2r_s(r-r_s)}{2r^2}, &\quad R_{T\vartheta r\vartheta} &= -\frac{cr_s}{2r}\sqrt{\frac{r_s}{r}},\\
  R_{T\varphi T\varphi} &= \frac{c^2r_s(r-r_s)\sin^2\vartheta}{2r^2}, & R_{T\varphi r\varphi} &= -\frac{cr_s}{2r}\sqrt{\frac{r_s}{r}}\sin^2\vartheta, & R_{r\vartheta r\vartheta} &= -\frac{r_s}{2r},\\
  R_{r\varphi r\varphi} &= -\frac{r_s\sin^2\vartheta}{2r}, & R_{\vartheta\varphi\vartheta\varphi} &= rr_s\sin^2\vartheta.
\end{alignat}
\end{subequations}

The Ricci tensor and the Ricci scalar vanish identically. 

%% -------------------- Kretschmann scalar --------------------
\SecKretsch
\begin{equation}
  \mathcal{K}=12 r_s^2/r^6.
\end{equation}

For the Painlev{\'e}-Gullstrand coordinates, we can define two natural local tetrads. 

%% -------------------- static local tetrad -------------------
\SecStatLocal
\begin{equation}
  \mathbf{\hat{e}}_{(T)} = \frac{1}{c\sqrt{1-r_s/r}}\partial_T, \quad \mathbf{\hat{e}}_{(r)} = \frac{\sqrt{r_s}}{c\sqrt{r-r_s}}\partial_T + \sqrt{1-\frac{r_s}{r}}\partial_r,\quad \mathbf{\hat{e}}_{(\vartheta)} = \frac{1}{r}\partial_{\vartheta},\quad\mathbf{\hat{e}}_{(\varphi)} = \frac{1}{r\sin\vartheta}\partial_{\varphi},
\end{equation}
Dual tetrad:
\begin{equation}
 \boldsymbol{\hat\theta}^{(T)}=c\sqrt{1-\frac{r_s}{r}}dT-\frac{dr}{\sqrt{r/r_s-1}},\quad \boldsymbol{\hat\theta}^{(r)}=\frac{dr}{\sqrt{1-r_s/r}},\quad \boldsymbol{\hat\theta}^{(\vartheta)}=r\,d\vartheta,\quad \boldsymbol{\hat\theta}^{(\varphi)} = r\sin\vartheta\,d\varphi.
\end{equation}

%% -------------------- Freely falling local tetrad -------------------
\SecFreeLocal
\begin{equation}
  \mathbf{e}_{(T)} = \frac{1}{c}\partial_T-\sqrt{\frac{r_s}{r}}\partial_r, \qquad\mathbf{e}_{(r)}=\partial_r,\qquad\mathbf{e}_{(\vartheta)} = \frac{1}{r}\partial_{\vartheta},\qquad\mathbf{e}_{(\varphi)} = \frac{1}{r\sin\vartheta}\partial_{\varphi}.
\end{equation}
Dual tetrad:
\begin{equation}
 \boldsymbol{\theta}^{(T)} = c\,dT,\qquad \boldsymbol{\theta}^{(r)} = c\sqrt{\frac{r_s}{r}}dT+dr,\qquad \boldsymbol{\theta}^{(\vartheta)} = r\,d\vartheta,\qquad \boldsymbol{\theta}^{(\varphi)} = r\sin\vartheta d\varphi.
\end{equation}

%% -------------------- Riemann tensor LT--------------------
\SecRiemannLT
\begin{subequations}
 \begin{align}
  R_{(T)(r)(T)(r)} &= -R_{(\vartheta)(\varphi)(\vartheta)(\varphi)} = -\frac{r_s}{r^3},\\
  R_{(T)(\vartheta)(T)(\vartheta)} &= R_{(T)(\varphi)(T)(\varphi)} = -R_{(r)(\vartheta)(r)(\vartheta)} = -R_{(r)(\varphi)(r)(\varphi)} = \frac{r_s}{2r^3}.
 \end{align}
\end{subequations}

}{

}

%% ------------------------------------------------------------------------
%%      I s r a e l
%% ------------------------------------------------------------------------
\ifthenelse{\boolean{isARXIV}}{
% ******** Start of file schwarzschildIsrael.tex *********
%
%  Copyright (c) 2009 Thomas Mueller,
%                     Universitaet Stuttgart, VISUS
%

\subsection{Israel coordinates}
The Schwarzschild metric in Israel coordinates $(x,y,\vartheta,\varphi)$ reads\cite{exact2003}
\metricEq{
  ds^2 = r_s^2\left[4dx\left(dy+\frac{y^2dx}{1+xy}\right)+(1+xy)^2\left(d\vartheta^2+\sin^2\vartheta d\varphi^2\right)\right],
}{isreal}
where the coordinates $x$ and $y$ follow from the Schwarzschild coordinates via
\begin{equation}
  t=r_s\left(1+xy+\ln\frac{y}{x}\right)\qquad\text{and}\qquad r=r_s(1+xy).
\end{equation}

%% -------------------- Christoffel symbols --------------------
\SecChristoffel
\begin{subequations}
\begin{alignat}{5}
  \Gamma_{xx}^x &= -\frac{y(2+xy)}{(1+xy)^2}, &\quad \Gamma_{xx}^y &= \frac{y^3(3+xy)}{(1+xy)^3}, &\quad \Gamma_{xy}^y &= \frac{y(2+xy)}{(1+xy)^2},\\
  \Gamma_{x\vartheta}^{\vartheta} &= \frac{y}{1+xy}, & \Gamma_{x\varphi}^{\varphi} &= \frac{y}{1+xy}, & \Gamma_{y\vartheta}^{\vartheta} &= \frac{x}{1+xy},\\
  \Gamma_{x\varphi}^{\varphi} &= \frac{x}{1+xy}, & \Gamma_{\vartheta\vartheta}^x &= -\frac{x}{2}(1+xy), & \Gamma_{\vartheta\vartheta}^y &= -\frac{y}{2}(1-xy),\\
  \Gamma_{\vartheta\varphi}^{\varphi} &= \cot\vartheta, & \Gamma_{\varphi\varphi}^x &= -\frac{x}{2}(1+xy)\sin^2\vartheta, & \Gamma_{\varphi\varphi}^y &= -\frac{y}{2}(1-xy)\sin^2\vartheta,\\
  \Gamma_{\varphi\varphi}^{\vartheta} &= -\sin\vartheta\cos\vartheta.
\end{alignat}
\end{subequations}

%% -------------------- Riemann tensor --------------------
\SecRiemann
\begin{subequations}
\begin{alignat}{5}
  R_{xyxy} &= -4\frac{r_s^ 2}{(1+xy)^3}, &\quad R_{x\vartheta x\vartheta} &= -2\frac{y^2r_s^2}{(1+xy)^2}, &\quad R_{x\vartheta y\vartheta} &= -\frac{r_s^2}{1+xy},\\ 
  R_{x\varphi x\varphi} &= -2\frac{r_s^2y^2\sin^2\vartheta}{(1+xy)^2}, &  R_{x\varphi y\varphi} &= -\frac{r_s^2\sin^2\vartheta}{1+xy}, & R_{\vartheta\varphi\vartheta\varphi} &= (1+xy)r_s^2\sin^2\vartheta.
\end{alignat}
\end{subequations}

\noindent The Ricci tensor as well as the Ricci scalar vanish identically. Hence, the Weyl tensor is identical to the Riemann tensor. The Kretschmann scalar reads
\begin{equation}
  \mathcal{K} = \frac{12}{r_s^4(1+xy)^6}.
\end{equation}

%% -------------------- Local tetrad --------------------
\SecLocal
\begin{subequations}
\begin{alignat}{3} 
  \mathbf{e}_{(0)} &= -\frac{\sqrt{1+xy}}{2r_sy}\partial_x+\frac{y}{r_s\sqrt{1+xy}}\partial_y, &\qquad \mathbf{e}_{(1)} &= \frac{\sqrt{1+xy}}{2r_sy}\partial_x,\\
  \mathbf{e}_{(2)} &= \frac{1}{r_s(1+xy)}\partial_{\vartheta}, &\qquad \mathbf{e}_{(3)} &= \frac{1}{r_s(1+xy)\sin\vartheta}\partial_{\varphi}.
\end{alignat}
\end{subequations}
Dual tetrad:
\begin{subequations}
\begin{alignat}{3}
  \boldsymbol{\theta}^{(0)} &= \frac{r_s\sqrt{1+xy}}{y}\,dy, &\qquad \boldsymbol{\theta}^{(1)} &= \frac{2r_sy}{\sqrt{1+xy}}dx+\frac{r_s\sqrt{1+xy}}{y}\,dy,\\
  \boldsymbol{\theta}^{(2)} &= r_s(1+xy)\,d\vartheta, &\qquad \boldsymbol{\theta}^{(3)} &= r_s(1+xy)\sin\vartheta\,d\varphi.
\end{alignat}
\end{subequations}

}{

}

%% ------------------------------------------------------------------------
%%     A L C U B I E R R E    W A R P
%% ------------------------------------------------------------------------
\clearpage
\section{Alcubierre Warp}
\setcounter{equation}{0}
\ifthenelse{\boolean{isARXIV}}{
% ******** Start of file alcubierre.tex *********
%
%  Copyright (c) 2009 Andreas Lemmer, Thomas Mueller
%                     Universitaet Stuttgart, VISUS
%

The Warp metric given by Miguel Alcubierre\cite{alcubierre1994} reads
\metricEq{
  ds^2 = -c^2dt^2+\left(dx-v_sf(r_s)dt\right)^2 + dy^2 + dz^2
}{alcubierre}
where
\begin{subequations}
\begin{align}
  v_s  &= \frac{dx_s(t)}{dt},\\
  r_s(t) &= \sqrt{(x-x_s(t))^2+y^2+z^2},\\
  f(r_s) &= \frac{\tanh (\sigma (r_s+R)) - \tanh (\sigma (r_s-R))}{2\tanh (\sigma R)}.
\end{align}
\end{subequations}
The parameter $R>0$ defines the radius of the warp bubble and the parameter $\sigma>0$ its thickness.

%% -------------------- Metric tensor --------------------
\SecMetric
\begin{equation}
 g_{tt} = -c^2+v_s^2f(r_s)^2,\qquad g_{tx} = -v_sf(r_s),\qquad g_{xx} = g_{yy} = g_{zz} = 1.
\end{equation}

%% -------------------- Christoffel symbols --------------------
\SecChristoffel
\begin{subequations}
\begin{alignat}{5}
 \Gamma^t_{tt} &= \frac{f^2f_xv_s^3}{c^2}, &\qquad
 \Gamma^z_{tt} &= -ff_zv_s^2, &\qquad
 \Gamma^y_{tt} &= -ff_yv_s^2,\\
 \Gamma^x_{tt} &= \frac{f^3f_xv_s^4-c^2ff_xv_s^2-c^2f_tv_s}{c^2}, &
 \Gamma^t_{tx} &= -\frac{ff_xv_s^2}{c^2}, &
 \Gamma^x_{tx} &= -\frac{f^2f_xv_s^3}{c^2},\\
 \Gamma^y_{tx} &= \frac{f_yv_s}{2}, &
 \Gamma^z_{tx} &= \frac{f_zv_s}{2}, &
 \Gamma^t_{ty} &= -\frac{ff_yv_s^2}{2c^2},\\
 \Gamma^x_{ty} &= -\frac{f^2f_yv_s^3+c^2f_yv_s}{2c^2}, &
 \Gamma^t_{tz} &= -\frac{ff_zv_s^2}{2c^2}, &
 \Gamma^x_{tz} &= -\frac{f^2f_zv_s^3+c^2f_zv_s}{2c^2},\\
 \Gamma^t_{xx} &= \frac{f_xv_s}{c^2}, &
 \Gamma^x_{xx} &= \frac{ff_xv_s^2}{c^2}, &
 \Gamma^t_{xy} &= \frac{f_yv_s}{2c^2},\\
 \Gamma^x_{xy} &= \frac{ff_yv_s^2}{2c^2}, &
 \Gamma^t_{xz} &= \frac{f_zv_s}{2c^2}, &
 \Gamma^x_{xz} &= \frac{ff_zv_s^2}{2c^2},
\end{alignat}
\end{subequations}
with derivatives
\begin{subequations}
\begin{alignat}{4}
 f_t &= \frac{df(r_s)}{dt} = \frac{-v_s\sigma \left( x-x_s(t)\right) }{2r_s\tanh (\sigma R)} \left[ \sech^2\left( \sigma (r_s+R)\right) - \sech^2\left( \sigma (r_s-R)\right) \right]\\
 f_x &= \frac{df(r_s)}{dx} = \frac{\sigma \left( x-x_s(t)\right) }{2r_s\tanh (\sigma R)} \left[ \sech^2\left( \sigma (r_s+R)\right) - \sech^2\left( \sigma (r_s-R)\right) \right]\\
 f_y &= \frac{df(r_s)}{dy} = \frac{\sigma y}{2r_s\tanh (\sigma R)}\left[ \sech^2\left( \sigma (r_s+R)\right) - \sech^2\left( \sigma (r_s-R)\right) \right]\\
 f_z &= \frac{df(r_s)}{dz} = \frac{\sigma z}{2r_s\tanh (\sigma R)} \left[ \sech^2\left( \sigma (r_s+R)\right) - \sech^2\left( \sigma (r_s-R)\right) \right]
\end{alignat}
\end{subequations}

Riemann- and Ricci-tensor as well as Ricci- and Kretschman-scalar are shown only in the Maple worksheet.

%% -------------------- Local tetrad --------------------
\SecComLocal
\begin{equation}
 \mathbf{e}_{(0)} = \frac{1}{c}\left(\partial_t + v_sf\partial_x\right), \quad
 \mathbf{e}_{(1)} = \partial_x,\quad
 \mathbf{e}_{(2)} = \partial_y,\quad
 \mathbf{e}_{(3)} = \partial_z.
\end{equation}

%% -------------------- Local tetrad --------------------
\SecStatLocal
\begin{equation}
 \mathbf{e}_{(0)} = \frac{1}{\sqrt{c^2-v_s^2f^2}}\partial_t,\quad
 \mathbf{e}_{(1)} = \frac{v_sf}{c\sqrt{c^2-v_s^2f^2}}\partial_t + \frac{\sqrt{c^2-v_s^2f^2}}{c}\partial_x,\quad
 \mathbf{e}_{(2)} = \partial_y,\quad
 \mathbf{e}_{(3)} = \partial_z.
\end{equation}

\FurtherReading

Pfenning\cite{pfenning1997}, Clark\cite{clark1999}, Van Den Broeck\cite{vanDenBroeck1999}

}{

}

%% ------------------------------------------------------------------------
%%     B a r r i o l a  -  V i l e n k i n
%% ------------------------------------------------------------------------
\clearpage
\section{Barriola-Vilenkin monopol}
\setcounter{equation}{0}
\ifthenelse{\boolean{isARXIV}}{
% ******** Start of file barriolavilenkin.tex *********
%
%  Copyright (c) 2009 Thomas Mueller,
%                     Universitaet Stuttgart, VISUS
%

The Barriola-Vilenkin metric describes the gravitational field of a global monopole\cite{barriola1989}. In spherical coordinates $(t,r,\vartheta,\varphi)$, the metric reads
\metricEq{
  ds^2 = -c^2dt^2 + dr^2 + k^2r^2\left(d\vartheta^2+\sin^2\!\vartheta\,d\varphi^2\right),
}{barriolaVilenkin}
where $k$ is the scaling factor responsible for the deficit/surplus angle.

%% -------------------- Christoffel symbols --------------------
\SecChristoffel
\begin{subequations}
  \begin{alignat}{3}
    \Gamma_{\vartheta\vartheta}^{r} &= -k^2r, &\qquad \Gamma_{\varphi\varphi}^{r} &= -k^2r\sin^2\vartheta, &\qquad \Gamma_{r\vartheta}^{\vartheta} &= \frac{1}{r},\\
    \Gamma_{\varphi\varphi}^{\vartheta} &= -\sin\vartheta\cos\vartheta, &\qquad \Gamma_{r\varphi}^{\varphi} &= \frac{1}{r}, & \Gamma_{\vartheta\varphi}^{\varphi} &= \cot\vartheta.
  \end{alignat}
\end{subequations}

Partial derivatives
\begin{subequations}
 \begin{alignat}{5}
    \Gamma_{r\vartheta,r}^{\vartheta} &= -\frac{1}{r^2}, &\qquad \Gamma_{r\varphi,r}^{\varphi} &= -\frac{1}{r^2}, &\qquad \Gamma_{\vartheta\vartheta,r}^r &= -k^2,\\    
    \Gamma_{\vartheta\varphi,\vartheta}^{\varphi} &= -\frac{1}{\sin^2\vartheta}, & \Gamma_{\varphi\varphi,r}^r &= -k^2\sin^2\vartheta, & \Gamma_{\varphi\varphi,\vartheta}^{\vartheta} &= -\cos(2\vartheta),\\
    \Gamma_{\varphi\varphi,\vartheta}^r &= -k^2r\sin(2\vartheta).
 \end{alignat}
\end{subequations}

%% -------------------- Riemann tensor --------------------
\SecRiemann
\begin{equation}
  R_{\vartheta\varphi\vartheta\varphi}=(1-k^2)k^2r^2\sin^2\vartheta.
\end{equation}

\noindent {\bf Ricci tensor, Ricci and Kretschmann scalar:}
\begin{equation}
  R_{\vartheta\vartheta} = (1-k^2),\qquad R_{\varphi\varphi} = (1-k^2)\sin^2\vartheta,\qquad \mathcal{R}=2\frac{1-k^2}{k^2r^2},\qquad \mathcal{K} = 4\frac{(1-k^2)^2}{k^4r^4}.
\end{equation}

%% -------------------- Weyl tensor --------------------
\SecWeyl
\begin{subequations}
 \begin{alignat}{5}
  C_{trtr} &= -\frac{c^2(1-k^2)}{3k^2r^2}, &\quad C_{t\vartheta t\vartheta} &= \frac{c^2}{6}(1-k^2), &\quad C_{t\varphi t\varphi} &= \frac{c^2}{6}(1-k^2)\sin^2\vartheta,\\
  C_{r\vartheta r\vartheta} &= -\frac{1}{6}(1-k^2), & C_{r\varphi r\varphi} &= -\frac{1}{6}(1-k^2)\sin^2\vartheta, & C_{\vartheta\varphi\vartheta\varphi} &= \frac{k^2r^2}{3}(1-k^2)\sin^2\vartheta.
 \end{alignat}
\end{subequations}

%% -------------------- Local tetrad --------------------
\SecLocal
\begin{equation}
  \mathbf{e}_{(t)} = \frac{1}{c}\partial_t, \qquad \mathbf{e}_{(r)} = \partial_r, \qquad \mathbf{e}_{(\vartheta)} =\frac{1}{kr}\partial_{\vartheta}, \qquad \mathbf{e}_{(\varphi)} = \frac{1}{kr\,\sin\vartheta}\partial_{\varphi}.
\end{equation}
Dual tetrad:
\begin{equation}
  \boldsymbol{\theta}^{(t)} = c\,dt,\qquad \boldsymbol{\theta}^{(r)} = dr,\qquad \boldsymbol{\theta}^{(\vartheta)} = kr\,d\vartheta,\qquad \boldsymbol{\theta}^{(\varphi)}=kr\sin\vartheta\,d\varphi.
\end{equation}

%% -------------------- Ricci rotation coefficients --------------------
\SecRicRotCoef
\begin{equation}
  \gamma_{(\vartheta)(r)(\vartheta)} = \gamma_{(\varphi)(r)(\varphi)} = \frac{1}{r},\qquad \gamma_{(\varphi)(\vartheta)(\varphi)} = \frac{\cot\vartheta}{kr}.
\end{equation}
The contractions of the Ricci rotation coefficients read
\begin{equation}
  \gamma_{(r)} = \frac{2}{r},\qquad \gamma_{(\vartheta)} = \frac{\cot\vartheta}{kr}.
\end{equation}

%% -------------------- Riemann tensor LT--------------------
\SecRiemannLT
\begin{equation}
  R_{(\vartheta)(\varphi)(\vartheta)(\varphi)} = \frac{1-k^2}{k^2r^2}.
\end{equation}

%% -------------------- Ricci tensor LT--------------------
\SecRicciLT
\begin{equation}
 R_{(\vartheta)(\vartheta)} = R_{(\varphi)(\varphi)} = \frac{1-k^2}{k^2r^2}.
\end{equation}

%% -------------------- Weyl tensor LT--------------------
\SecWeylLT
\begin{subequations}
 \begin{align}
  C_{(t)(r)(t)(r)} &= -C_{(\vartheta)(\varphi)(\vartheta)(\varphi)} = -\frac{1-k^2}{3k^2r^2},\\
  C_{(t)(\vartheta)(t)(\vartheta)} &= C_{(t)(\varphi)(t)(\varphi)} = -C_{(r)(\vartheta)(r)(\vartheta)} = -C_{(r)(\varphi)(r)(\varphi)} = \frac{1-k^2}{6k^2r^2}.
 \end{align}
\end{subequations}

%% -------------------- Embedding --------------------
\SecEmbedding

The embedding function, see Sec.~\ref{sec:embedding}, for $k<1$ reads
\begin{equation}
  \label{eq:bvEmb}
  z = \sqrt{1-k^2}\,r.
\end{equation}

%% -------------------- Euler Lagrange --------------------
\SecEulLag

The Euler-Lagrangian formalism, Sec.~\ref{subsec:EL}, for geodesics in the $\vartheta=\pi/2$ hyperplane yields \begin{equation} 
 \frac{1}{2}\dot{r}^2 + V_{\text{eff}} = \frac{1}{2}\frac{h_1^2}{c^2},\qquad V_{\text{eff}}=\frac{1}{2}\left(\frac{h_2^2}{k^2r^2}-\kappa c^2\right),
\end{equation}
with the constants of motion $h_1=c^2\dot{t}$ and $h_2=k^2r^2\dot{\varphi}$.\\[0.5em]
The point of closest approach $r_{\text{pca}}$ for a null geodesic that starts at $r=r_i$ with $\mathbf{y}=\pm\Clt{t}+\cos\xi\Clt{r}+\sin\xi\Clt{\varphi}$ is given by $r=r_i\sin\xi$. Hence, the $r_{\text{pca}}$ is independent of $k$. The same is also true for timelike geodesics.
\vspace*{0.5cm}

%% -------------------- Further reading --------------------
\FurtherReading

Barriola and Vilenkin\cite{barriola1989}, Perlick\cite{perlick2004}.

}{

}

%% ------------------------------------------------------------------------
%%    B e r t o t t i  -  K a s n e r
%% ------------------------------------------------------------------------
\clearpage
\section{Bertotti-Kasner}
\setcounter{equation}{0}
\ifthenelse{\boolean{isARXIV}}{
% ******** Start of file bertottikasner.tex *********
%
%  Copyright (c) 2009 Thomas Mueller,
%                     Universitaet Stuttgart, VISUS
%

The Bertotti-Kasner spacetime in spherical coordinates $(t,r,\vartheta,\varphi)$ reads\cite{rindler1998}
\metricEq{
  ds^2 = -c^2dt^2+e^{2\sqrt{\Lambda}ct}dr^2+\frac{1}{\Lambda}\left(d\vartheta^2+\sin^2\vartheta d\varphi^2\right),
}{bertottikasner}
where the cosmological constant $\Lambda$ must be positive.

%% -------------------- Christoffel symbols --------------------
\SecChristoffel
\begin{equation}
  \Gamma_{tr}^r = c\sqrt{\Lambda}, \qquad\Gamma_{rr}^t = \frac{\sqrt{\Lambda}}{c}e^{2\sqrt{\Lambda}ct}, \qquad\Gamma_{\vartheta\varphi}^{\varphi} = \cot\vartheta, \qquad \Gamma_{\varphi\varphi}^{\vartheta} = -\sin\vartheta\cos\vartheta.
\end{equation}

Partial derivatives
\begin{equation}
  \Gamma_{rr,t}^t = 2\Lambda e^{2\sqrt{\Lambda}ct}, \qquad \Gamma_{\vartheta\varphi,\vartheta}^{\varphi} = -\frac{1}{\sin^2\vartheta},\qquad \Gamma_{\varphi\varphi,\vartheta}^{\vartheta} = -\cos(2\vartheta).
\end{equation}

%% -------------------- Riemann tensor --------------------
\SecRiemann
\begin{equation}
  R_{trtr} = -\Lambda c^2 e^{2\sqrt{\Lambda}ct},\qquad R_{\vartheta\varphi\vartheta\varphi} = \frac{\sin^2\vartheta}{\Lambda}.
\end{equation}

%% -------------------- Ricci tensor --------------------
\SecRicci
\begin{equation}
  R_{tt} = -\Lambda c^2,\qquad R_{rr} = \Lambda e^{2\sqrt{\Lambda}ct}, \qquad R_{\vartheta\vartheta} = 1,\qquad R_{\varphi\varphi} = \sin^2\vartheta.
\end{equation}

The Ricci and Kretschmann scalars read
\begin{equation}
 \mathcal{R} = 4\Lambda,\qquad \mathcal{K} = 8\Lambda^2.
\end{equation}

%% -------------------- Weyl tensor --------------------
\SecWeyl
\begin{subequations}
 \begin{alignat}{5}
   C_{trtr} &= -\frac{2}{3}\Lambda c^2 e^{2\sqrt{\Lambda}ct}, &\qquad C_{t\vartheta t\vartheta} &= \frac{c^2}{3}, &\qquad C_{t\varphi t\varphi} &= -\frac{1}{3} e^{2\sqrt{\Lambda} ct},\\
  C_{r\vartheta r\vartheta} &= -\frac{1}{3} e^{2\sqrt{\Lambda} ct}, & C_{r\varphi r\varphi} &= -\frac{1}{3}e^{2\sqrt{\Lambda} ct}\sin^2\vartheta, & C_{\vartheta\varphi\vartheta\varphi} &= \frac{2}{3}\frac{\sin^2\vartheta}{\Lambda}.
 \end{alignat}
\end{subequations}

%% -------------------- Local tetrad --------------------
\SecLocal
\begin{equation}
  \mathbf{e}_{(t)} = \frac{1}{c}\partial_t, \qquad \mathbf{e}_{(r)} = e^{-\sqrt{\Lambda}ct}\partial_r,\qquad \mathbf{e}_{(\vartheta)} = \sqrt{\Lambda}\partial_{\vartheta},\qquad \mathbf{e}_{(\varphi)} = \frac{\sqrt{\Lambda}}{\sin\vartheta}\partial_{\varphi}.
\end{equation}
Dual tetrad:
\begin{equation}
  \boldsymbol{\theta}^{(t)} = c\,dt,\qquad \boldsymbol{\theta}^{(r)} = e^{\sqrt{\Lambda}ct}dr,\qquad \boldsymbol{\theta}^{(\vartheta)} = \frac{1}{\sqrt{\Lambda}}d\vartheta,\qquad \boldsymbol{\theta}^{(\varphi)} = \frac{\sin\vartheta}{\sqrt{\Lambda}}d\varphi.
\end{equation}

%% -------------------- Ricci rotation coefficients --------------------
\SecRicRotCoef
\begin{equation}
  \boldsymbol{\gamma}_{(t)(r)(r)} = \sqrt{\Lambda},\qquad \boldsymbol{\gamma}_{(\vartheta)(\varphi)(\varphi)} = -\sqrt{\Lambda}\cot\vartheta.
\end{equation}
The contractions of the Ricci rotation coefficients read
\begin{equation}
  \boldsymbol{\gamma}_{(t)} = -\sqrt{\Lambda},\qquad \boldsymbol{\gamma}_{(\vartheta)} = \sqrt{\Lambda}\cot\vartheta.
\end{equation}

%% -------------------- Riemann tensor LT--------------------
\SecRiemannLT
\begin{equation}
  R_{(t)(r)(t)(r)} = -R_{(\vartheta)(\varphi)(\vartheta)(\varphi)} = -\Lambda.
\end{equation}

%% -------------------- Ricci tensor LT--------------------
\SecRicciLT
\begin{equation}
  R_{(t)(t)} = -R_{(r)(r)} = -R_{(\vartheta)(\vartheta)} = -R_{(\varphi)(\varphi)} = -\Lambda.
\end{equation}

%% -------------------- Weyl tensor LT--------------------
\SecWeylLT
\begin{subequations}
 \begin{align}
   C_{(t)(r)(t)(r)} &= -C_{(\vartheta)(\varphi)(\vartheta)(\varphi)} = -\frac{2\Lambda}{3},\\
   C_{(t)(\vartheta)(t)(\vartheta)} &= C_{(t)(\varphi)(t)(\varphi)} = -C_{(r)(\vartheta)(r)(\vartheta)} = -C_{(r)(\varphi)(r)(\varphi)} = \frac{\Lambda}{3}.
 \end{align}
\end{subequations}

%% -------------------- Euler Lagrange --------------------
\SecEulLag

The Euler-Lagrangian formalism, Sec.~\ref{subsec:EL}, for geodesics in the $\vartheta=\pi/2$ hyperplane yields
\begin{equation}
 c^2\dot{t}^2 = h_1^2e^{-2\sqrt{\Lambda}\,ct}+\Lambda h_2^2-\kappa
\end{equation}
with the constants of motion $h_1=\dot{r}e^{2\sqrt{\Lambda}\,ct}$ and $h_2=\dot{\varphi}/\Lambda$. Thus,
\begin{equation}
 \lambda = \frac{1}{c\sqrt{\Lambda}\sqrt{\Lambda h_2^2-\kappa}}\ln\left(\frac{1+q(t)}{1-q(t)}\frac{1-q(t_i)}{1+q(t_i)}\right), \qquad q(t)=\frac{h_1^2e^{-2\sqrt{\Lambda}\,ct}}{\Lambda h_2^2-\kappa}+1,
\end{equation}
where $t_i$ is the initial time. We can also solve the orbital equation:
\begin{equation}
  r(t) = w(t)-w(t_i)+r_i,\qquad w(t) = -\frac{\sqrt{h_1^2e^{-2\sqrt{\Lambda}\,ct}+\Lambda h_2^2-\kappa}}{h_1\sqrt{\Lambda}},
\end{equation}
where $r_i$ is the initial radial position.
\vspace*{1cm}

%% -------------------- Further reading --------------------
\FurtherReading

Rindler\cite{rindler1998}: \textit{``Every spherically symmetric solution of the generalized vacuum field equations $R_{ij}=\Lambda g_{ij}$ is either equivalent to Kottler's generalization of Schwarzschild space or to the [...] Bertotti-Kasner space (for which $\Lambda$ must be necessarily be positive).''}
}{

}

%% ------------------------------------------------------------------------
%%     B e s s e l - W a v e
%% ------------------------------------------------------------------------
\clearpage
\section{Bessel gravitational wave}
\setcounter{equation}{0}
\ifthenelse{\boolean{isARXIV}}{
% ******** Start of file besselWave.tex *********
%
%  Copyright (c) 2010 Thomas Mueller, Heiko Munz
%                     Universitaet Stuttgart, VISUS
%

D. Kramer introduced in \cite{kramer1999} an exact gravitational wave solution of Einstein's vacuum field equations.
According to \cite{stephani2003} we execute the substitution $x\rightarrow t$ and $y\rightarrow z$.

%% ------------------------------------------------------
%%    Cylindrical coordinates
%% ------------------------------------------------------
\subsection{Cylindrical coordinates}
The metric of the Bessel wave in cylindrical coordinates reads
\metricEq{
  ds^{2} = \mathrm{e}^{-2U}\left[\mathrm{e}^{2K}\left(d\rho^{2}-d t^{2}\right) + \rho^{2}d\varphi^{2}\right] 
         + \mathrm{e}^{2U}dz^{2}.
}{besselWaveCylindric}
The functions $U$ and $K$ are given by 
\begin{align}
  U &:= CJ_{0}\left(\rho\right)\cos\left(t\right), \\
  K &:= \frac{1}{2}C^{2}\rho\left\{\rho\left[J_{0}\left(\rho\right)^{2}+J_{1}\left(\rho\right)^{2}\right]
       -2J_{0}\left(\rho\right)J_{1}\left(\rho\right)\cos^{2}\left(t\right)\right\},
\end{align}
where $J_{n}\left(\rho\right)$ are the Bessel functions of the first kind.

%% -------------------- Christoffel symbols --------------------
\SecChristoffel
\begin{subequations}
\begin{align}
  &\Cchris{tt}{t} = \Cchris{t\rho}{\rho} = \Cchris{\rho\rho}{t} = -\pdiff{U}{t}+\pdiff{K}{t},& 
  &\Cchris{t\varphi}{\varphi} = \Cchris{tz}{z} = -\pdiff{U}{t},&
  &\Cchris{\varphi\varphi}{t} = -\mathrm{e}^{-2K}\rho^{2}\pdiff{U}{t},\\
  &\Cchris{tt}{\rho} = \Cchris{t\rho}{t} = \Cchris{\rho\rho}{\rho} = -\pdiff{U}{\rho}+\pdiff{K}{\rho},&
  &\Cchris{\rho\varphi}{\varphi} = \frac{1}{\rho}-\pdiff{U}{\rho},&
  &\Cchris{zz}{\rho} = -\mathrm{e}^{4U-2K}\pdiff{U}{\rho},\\
  &\Cchris{\varphi\varphi}{\rho} = \rho\mathrm{e}^{-2K}\left(\rho\pdiff{U}{\rho}-1\right),& 
  &\Cchris{\rho z}{z} = \pdiff{U}{\rho},&
  &\Cchris{zz}{t} = \mathrm{e}^{4U-2K}\pdiff{U}{t}.
\end{align}
\end{subequations}

%% -------------------- Local tetrad --------------------
\SecLocal
\begin{equation}
  \Clt{t} = \mathrm{e}^{U-K}\partial_{t}, \quad \Clt{\rho} = \mathrm{e}^{U-K}\partial_{\rho}, \quad 
  \Clt{\varphi} = \frac{1}{\rho}\mathrm{e}^{U}\partial_{\varphi}, \quad \Clt{z} = \mathrm{e}^{-U}\partial_{z}.
\end{equation}
Dual tetrad:
\begin{equation}
  \Cdlt{t} = \mathrm{e}^{K-U} dt, \quad \Cdlt{\rho} = \mathrm{e}^{K-U} d\rho, \quad 
  \Cdlt{\varphi} = \rho\mathrm{e}^{-U} d\varphi, \quad \Cdlt{z} = \mathrm{e}^{U} dz.
\end{equation}

%% ------------------------------------------------------
%%    Cartesian coordinates
%% ------------------------------------------------------
\subsection{Cartesian coordinates}
In Cartesian coordinates with $\rho=\sqrt{x^2+y^2}$ the metric (\ref{eqM:besselWaveCylindric}) reads
\metricEq{
  \begin{split}
  ds^{2}& = -\mathrm{e}^{2\left(K-U\right)} dt^{2} + \frac{\mathrm{e}^{-2U}}{x^{2}+y^{2}} 
               \bigg[ \left(\mathrm{e}^{2K}x^{2}+y^{2}\right) dx^{2} + 2xy\left(\mathrm{e}^{2K}-1\right) dx dy \\
               &\quad + \left(x^{2}+\mathrm{e}^{2K}y^{2}\right) dy^{2} \bigg] + \mathrm{e}^{2U} dz^{2}.
\end{split}
}{besselWaveCartesian}

%% -------------------- Local tetrad --------------------
\SecLocal
\begin{equation}
\begin{split}
  \Clt{t} &= \mathrm{e}^{U-K}\partial_{t}, \qquad 
  \Clt{x} = \mathrm{e}^{U}\sqrt{\frac{x^2+y^2}{\mathrm{e}^{2K}x^2+y^2}} \partial_{x}, \\ 
  \Clt{y} &= \mathrm{e}^{U-K}\sqrt{\frac{\mathrm{e}^{2K}x^2+y^2}{x^2+y^2}} \partial_{y} 
          + xy\frac{\mathrm{e}^{U-K}\left(\mathrm{e}^{2K}-1\right)}
                   {\sqrt{\left(x^2+y^2\right)\left(\mathrm{e}^{2K}x^2+y^2\right)}} \partial_{x}, \qquad
  \Clt{z} = \mathrm{e}^{-U}\partial_{z}
\end{split}
\end{equation}

}{

}

%% ------------------------------------------------------------------------
%%    Cosmic string in Schwarzschild spacetime
%% ------------------------------------------------------------------------
\clearpage
\section{Cosmic string in Schwarzschild spacetime}
\setcounter{equation}{0}
\ifthenelse{\boolean{isARXIV}}{
% ******** Start of file schwarzschildString.tex *********
%
%  Copyright (c) 2009 Thomas Mueller,
%                     Universitaet Stuttgart, VISUS
%

A cosmic string in the Schwarzschild spacetime represented by Schwarzschild coordinates $(t,r,\vartheta,\varphi)$ reads
\metricEq{
  ds^2 = -\left(1-\frac{r_s}{r}\right)c^2dt^2+\frac{1}{1-r_s/r}dr^2 + r^2\left(d\vartheta^2+\beta^2\sin^2\vartheta d\varphi^2\right),
}{schwarzschildString}
where $r_s=2GM/c^2$ is the Schwarzschild radius, $G$ is Newton's constant, $c$ is the speed of light, $M$ is the mass of the black hole, and $\beta$ is the string parameter, compare  Aryal et al\cite{aryal1986}.

%% -------------------- Christoffel symbols --------------------
\SecChristoffel
\begin{subequations}
\begin{alignat}{5}
  \Gamma_{tt}^r &= \frac{c^2r_s(r-r_s)}{2r^3}, &\qquad\Gamma_{tr}^t &= \frac{r_s}{2r(r-r_s)}, &\qquad\Gamma_{rr}^r &= -\frac{r_s}{2r(r-r_s)},\\
  \Gamma_{r\vartheta}^{\vartheta} &= \frac{1}{r}, &\qquad \Gamma_{r\varphi}^{\varphi} &= \frac{1}{r}, &\qquad \Gamma_{\vartheta\vartheta}^{r}&=-(r-r_s),\\
   \Gamma_{\vartheta\varphi}^{\varphi} &= \cot\vartheta, &\qquad \Gamma_{\varphi\varphi}^{r}&=-(r-r_s)\beta^2\sin^2\vartheta,&\quad\Gamma_{\varphi\varphi}^{\vartheta}&=-\beta^2\sin\vartheta\cos\vartheta.
\end{alignat}
\end{subequations}

Partial derivatives
\begin{subequations}
 \begin{alignat}{5}
    \Gamma_{tt,r}^r &= -\frac{(2r-3r_s)c^2r_s}{2r^4}, &\quad \Gamma_{tr,r}^t &= -\frac{(2r-r_s)r_s}{2r^2(r-r_s)^2}, &\quad \Gamma_{rr,r}^r &= \frac{(2r-r_s)r_s}{2r^2(r-r_s)^2},\\
    \Gamma_{r\vartheta,r}^{\vartheta} &= -\frac{1}{r^2}, & \Gamma_{r\varphi,r}^{\varphi} &= -\frac{1}{r^2}, & \Gamma_{\vartheta\vartheta,r}^r &= -1,\\
    \Gamma_{\vartheta\varphi,\vartheta}^{\varphi} &= -\frac{1}{\sin^2\vartheta}, & \Gamma_{\varphi\varphi,r}^r &= -\beta^2\sin^2\vartheta, & \Gamma_{\varphi\varphi,\vartheta}^{\vartheta} &= -\beta^2\cos(2\vartheta),\\
    \Gamma_{\varphi\varphi,\vartheta}^r &= -(r-r_s)\beta^2\sin(2\vartheta).
 \end{alignat}
\end{subequations}

%% -------------------- Riemann tensor --------------------
\SecRiemann
\begin{subequations}
\begin{alignat}{3}
  R_{trtr} &= -\frac{c^2r_s}{r^3}, &\quad R_{t\vartheta t\vartheta} &= \frac{1}{2}\frac{c^2\left(r-r_s\right)r_s}{r^2}, &\quad R_{t\varphi t\varphi} &= \frac{1}{2}\frac{c^2\left(r-r_s\right)r_s\beta^2\sin^2\vartheta}{r^2},\\
  R_{r\vartheta r\vartheta} &= -\frac{1}{2}\frac{r_s}{r-r_s}, &\quad R_{r\varphi r\varphi} &= -\frac{1}{2}\frac{r_s\beta^2\sin^2\vartheta}{r-r_s}, & R_{\vartheta\varphi\vartheta\varphi} &= rr_s\beta^2\sin^2\vartheta.
\end{alignat}
\end{subequations}

\noindent The Ricci tensor as well as the Ricci scalar vanish identically. Hence, the Weyl tensor is identical to the Riemann tensor. The Kretschmann scalar reads
\begin{equation}
  \mathcal{K} = 12\frac{r_s^2}{r^6}.
\end{equation}

%% -------------------- Local tetrad --------------------
\SecLocal
\begin{equation}
  \mathbf{e}_{(t)} = \frac{1}{c\sqrt{1-r_s/r}}\partial_t, \qquad \mathbf{e}_{(r)} = \sqrt{1-\frac{r_s}{r}}\partial_r,\qquad \mathbf{e}_{(\vartheta)} = \frac{1}{r}\partial_{\vartheta}, \qquad \mathbf{e}_{(\varphi)} = \frac{1}{r\beta\sin\vartheta}\partial_{\varphi}.
\end{equation}
Dual tetrad:
\begin{equation}
 \boldsymbol{\theta}^{(t)} = c\sqrt{1-\frac{r_s}{r}}\,dt,\qquad \boldsymbol{\theta}^{(r)} = \frac{dr}{\sqrt{1-r_s/r}},\qquad \boldsymbol{\theta}^{(\vartheta)}=r\,d\vartheta,\qquad \boldsymbol{\theta}^{(\varphi)} = r\beta\sin\vartheta\,d\varphi.
\end{equation}

%% -------------------- Ricci rotation coefficients --------------------
\SecRicRotCoef
\begin{equation}
  \gamma_{(r)(t)(t)} = \frac{r_s}{2r^2\sqrt{1-r_s/r}},\quad \gamma_{(\vartheta)(r)(\vartheta)} = \gamma_{(\varphi)(r)(\varphi)} = \frac{1}{r}\sqrt{1-\frac{r_s}{r}},\quad \gamma_{(\varphi)(\vartheta)(\varphi)} = \frac{\cot\vartheta}{r}.
\end{equation}
The contractions of the Ricci rotation coefficients read
\begin{equation}
   \gamma_{(r)} = \frac{4r-3r_s}{2r^2\sqrt{1-r_s/r}},\qquad \gamma_{(\vartheta)} = \frac{\cot\vartheta}{r}.
\end{equation}

%% -------------------- Riemann tensor LT--------------------
\SecRiemannLT
\begin{subequations}
 \begin{align}
  R_{(t)(r)(t)(r)} &= -R_{(\vartheta)(\varphi)(\vartheta)(\varphi)} = -\frac{r_s}{r^3},\\
  R_{(t)(\vartheta)(t)(\vartheta)} &= R_{(t)(\varphi)(t)(\varphi)} = -R_{(r)(\vartheta)(r)(\vartheta)} = -R_{(r)(\varphi)(r)(\varphi)} = \frac{r_s}{2r^3}.
 \end{align}
\end{subequations}

%% -------------------- Embedding --------------------
\SecEmbedding

The embedding function for $\beta^2<1$ reads
\begin{equation}
  z = (r-r_s)\sqrt{\frac{r}{r-r_s}-\beta^2}-\frac{r_s}{2\sqrt{1-\beta^2}}\ln\frac{\sqrt{r/(r-r_s)-\beta^2}-\sqrt{1-\beta^2}}{\sqrt{r/(r-r_s)-\beta^2}+\sqrt{1-\beta^2}}.
\end{equation}
If $\beta^2=1$, we have the embedding function of the standard Schwarzschild metric, compare Eq.(\ref{eq:schwEmb}).
\vspace*{0.5cm}

%% -------------------- Euler Lagrange --------------------
\SecEulLag

The Euler-Lagrangian formalism, Sec.~\ref{subsec:EL}, for geodesics in the $\vartheta=\pi/2$ hyperplane yields \begin{equation}
  \frac{1}{2}\dot{r}^2+V_{\text{eff}} = \frac{1}{2}\frac{k^2}{c^2},\qquad V_{\text{eff}}=\frac{1}{2}\left(1-\frac{r_s}{r}\right)\left(\frac{h^2}{r^2\beta^2}-\kappa c^2\right)
\end{equation}
with the constants of motion $k=(1-r_s/r)c^2\dot{t}$ and $h=r^2\beta^2\dot{\varphi}$. The maxima of the effective potential $V_{\text{eff}}$ lead to the same critical orbits $r_{\text{po}}=\frac{3}{2}r_s$ and $r_{\text{itcg}}=3r_s$ as in the standard Schwarzschild metric.

}{

}

%% ------------------------------------------------------------------------
%%    Ernst spacetime
%% ------------------------------------------------------------------------
\clearpage
\section{Ernst spacetime}
\setcounter{equation}{0}
\ifthenelse{\boolean{isARXIV}}{
% ******** Start of file ernst.tex *********
%
%  Copyright (c) 2010 Thomas Mueller,
%                     Universitaet Stuttgart, VISUS
%

\textit{``The Ernst metric is a static, axially symmetric, electro-vacuum solution of the Einstein-Maxwell equations with a black hole immersed in a magnetic field.''}\cite{karas1992}

In spherical coordinates $(t,r,\vartheta,\varphi)$, the Ernst metric reads\cite{ernst1976} $(G=c=1)$
\metricEq{
  ds^2 = \Lambda^2\left[-\left(1-\frac{2M}{r}\right)dt^2+\frac{dr^2}{1-2M/r}+r^2\,d\vartheta^2\right]+\frac{r^2\sin^2\vartheta}{\Lambda^2}d\varphi^2,
}{Ernst}
where $\Lambda = 1+B^2r^2\sin^2\vartheta$. Here, $M$ is the mass of the black hole and $B$ the magnetic field strength.

%% -------------------- Christoffel symbols --------------------
\SecChristoffel
\begin{subequations}
 \begin{alignat}{5}
   \Cchris{tt}{r} &= \frac{\left(2B^2r^3\sin^2\vartheta-3MB^2r^2\sin^2\vartheta+M\right)\left(r-2M\right)}{r^3\Lambda}, &
  \Cchris{tt}{\vartheta} &= \frac{2\left(r-2M\right)B^2\sin\vartheta\cos\vartheta}{r\Lambda},\\
  \Cchris{tr}{t} &= \frac{2B^2r^3\sin^2\vartheta-3MB^2r^2\sin^2\vartheta+M}{r\left(r-2M\right)\Lambda}, & \Cchris{t\vartheta}{t} &= \frac{2B^2r^2\sin\vartheta\cos\vartheta}{\Lambda},\\
  \Cchris{rr}{r} &= \frac{2B^2r^3\sin^2\vartheta-5MB^2r^2\sin^2\vartheta-M}{r\left(r-2M\right)\Lambda}, & \Cchris{rr}{\vartheta} &= -\frac{2B^2r\sin\vartheta\cos\vartheta}{\left(r-2M\right)\Lambda},\\
  \Cchris{r\vartheta}{r} &= \frac{2B^2r^2\sin\vartheta\cos\vartheta}{\Lambda}, & \Cchris{r\vartheta}{\vartheta} &= \frac{3B^2r^2\sin^2\vartheta+1}{r\Lambda},\\
  \Cchris{r\varphi}{\varphi} &= \frac{1-B^2r^2\sin^2\vartheta}{r\Lambda}, & \Cchris{\vartheta\vartheta}{r} &= \frac{\left(3B^2r^2\sin^2\vartheta+1\right)\left(r-2M\right)}{\Lambda},\\
  \Cchris{\vartheta\vartheta}{\vartheta} &= \frac{2B^2r^2\sin\vartheta\cos\vartheta}{\Lambda}, & \Cchris{\vartheta\varphi}{\varphi} &= \frac{\Xi\cos\vartheta}{\Lambda},\\
  \Cchris{\varphi\varphi}{r} &= \frac{\left(r-2M\right)\Xi\sin^2\vartheta}{\Lambda^5},\\
  \Cchris{\varphi\varphi}{\vartheta} &= \frac{\Xi\sin\vartheta\cos\vartheta}{\Lambda^5}.
 \end{alignat}
\end{subequations}
with $\Xi=1-B^2r^2\sin^2\vartheta$.

%% -------------------- Riemann tensor --------------------
\SecRiemann
\begin{subequations}
 \begin{align}
   R_{trtr} &= \frac{2}{r^3}\left[B^4r^4\sin^4\vartheta\left(3M-r\right)-M+2r^5B^4\sin^2\vartheta\cos^2\vartheta+B^2r^2\sin^2\vartheta\left(r-2M\right)\right],\\
   R_{trt\vartheta} &= 2B^2\sin\vartheta\cos\vartheta\left[(3B^2r^2\sin^2\vartheta\left(2M-3r\right)+r-2M\right],\\
   R_{t\vartheta t\vartheta} &= \frac{1}{r^2}\left[B^4r^4(r-2M)(4r-9M)\sin^4\vartheta + 2\Xi B^2r^3(r-2M)\cos^2\vartheta + M(r-2M)\right],\\
   R_{t\varphi t\varphi} &= \frac{1}{\Lambda^4r^2}\left[(2B^2r^3-3B^2Mr^2\sin^2\vartheta+M)\Xi(r-2M)\sin^2\vartheta\right],\\
   R_{r\vartheta r\vartheta} &= -\frac{(2B^2r^3-3B^2Mr^2\sin^2\vartheta+M)\Xi}{r-2M},\\
   R_{r\varphi r\varphi} &= -\frac{\sin^2\vartheta}{\Lambda^4(r-2M)}\left[B^4r^4(4r-9M)\sin^4\vartheta+2B^2r^2(8M-4r\vartheta)\sin^2\vartheta+2\Xi B^2r^3\cos^2\vartheta+M\right],\\
   R_{r\varphi\vartheta\varphi} &= -\frac{2B^2r^3\sin^3\vartheta\cos\vartheta\left(3B^2r^2\sin^2\vartheta-5\right)}{\Lambda^4},\\
   R_{\vartheta\varphi\vartheta\varphi} &= \frac{r\sin^2\vartheta}{\Lambda^4}\left[2B^4r^4(r-3M)\sin^4\vartheta+4B^2r^3\cos^2\vartheta(1+\Xi)+2B^2r^2\sin^2\vartheta(2M-r)+2M\right].
 \end{align}
\end{subequations}

%% -------------------- Rricci tensor --------------------
\SecRicci
\begin{subequations}
 \begin{alignat}{5}
   R_{tt} &= \frac{4B^2(r-2M)(r+2M\sin^2\vartheta)}{r^2\Lambda^2}, &\quad R_{rr} &= -\frac{4B^2[r\cos^2\vartheta-(r-2M)\sin^2\vartheta]}{(r-2M)\Lambda^2},\\
   R_{r\vartheta} &= \frac{8B^2r\sin\vartheta\cos\vartheta}{\Lambda^2}, & R_{\vartheta\vartheta} &= \frac{4B^2r\left[r\cos^2\vartheta+(r-2M)\sin^2\vartheta\right]}{\Lambda^2},\\
   R_{\varphi\varphi} &= \frac{4B^2r\sin^2\vartheta\left(r+2M\sin^2\vartheta\right)}{\Lambda^6}.
 \end{alignat}
\end{subequations}

\noindent {\bf Ricci and Kretschmann scalars:}
\begin{subequations}
 \begin{align}
  R &= 0,\\
  \nonumber\mathcal{K} &= \frac{16}{r^6\Lambda^8}\bigg[3B^8r^8\left(4r^2-18Mr+21M^2\right)\sin^8\vartheta\\
  \nonumber &\quad + 2B^4r^4\left(31M^2-37Mr-24B^2r^4\cos^2\vartheta+42B^2Mr^3\cos^2\vartheta+10r^2+6B^4r^6\cos^4\vartheta\right)\sin^6\vartheta\\
  \nonumber&\quad + 2B^2r^2\left(-3Mr+20B^2r^4\cos^2\vartheta+6M^2-46B^2Mr^3\cos^2\vartheta-12B^4r^6\cos^4\vartheta\right)\sin^4\vartheta\\
  \nonumber&\quad - 6B^6r^6\left(6B^2Mr^3\cos^2\vartheta+4r^2-4B^2r^4\cos^2\vartheta+18M^2-17Mr\right)\\
   &\quad + 20B^4r^6\cos^4\vartheta+12B^2Mr^3\cos^2\vartheta+3M^2\bigg].
 \end{align}
\end{subequations}

%% -------------------- static local tetrad --------------------
\SecStatLocal
\begin{equation}
  \Clt{t} = \frac{1}{\Lambda\sqrt{1-2m/r}}\partial_t, \quad \Clt{r} = \frac{\sqrt{1-2m/r}}{\Lambda}\partial_r,\quad \Clt{\vartheta} = \frac{1}{\Lambda r}\partial_{\vartheta}, \quad \Clt{\varphi} = \frac{\Lambda}{r\sin\vartheta}\partial_{\varphi}.
\end{equation}
Dual tetrad:
\begin{equation}
 \Cdlt{t} = \Lambda\sqrt{1-\frac{2m}{r}}dt,\quad \Cdlt{r} = \frac{\Lambda}{\sqrt{1-2m/r}}dr,\quad \Cdlt{\vartheta} = \Lambda r\,d\vartheta,\quad \Cdlt{\varphi} = \frac{r\sin\vartheta}{\Lambda}\,d\varphi.
\end{equation}

%% -------------------- Euler Lagrange --------------------
\SecEulLag

The Euler-Lagrangian formalism, Sec.~\ref{subsec:EL}, for geodesics in the $\vartheta=\pi/2$ hyperplane yields \begin{equation}                                                                                               
 \dot{r}^2 + \frac{h^2(1-r_s/r)}{r^2} - \frac{k^2}{\Lambda^4} + \kappa\frac{1-r_s/r}{\Lambda^2} = 0
\end{equation}
with constants of motion $k=\Lambda^2(1-r_s/r)\dot{t}$ and $h=(r^2/\Lambda^2)\dot{\varphi}$.
\vspace*{0.5cm}

%% -------------------- Further reading --------------------
\FurtherReading

Ernst\cite{ernst1976}, Dhurandhar and Sharma\cite{dhurandhar1983}, Karas and Vokrouhlicky\cite{karas1992}, Stuchl{\'i}k and Hled{\'i}k\cite{stucklik1999}.
}{

}

%% ------------------------------------------------------------------------
%%    F r i e d m a n  -  R o b e r t s o n  -  W a l k e r
%% ------------------------------------------------------------------------
\clearpage
\section{Friedman-Robertson-Walker}
\setcounter{equation}{0}
\ifthenelse{\boolean{isARXIV}}{
% ******** Start of file FRW.tex *********
%
%  Copyright (c) 2009 Thomas Mueller,
%                     Universitaet Stuttgart, VISUS
%

The Friedman-Robertson-Walker metric describes a general homogeneous and isotropic universe. In a general form it reads:
\begin{equation}
 \label{eq:FRW}
  ds^2 = -c^2dt^2 + R^2d\sigma^2
\end{equation}
with $R=R(t)$ being an arbitrary function of time only and $d\sigma^2$ being a metric of a 3-space of constant curvature for which three explicit forms will be described here.
\\
In all formulas in this section a dot denotes differentiation with respect to $t$, e.g.  $\dot{R}=dR(t)/dt$.

%% ------------------------------------------------------
%%    Form 1
%% ------------------------------------------------------
\subsection{Form 1}
\metricEq{
  ds^2 = -c^2dt^2 + R^2\left\{\frac{d\eta^2}{1-k\eta^2} + \eta^2\left(d\vartheta^2 + \sin^2\vartheta d\varphi^2\right)\right\}
}{FRW1}

\SecChristoffel
\begin{subequations}
\begin{alignat}{5}
\Gamma_{t\eta}^\eta &= \dfrac{\dot{R}}{R}, &\, \Gamma_{t\vartheta}^\vartheta &= \frac{\dot{R}}{R}, &\, \Gamma_{t\varphi}^\varphi &= \frac{\dot{R}}{R}, \\
\Gamma_{\eta\eta}^t &= \frac{R\dot{R}}{c^2(1-k\eta^2)}, &\, \Gamma_{\eta\eta}^\eta &= \frac{k\eta}{1-k\eta^2}, &\, \Gamma_{\eta\vartheta}^\vartheta &= \frac{1}{\eta}, \\
\Gamma_{\eta\varphi}^\varphi &= \frac{1}{\eta}, &\, \Gamma_{\vartheta\vartheta}^t &= \frac{R\eta^2\dot{R}}{c^2}, &\, \Gamma_{\vartheta\vartheta}^\eta &= (k\eta^2 -1)\eta, \\
\Gamma_{\vartheta\varphi}^\varphi &= \cot\vartheta, &\, \Gamma_{\varphi\varphi}^t &= \frac{R\eta^2\sin^2\vartheta\dot{R}}{c^2}, &\qquad \Gamma_{\varphi\varphi}^\eta &= (k\eta^2 - 1)\eta\sin^2\vartheta, \\
\Gamma_{\varphi\varphi}^\vartheta &= -\sin\vartheta\cos\vartheta.
\end{alignat}
\end{subequations}

\SecRiemann
\begin{subequations}
\begin{alignat}{3}
  R_{t\eta t\eta} &= \frac{R\ddot{R}}{k\eta^2 - 1}, &\qquad R_{t\vartheta t\vartheta} &= -R\eta^2\ddot{R}, \\
  R_{t\varphi t\varphi} &= -R\eta^2\sin^2\vartheta\ddot{R}, &\qquad R_{\eta\vartheta\eta\vartheta} &= -\dfrac{R^2\eta^2\left(\dot{R}^2+kc^2\right)}{c^2(k\eta^2 - 1)}, \\
  R_{\eta\varphi\eta\varphi} &= -\dfrac{R^2\eta^2\sin^2\vartheta\left(\dot{R}^2+kc^2\right)}{c^2(k\eta^2 - 1)}, &\qquad R_{\vartheta\varphi\vartheta\varphi} &= \dfrac{R^2\eta^4\sin^2\vartheta\left(\dot{R}^2+kc^2\right)}{c^2}.
\end{alignat}
\end{subequations}

\SecRicci
\begin{subequations}
\begin{alignat}{3}
  R_{tt} &= -3\frac{\ddot{R}}{R},&\qquad R_{\eta\eta} &=\frac{R\ddot{R} + 2(\dot{R}^2 + kc^2)}{c^2(1-k\eta^2)}, \\
  R_{\vartheta\vartheta} &= \eta^2\frac{R\ddot{R} + 2(\dot{R}^2 + kc^2)}{c^2},&\qquad R_{\varphi\varphi} &= \eta^2\sin^2\vartheta\frac{R\ddot{R} + 2(\dot{R}^2 + kc^2)}{c^2}.
\end{alignat}
\end{subequations}

\noindent The {\sl Ricci scalar and Kretschmann scalar read:}

\begin{equation}
  \mathcal{R} = 6\frac{R\ddot{R} + \dot{R}^2 + kc^2}{R^2c^2},\qquad \mathcal{K}=12\frac{\ddot{R}^2R^2 + \dot{R}^4 + 2\dot{R}^2kc^2 + k^2c^4}{R^4c^4}.
\end{equation}

%% -------------------- Local tetrad --------------------

\SecLocal
\begin{equation}
 e_{(t)} = \frac{1}{c}\partial_{t}, \qquad e_{(\eta)} = \frac{\sqrt{1-k\eta^2}}{R}\partial_{\eta}, \qquad e_{\vartheta} = \frac{1}{R\eta}\partial_{\vartheta}, \qquad e_{\varphi} = \frac{1}{R\eta\sin{\vartheta}}\partial_{\varphi}.
\end{equation}

%% -------------------- Ricci rotation coefficients --------------------
\SecRicRotCoef
\begin{equation}
\begin{aligned}
  &\gamma_{(\eta)(t)(\eta)} = \gamma_{(\vartheta)(t)(\vartheta)} = \gamma_{(\varphi)(t)(\varphi)} = \frac{\dot{R}}{Rc}\qquad \gamma_{(\vartheta)(\eta)(\vartheta)} = \gamma_{(\varphi)(\eta)(\varphi)} = \frac{\sqrt{1-k\eta^2}}{R\eta}, \\
  &\gamma_{(\varphi)(\vartheta)(\varphi)} = \frac{\cot\vartheta}{R\eta}.
\end{aligned}
\end{equation}
The contractions of the Ricci rotation coefficients read
\begin{equation}
  \gamma_{(t)} = \frac{3\dot{R}}{Rc},\qquad \gamma_{(r)} = \frac{2\sqrt{1-k\eta^2}}{R\eta},\qquad \gamma_{(\vartheta)} = \frac{\cot\vartheta}{R\eta}.
\end{equation}

%% -------------------- Riemann tensor LT--------------------
\SecRiemannLT
\begin{subequations}
\begin{align}
   R_{(t)(\eta)(t)(\eta)} &= R_{(t)(\vartheta)(t)(\vartheta)} = R_{(t)(\varphi)(t)(\varphi)} = -\frac{\ddot{R}}{Rc^2} \\
   R_{(\eta)(\vartheta)(\eta)(\vartheta)} &= R_{(\eta)(\varphi)(\eta)(\varphi)} = R_{(\vartheta)(\varphi)(\vartheta)(\varphi)} = \frac{\dot{R}^2 + kc^2}{R^2c^2} .
\end{align}
\end{subequations}

%% -------------------- Ricci tensor LT--------------------
\SecRicciLT
\begin{equation} 
  R_{(t)(t)} = -\frac{3\ddot{R}}{Rc^2},\qquad R_{(r)(r)} = R_{(\vartheta)(\vartheta)} = R_{(\varphi)(\varphi)} = \frac{R\ddot{R} + 2\dot{R}^2 + 2kc^2}{R^2c^2}.
\end{equation}

%% ------------------------------------------------------
%%    Form 2
%% ------------------------------------------------------
\subsection{Form 2}
\metricEq{
  ds^2 = -c^2dt^2 + \dfrac{R^2}{(1+\frac{k}{4}r^2)^2}\left\{dr^2 + r^2(d\vartheta^2 + \sin^2\vartheta d\varphi^2)\right\}
}{FRW2}

%% -------------------- Christoffel symbols --------------------
\SecChristoffel
\begin{subequations}
\begin{alignat}{5}
\Gamma_{tr}^r &= \dfrac{\dot{R}}{R}, &\, \Gamma_{t\vartheta}^\vartheta &= \frac{\dot{R}}{R}, &\, \Gamma_{t\varphi}^\varphi &= \frac{\dot{R}}{R}, \\
\Gamma_{rr}^t &= 16\frac{R\dot{R}}{c^2(4+kr^2)^2}, &\, \Gamma_{rr}^r &= -\frac{2kr}{4+kr^2}, &\, \Gamma_{r\vartheta}^\vartheta &= \frac{4-kr^2}{(4+kr^2)r}, \\
\Gamma_{r\varphi}^\varphi &= \frac{4-kr^2}{(4+kr^2)r}, &\,\Gamma_{\vartheta\vartheta}^t &= 16\frac{Rr^2\dot{R}}{c^2(4+kr^2)^2}, &\, \Gamma_{\vartheta\vartheta}^r &= \frac{r(kr^2-4)}{4+kr^2}, \\
\Gamma_{\vartheta\varphi}^\varphi &= \cot\vartheta, &\,\Gamma_{\varphi\varphi}^t &= 16\frac{Rr^2\sin^2\vartheta\dot{R}}{c^2(4+kr^2)^2}, &\quad \Gamma_{\varphi\varphi}^\vartheta &= -\sin\vartheta\cos\vartheta, \\
\Gamma_{\varphi\varphi}^r &= \frac{r\sin^2\vartheta(kr^2-4)}{4+kr^2}.
\end{alignat}
\end{subequations}

%% -------------------- Riemann tensor --------------------
\SecRiemann
\begin{subequations}
\begin{alignat}{3}
  R_{trtr} &= -16\frac{R\ddot{R}}{(4+kr^2)^2}, &\quad R_{t\vartheta t\vartheta} &= -16\frac{Rr^2\ddot{R}}{(4+kr^2)^2}, \\
  R_{t\varphi t\varphi} &= -16\frac{Rr^2\sin^2\vartheta\ddot{R}}{(4+kr^2)^2}, &\quad R_{r\vartheta r\vartheta} &= 256\frac{R^2r^2\left(\dot{R}^2+kc^2\right)}{c^2(4+kr^2)^4}, \\
  R_{r\varphi r\varphi} &= 256\frac{R^2r^2\sin^2\vartheta\left(\dot{R}^2+kc^2\right)}{c^2(4+kr^2)^4}, &\quad R_{\vartheta\varphi\vartheta\varphi} &= 256\frac{R^2r^4\sin^2\vartheta\left(\dot{R}^2+kc^2\right)}{c^2(4+kr^2)^4}.
\end{alignat}
\end{subequations}

%% -------------------- Ricci tensor --------------------
\SecRicci
\begin{subequations}
\begin{alignat}{3}
  R_{tt} &= -3\frac{\ddot{R}}{R},&\qquad R_{rr} &= 16\frac{R\ddot{R} + 2(\dot{R}^2 + kc^2)}{c^2(4+kr^2)^2}, \\ 
  R_{\vartheta\vartheta} &= 16r^2\frac{R\ddot{R} + 2(\dot{R}^2 + kc^2)}{c^2(4+kr^2)^2}, &\qquad R_{\varphi\varphi} &= 16r^2\sin^2\vartheta\frac{R\ddot{R} + 2(\dot{R}^2 + kc^2)}{c^2(4+kr^2)^2}.
\end{alignat}
\end{subequations}

\noindent The {\sl Ricci scalar and Kretschmann scalar read:}

\begin{equation}
  \mathcal{R} = 6\frac{R\ddot{R} + \dot{R}^2 + kc^2}{R^2c^2},\qquad \mathcal{K}=12\frac{\ddot{R}^2R^2 + \dot{R}^4 + 2\dot{R}^2kc^2 + k^2c^4}{R^4c^4}.
\end{equation}

%% -------------------- Local tetrad --------------------

\SecLocal
\begin{equation}
e_{(t)} = \frac{1}{c}\partial_{t}, \qquad e_{(r)} = \frac{1+\frac{k}{4}r^2}{R}\partial_{r}, \qquad e_{\vartheta} = \frac{1+\frac{k}{4}r^2}{Rr}\partial_{\vartheta}, \qquad e_{\varphi} = \frac{1+\nicefrac{k}{4}r^2}{Rr\sin{\vartheta}}\partial_{\varphi}.
\end{equation}

%% -------------------- Ricci rotation coefficients --------------------
\SecRicRotCoef
\begin{subequations}
\begin{align}
  \gamma_{(r)(t)(r)} &= \gamma_{(\vartheta)(t)(\vartheta)} = \gamma_{(\varphi)(t)(\varphi)} = \frac{\dot{R}}{Rc}\qquad \gamma_{(\vartheta)(r)(\vartheta)} = \gamma_{(\varphi)(r)(\varphi)} = -\frac{\frac{k}{4}r^2 - 1}{Rr}, \\
  \gamma_{(\varphi)(\vartheta)(\varphi)} &= \frac{(\frac{k}{4}r^2 + 1)\cot\vartheta}{Rr}.
\end{align}
\end{subequations}
The contractions of the Ricci rotation coefficients read
\begin{equation}
  \gamma_{(t)} = \frac{3\dot{R}}{Rc},\qquad \gamma_{(r)} = 2\frac{1-\frac{k}{4}r^2}{Rr},\qquad \gamma_{(\vartheta)} = \frac{(\frac{k}{4}r^2+1)\cot\vartheta}{Rr}.
\end{equation}

%% -------------------- Riemann tensor LT--------------------
\SecRiemannLT
\begin{subequations}
\begin{alignat}{3}
  &R_{(t)(\eta)(t)(\eta)} = R_{(t)(\vartheta)(t)(\vartheta)} = R_{(t)(\varphi)(t)(\varphi)} = -\frac{\ddot{R}}{Rc^2}\\
   &R_{(\eta)(\vartheta)(\eta)(\vartheta)} = R_{(\eta)(\varphi)(\eta)(\varphi)} = R_{(\vartheta)(\varphi)(\vartheta)(\varphi)} = \frac{\dot{R}^2 + kc^2}{R^2c^2} .
\end{alignat}
\end{subequations}

%% -------------------- Ricci tensor LT--------------------
\SecRicciLT
\begin{equation} 
  R_{(t)(t)} = -\frac{3\ddot{R}}{Rc^2},\qquad R_{(r)(r)} = R_{(\vartheta)(\vartheta)} = R_{(\varphi)(\varphi)} = \frac{R\ddot{R} + 2\dot{R}^2 + 2kc^2}{R^2c^2}.
\end{equation}

%% ------------------------------------------------------
%%    Form 3
%% ------------------------------------------------------
\subsection{Form 3}
The following forms of the metric are obtained from \ref{eqM:FRW1} by setting $\eta = \sin\psi,\psi,\sinh\psi$ for $k=1,0,-1$ respectively.
\subsubsection{Positive Curvature}
\metricEq{
  ds^2 = -c^2dt^2 + R^2\left\{d\psi^2 + \sin^2\psi \left(d\vartheta^2 + \sin^2\vartheta d\varphi^2\right)\right\}
}{FRW3p1}

%% -------------------- Christoffel symbols --------------------
\SecChristoffel
\begin{subequations}
\begin{alignat}{5}
\Gamma_{t\psi}^\psi &= \dfrac{\dot{R}}{R}, &\, \Gamma_{t\vartheta}^\vartheta &= \frac{\dot{R}}{R}, &\, \Gamma_{t\varphi}^\varphi &= \frac{\dot{R}}{R}, \\
\Gamma_{\psi\psi}^t &= \frac{R\dot{R}}{c^2}, &\, \Gamma_{\psi\vartheta}^\vartheta &= \cot\psi, &\, \Gamma_{\psi\varphi}^\varphi &= \cot\psi, \\
\Gamma_{\vartheta\vartheta}^t &= \frac{R\sin^2\psi \dot{R}}{c^2}, &\, \Gamma_{\vartheta\vartheta}^\psi &= -\sin\psi\cos\psi, &\, \Gamma_{\vartheta\varphi}^\varphi &= \cot(\vartheta), \\
\Gamma_{\varphi\varphi}^t &= \frac{R\sin^2\psi \sin^2\vartheta\dot{R}}{c^2}, &\, \Gamma_{\varphi\varphi}^\psi &= -\sin\psi\cos\psi\sin^2\vartheta, &\, \Gamma_{\varphi\varphi}^\vartheta &= -\sin\vartheta\cos\vartheta.
\end{alignat}
\end{subequations}

%% -------------------- Riemann tensor --------------------
\SecRiemann
\begin{subequations}
\begin{alignat}{3}
  R_{t\psi t\psi} &= -R\ddot{R}, &\quad R_{t\vartheta t\vartheta} &= -R\sin^2\psi \ddot{R}, \\
  R_{t\varphi t\varphi} &= -R\sin^2\psi \sin^2\vartheta\ddot{R}, &\quad R_{\psi\vartheta\psi\vartheta} &= \dfrac{R^2\sin^2\psi \left(\dot{R}^2+c^2\right)}{c^2}, \\
  R_{\psi\varphi\psi\varphi} &= \dfrac{R^2\sin^2\psi \sin^2\vartheta\left(\dot{R}^2+c^2\right)}{c^2}, &\, R_{\vartheta\varphi\vartheta\varphi} &= \dfrac{R^2\sin^4\psi\sin^2\vartheta\left(\dot{R}^2+c^2\right)}{c^2}.
\end{alignat}
\end{subequations}

%% -------------------- Ricci tensor --------------------
\SecRicci
\begin{subequations}
\begin{alignat}{3}
  R_{tt} &= -3\frac{\ddot{R}}{R},&\qquad R_{\psi\psi} &=\frac{R\ddot{R} + 2(\dot{R}^2 + c^2)}{c^2}, \\
  R_{\vartheta\vartheta} &= \sin^2\psi \frac{R\ddot{R} + 2(\dot{R}^2 + c^2)}{c^2},&\qquad R_{\varphi\varphi} &= \sin^2\vartheta\sin^2\psi \frac{R\ddot{R} + 2(\dot{R}^2 + c^2)}{c^2}.
\end{alignat}
\end{subequations}

\noindent The {\sl Ricci scalar and Kretschmann read}
\begin{equation}
  \mathcal{R} = 6\frac{R\ddot{R} + \dot{R}^2 + c^2}{R^2c^2},\qquad \mathcal{K}=12\frac{\ddot{R}^2R^2 + \dot{R}^4 + 2\dot{R}^2c^2 + c^4}{R^4c^4}.
\end{equation}

%% -------------------- Local tetrad --------------------

\SecLocal
\begin{equation}
 e_{(t)} = \frac{1}{c}\partial_{t}, \qquad e_{(\psi)} = \frac{1}{R}\partial_{\psi}, \qquad e_{\vartheta} = \frac{1}{R\sin\psi}\partial_{\vartheta}, \qquad e_{\varphi} = \frac{1}{R\sin\psi\sin{\vartheta}}\partial_{\varphi}.
\end{equation}

%% -------------------- Ricci rotation coefficients --------------------
\SecRicRotCoef
\begin{subequations}
\begin{align}
  \gamma_{(\psi)(t)(\psi)} &= \gamma_{(\vartheta)(t)(\vartheta)} = \gamma_{(\varphi)(t)(\varphi)} = \frac{\dot{R}}{Rc}\qquad \gamma_{(\vartheta)(\psi)(\vartheta)} = \gamma_{(\varphi)(\psi)(\varphi)} = \frac{\cot\psi}{R}, \\
  \gamma_{(\varphi)(\vartheta)(\varphi)} &= \frac{\cot\theta}{R\sin\psi}.
\end{align}
\end{subequations}
The contractions of the Ricci rotation coefficients read
\begin{equation}
  \gamma_{(t)} = \frac{3\dot{R}}{Rc},\qquad \gamma_{(r)} = 2\frac{\cot\psi}{R},\qquad \gamma_{(\vartheta)} = \frac{\cot\vartheta}{R\sin\psi}.
\end{equation}

%% -------------------- Riemann tensor LT--------------------
\SecRiemannLT
\begin{subequations}
\begin{align}
   R_{(t)(\psi)(t)(\psi)} &= R_{(t)(\vartheta)(t)(\vartheta)} = R_{(t)(\varphi)(t)(\varphi)} = -\frac{\ddot{R}}{Rc^2}, \\
   R_{(\psi)(\vartheta)(\psi)(\vartheta)} &= R_{(\psi)(\varphi)(\psi)(\varphi)} = R_{(\vartheta)(\varphi)(\vartheta)(\varphi)} = \frac{\dot{R}^2 + c^2}{R^2c^2} .
\end{align}
\end{subequations}

%% -------------------- Ricci tensor LT--------------------
\SecRicciLT
\begin{equation} 
  R_{(t)(t)} = -\frac{3\ddot{R}}{Rc^2},\qquad R_{(\psi)(\psi)} = R_{(\vartheta)(\vartheta)} = R_{(\varphi)(\varphi)} = \frac{R\ddot{R} + 2(\dot{R}^2 + c^2)}{R^2c^2}.
\end{equation}

\subsubsection{Vanishing Curvature}
\metricEq{
  ds^2 = -c^2dt^2 + R^2\left\{d\psi^2 + \psi^2\left(d\vartheta^2 + \sin^2\vartheta d\varphi^2\right)\right\}
}{FRW3p2}

%% -------------------- Christoffel symbols --------------------
\SecChristoffel
\begin{subequations}
\begin{alignat}{5}
\Gamma_{t\psi}^\psi &= \dfrac{\dot{R}}{R}, &\quad \Gamma_{t\vartheta}^\vartheta &= \frac{\dot{R}}{R}, &\quad \Gamma_{t\varphi}^\varphi &= \frac{\dot{R}}{R}, \\
\Gamma_{\psi\psi}^t &= \frac{R\dot{R}}{c^2}, &\quad\Gamma_{\psi\vartheta}^\vartheta &= \frac{1}{\psi}, &\quad\Gamma_{\psi\varphi}^\varphi &= \frac{1}{\psi}, \\
\Gamma_{\vartheta\vartheta}^t &= \frac{R\psi^2\dot{R}}{c^2}, &\qquad \Gamma_{\vartheta\vartheta}^\psi &= -\psi, &\quad\Gamma_{\vartheta\varphi}^\varphi &= \cot(\vartheta), \\
\Gamma_{\varphi\varphi}^t &= \frac{R\psi^2\sin^2\vartheta\dot{R}}{c^2}, &\quad \Gamma_{\varphi\varphi}^\psi &= -\psi\sin^2\vartheta, &\quad \Gamma_{\varphi\varphi}^\vartheta &= -\sin\vartheta\cos\vartheta.
\end{alignat}
\end{subequations}

%% -------------------- Riemann tensor --------------------
\SecRiemann
\begin{subequations}
\begin{alignat}{3}
  R_{t\psi t\psi} &= -R\ddot{R}, &\qquad R_{t\vartheta t\vartheta} &= -R\psi^2\ddot{R}, \\
  R_{t\varphi t\varphi} &= -R\psi^2\sin^2\vartheta\ddot{R}, &\qquad R_{\psi\vartheta\psi\vartheta} &= \dfrac{R^2\psi^2\dot{R}^2}{c^2}, \\
  R_{\psi\varphi\psi\varphi} &= \dfrac{R^2\psi^2\sin^2\vartheta\dot{R}^2}{c^2}, &\qquad R_{\vartheta\varphi\vartheta\varphi} &= \dfrac{R^2\psi^4\sin^2\vartheta\dot{R}^2}{c^2}.
\end{alignat}
\end{subequations}

%% -------------------- Ricci tensor --------------------
\SecRicci
\begin{subequations}
\begin{alignat}{3}
  R_{tt} &= -3\frac{\ddot{R}}{R},&\qquad R_{\psi\psi} &=\frac{R\ddot{R} + 2\dot{R}^2}{c^2}, \\
  R_{\vartheta\vartheta} &= \psi^2\frac{R\ddot{R} + 2\dot{R}^2}{c^2},&\qquad R_{\varphi\varphi} &= \sin^2\vartheta\psi^2\frac{R\ddot{R} + 2\dot{R}^2}{c^2}.
\end{alignat}
\end{subequations}

\noindent The {\sl Ricci scalar and Kretschmann read}
\begin{equation}
  \mathcal{R} = 6\frac{R\ddot{R} + \dot{R}^2}{R^2c^2},\qquad \mathcal{K}=12\frac{\ddot{R}^2R^2 + \dot{R}^4}{R^4c^4}.
\end{equation}

%% -------------------- Local tetrad --------------------

\SecLocal
\begin{equation}
 e_{(t)} = \frac{1}{c}\partial_{t}, \qquad e_{(\psi)} = \frac{1}{R}\partial_{\psi}, \qquad e_{\vartheta} = \frac{1}{R\psi}\partial_{\vartheta}, \qquad e_{\varphi} = \frac{1}{R\psi\sin{\vartheta}}\partial_{\varphi}.
\end{equation}

%% -------------------- Ricci rotation coefficients --------------------
\SecRicRotCoef
\begin{subequations}
\begin{align}
 \gamma_{(\psi)(t)(\psi)} &= \gamma_{(\vartheta)(t)(\vartheta)} = \gamma_{(\varphi)(t)(\varphi)} = \frac{\dot{R}}{Rc}\qquad \gamma_{(\vartheta)(\psi)(\vartheta)} = \gamma_{(\varphi)(\psi)(\varphi)} = \frac{1}{R\psi}, \\
 \gamma_{(\varphi)(\vartheta)(\varphi)} &= \frac{\cot(\vartheta)}{R\psi}.
\end{align}
\end{subequations}
The contractions of the Ricci rotation coefficients read
\begin{equation}
  \gamma_{(t)} = \frac{3\dot{R}}{Rc},\qquad \gamma_{(r)} = \frac{2}{R\psi}, \qquad \gamma_{(\vartheta)} = \frac{\cot\vartheta}{R\psi}.
\end{equation}

%% -------------------- Riemann tensor LT--------------------
\SecRiemannLT
\begin{subequations}
\begin{align}
   R_{(t)(\psi)(t)(\psi)} &= R_{(t)(\vartheta)(t)(\vartheta)} = R_{(t)(\varphi)(t)(\varphi)} = -\frac{\ddot{R}}{Rc^2}, \\
   R_{(\psi)(\vartheta)(\psi)(\vartheta)} &= R_{(\psi)(\varphi)(\psi)(\varphi)} = R_{(\vartheta)(\varphi)(\vartheta)(\varphi)} = \frac{\dot{R}^2}{R^2c^2} .
\end{align}
\end{subequations}

%% -------------------- Ricci tensor LT--------------------
\SecRicciLT
\begin{equation} 
  R_{(t)(t)} = -\frac{3\ddot{R}}{Rc^2},\qquad R_{(\psi)(\psi)} = R_{(\vartheta)(\vartheta)} = R_{(\varphi)(\varphi)} = \frac{R\ddot{R} + 2\dot{R}^2}{R^2c^2}.
\end{equation}

\subsubsection{Negative Curvature}
\metricEq{
  ds^2 = -c^2dt^2 + R^2\left\{d\psi^2 + \sinh^2\psi\left(d\vartheta^2 + \sin^2\vartheta d\varphi^2\right)\right\}
}{FRW3p3}

%% -------------------- Christoffel symbols --------------------
\SecChristoffel
\begin{subequations}
\begin{alignat}{5}
\Gamma_{t\psi}^\psi &= \dfrac{\dot{R}}{R}, &\, \Gamma_{t\vartheta}^\vartheta &= \frac{\dot{R}}{R}, &\, \Gamma_{t\varphi}^\varphi &= \frac{\dot{R}}{R}, \\
\Gamma_{\psi\psi}^t &= \frac{R\dot{R}}{c^2}, &\, \Gamma_{\psi\vartheta}^\vartheta &= \coth\psi, &\, \Gamma_{\psi\varphi}^\varphi &= \coth\psi, \\
\Gamma_{\vartheta\vartheta}^t &= \frac{R\sinh^2\psi\dot{R}}{c^2}, &\, \Gamma_{\vartheta\vartheta}^\psi &= -\sinh\psi\cosh\psi, &\, \Gamma_{\vartheta\varphi}^\varphi &= \cot\vartheta, \\
\Gamma_{\varphi\varphi}^t &= \frac{R\sinh^2\psi\sin^2\vartheta\dot{R}}{c^2}, &\quad \Gamma_{\varphi\varphi}^\psi &= -\sinh\psi\cosh\psi\sin^2\vartheta, &\quad \Gamma_{\varphi\varphi}^\vartheta &= -\sin\vartheta\cos\vartheta.
\end{alignat}
\end{subequations}

%% -------------------- Riemann tensor --------------------
\SecRiemann
\begin{subequations}
\begin{alignat}{3}
  R_{t\psi t\psi} &= -R\ddot{R}, &\quad R_{t\vartheta t\vartheta} &= -R\sinh^2\psi\ddot{R}, \\
  R_{t\varphi t\varphi} &= -R\sinh^2\psi\sin^2\vartheta\ddot{R}, &\quad R_{\psi\vartheta\psi\vartheta} &= \dfrac{R^2\sinh^2\psi\left(\dot{R}^2-c^2\right)}{c^2}, \\
  R_{\psi\varphi\psi\varphi} &= \dfrac{R^2\sinh^2\psi\sin^2\vartheta\left(\dot{R}^2-c^2\right)}{c^2}, &\quad R_{\vartheta\varphi\vartheta\varphi} &= \dfrac{R^2\sinh\psi^4\sin^2\vartheta\left(\dot{R}^2-c^2\right)}{c^2}.
\end{alignat}
\end{subequations}

%% -------------------- Ricci tensor --------------------
\SecRicci
\begin{subequations}
\begin{alignat}{3}
  R_{tt} &= -3\frac{\ddot{R}}{R},&\quad R_{\psi\psi} &=\frac{R\ddot{R} + 2(\dot{R}^2 - c^2)}{c^2}, \\
  R_{\vartheta\vartheta} &= \sinh^2\psi\frac{R\ddot{R} + 2(\dot{R}^2 - c^2)}{c^2},&\quad R_{\varphi\varphi} &= \sin^2\vartheta\sin^2\psi \frac{R\ddot{R} + 2(\dot{R}^2 - c^2)}{c^2}.
\end{alignat}
\end{subequations}

\noindent The {\sl Ricci scalar and Kretschmann read}
\begin{equation}
  \mathcal{R} = 6\frac{R\ddot{R} + \dot{R}^2 - c^2}{R^2c^2},\qquad \mathcal{K}=12\frac{\ddot{R}^2R^2 + \dot{R}^4 - 2\dot{R}^2c^2 + c^4}{R^4c^4}.
\end{equation}

%% -------------------- Local tetrad --------------------

\SecLocal
\begin{equation}
 e_{(t)} = \frac{1}{c}\partial_{t}, \qquad e_{(\psi)} = \frac{1}{R}\partial_{\psi}, \qquad e_{\vartheta} = \frac{1}{R\sinh\psi}\partial_{\vartheta}, \qquad e_{\varphi} = \frac{1}{R\sinh\psi\sin{\vartheta}}\partial_{\varphi}.
\end{equation}

%% -------------------- Ricci rotation coefficients --------------------
\SecRicRotCoef
\begin{subequations}
\begin{align}
  \gamma_{(\psi)(t)(\psi)} &= \gamma_{(\vartheta)(t)(\vartheta)} = \gamma_{(\varphi)(t)(\varphi)} = \frac{\dot{R}}{Rc}\qquad \gamma_{(\vartheta)(\psi)(\vartheta)} = \gamma_{(\varphi)(\psi)(\varphi)} = \frac{\coth\psi}{R}, \\
  \gamma_{(\varphi)(\vartheta)(\varphi)} &= \frac{\cot\theta}{R\sinh\psi}.
\end{align}
\end{subequations}
The contractions of the Ricci rotation coefficients read
\begin{equation}
  \gamma_{(t)} = \frac{3\dot{R}}{Rc},\qquad \gamma_{(r)} = 2\frac{\coth\psi}{R},\qquad \gamma_{(\vartheta)} = \frac{\cot\vartheta}{R\sinh\psi}.
\end{equation}

%% -------------------- Riemann tensor LT--------------------
\SecRiemannLT
\begin{subequations} 
\begin{align}
  R_{(t)(\psi)(t)(\psi)} &= R_{(t)(\vartheta)(t)(\vartheta)} = R_{(t)(\varphi)(t)(\varphi)} = -\frac{\ddot{R}}{Rc^2}, \\
  R_{(\psi)(\vartheta)(\psi)(\vartheta)} &= R_{(\psi)(\varphi)(\psi)(\varphi)} = R_{(\vartheta)(\varphi)(\vartheta)(\varphi)} = \frac{\dot{R}^2 - c^2}{R^2c^2} .
\end{align}
\end{subequations}

%% -------------------- Ricci tensor LT--------------------
\SecRicciLT
\begin{equation} 
  R_{(t)(t)} = -\frac{3\ddot{R}}{Rc^2},\qquad R_{(\psi)(\psi)} = R_{(\vartheta)(\vartheta)} = R_{(\varphi)(\varphi)} = \frac{R\ddot{R} + 2(\dot{R}^2 - c^2)}{R^2c^2}.
\end{equation}

\FurtherReading

Rindler\cite{rindler}

}{

}

%% ------------------------------------------------------------------------
%%    G O E D E L
%% ------------------------------------------------------------------------
\clearpage
\section{G\"odel Universe}
\setcounter{equation}{0}
\ifthenelse{\boolean{isARXIV}}{
% ******** Start of file goedel.tex *********
%
%  Copyright (c) 2009 Frank Grave,
%                     Universitaet Stuttgart, VISUS
%

G\"odel introduced a homogeneous and rotating universe model in \cite{goedel1949}. We follow the notation of \cite{kajari2004}

\subsection{Cylindrical coordinates}
The G\"odel metric in cylindrical coordinates is
\metricEq{
 d s^2 = -c^2d t^2 + \frac{d r^2}{1+[r/(2a)]^2}+r^2\left[1-\left(\frac{r}{2a}\right)^2\right]d \varphi^2+d z^2
 -2r^2\frac{c}{\sqrt{2}a}d td \varphi,
}{goedelCyl}
where $2a$ is the G\"odel radius.

%% -------------------- Christoffel symbols --------------------
\SecChristoffel
\begin{subequations}
\begin{alignat}{3}
  \Gamma_{tr}^t & = \frac{r}{2a^2}\frac{1}{1+[r/(2a)]^2}, &\quad   \Gamma_{tr}^\varphi & = -\frac{c}{\sqrt{2}ar}\frac{1}{1+[r/(2a)]^2},\\
  \Gamma_{t\varphi}^r & = \frac{cr}{\sqrt{2}a} \left[1+\left(\frac{r}{2a}\right)\right]^2, &\quad     \Gamma_{rr}^r & = -\frac{r}{4a^2}\frac{1}{1+[r/(2a)]^2},\\
  \Gamma_{r\varphi}^t & = \frac{r^3}{4\sqrt{2}ca^3}\frac{1}{1+[r/(2a)]^2}, &\quad     \Gamma_{r\varphi}^\varphi & = \frac{1}{r}\frac{1}{1+[r/(2a)]^2},\\
  \Gamma_{\varphi\varphi}^r & = r\left[1+\left(\frac{r}{2a}\right)^2\right]\left[1-\frac{1}{2}\left(\frac{r}{a}\right)^2\right].\phantom{,} &\quad&
\end{alignat}
\end{subequations}

%% -------------------- Riemann tensor --------------------
\SecRiemann
\begin{subequations}
\begin{alignat}{3}
 R_{trtr} &= \frac{c^2}{2a^2}\frac{1}{1+[r/(2a)]^2}   , &\quad   R_{trr\varphi} &=  -\frac{cr^2}{2\sqrt{2}a^3}\frac{1}{1+[r/(2a)]^2},\\
 R_{t\varphi t\varphi} &= \frac{c^2r^2}{2a^2}\frac{1}{1+[r/(2a)]^2}, &\quad R_{r\varphi r\varphi} &= \frac{r^2}{2a^2}\frac{1+3[r/(2a)]^2}{1+[r/(2a)]^2}.
\end{alignat}
\end{subequations}

%% -------------------- Ricci tensor --------------------
\SecRicci
\begin{equation}
 R_{tt}            = \frac{c^2}{a^2}, \quad   
 R_{t\varphi}      = \frac{r^2 c}{\sqrt{2}a^3}, \quad
 R_{\varphi\varphi} = \frac{r^4}{2a^4}.
\end{equation}

\noindent {\bf Ricci and Kretschmann scalar}
\begin{equation}
  \mathcal{R} = -\frac{1}{a^2},\qquad \mathcal{K} = \frac{3}{a^4}.
\end{equation}

\noindent {\bf cosmological constant}:
\begin{equation}
 \Lambda = \frac{R}{2} 
\end{equation}

\SecKilling\\
An infinitesimal isometric transformation $x'^\mu=x^\mu+\epsilon\xi^\mu(x^\nu)$ leaves the metric unchanged, that is $g'_{\mu\nu}(x'^\sigma)=g_{\mu\nu}(x'^\sigma)$. A killing vector field $\xi^\mu$ is solution to the killing equation $\xi_{\mu;\nu}+\xi_{\nu;\mu}=0$. There exist five killing vector fields in G\"odel's spacetime:
\begin{subequations}
\begin{alignat}{1}
 \underset{a}{\xi}^\mu = \left(\begin{array}{c}1\\0\\0\\0\end{array}\right),&\quad
 \underset{b}{\xi}^\mu = \frac{1}{\sqrt{1+[r/(2a)]^2}}\left(\begin{array}{c}\frac{r}{\sqrt{2}c}\cos\varphi\\a\left(1+[r/(2a)]^2\right)\sin\varphi\\\frac{a}{r}\left(1+2[r/(2a)]^2\right)\cos\varphi\\0 \end{array}\right),\quad
  \underset{c}{\xi}^\mu = \left(\begin{array}{c}0\\0\\1\\0\end{array}\right),\\
  \underset{d}{\xi}^\mu = \left(\begin{array}{c}0\\0\\0\\1\end{array}\right),&\quad
  \underset{e}{\xi}^\mu = \frac{1}{\sqrt{1+[r/(2a)]^2}}\left(\begin{array}{c}\frac{r}{\sqrt{2}c}\sin\varphi\\-a\left(1+[r/(2a)]^2\right)\cos\varphi\\\frac{a}{r}\left(1+2[r/(2a)]^2\right)\sin\varphi\\0 \end{array}\right).
\end{alignat}
\end{subequations}
An arbitrary linear combination of killing vector fields is again a killing vector field.

%% -------------------- Local tetrad --------------------
\SecLocal\\
For the local tetrad in G\"odel's spacetime an ansatz similar to the local tetrad of a rotating spacetime in spherical coordinates (Sec. \ref{subsec:axisymST}) can be used. After substituting $\vartheta\rightarrow z$ and swapping base vectors $\mathbf{e}_{(2)}$ and $\mathbf{e}_{(3)}$ an orthonormalized and right-handed local tetrad is obtained.
\begin{subequations}
\begin{alignat}{1}
 \mathbf{e}_{(0)} &=\Gamma\left(\partial_t+\zeta\partial_{\varphi}\right), \quad \mathbf{e}_{(1)} = \sqrt{1+[r/(2a)]^2}\partial_r,\quad
 \mathbf{e}_{(2)} =\Delta\Gamma\left(A\partial_t+B\partial_\varphi\right) ,   \quad \mathbf{e}_{(3)} = \partial_z,
\end{alignat}
\end{subequations}
where
\begin{subequations}
 \begin{alignat}{3}
  A &= -\frac{r^2c}{\sqrt{2}a}+\zeta r^2\left(1-[r/(2a)]^2\right),&\qquad B &= c^2+\frac{\zeta r^2 c}{\sqrt{2}a}, \\
  \Gamma &= \frac{1}{\sqrt{c^2+\zeta r^2c\sqrt{2}/a - \zeta^2r^2\left(1-[r/(2a)]^2\right)}},&\qquad\Delta&=\frac{1}{rc\sqrt{1+[r/(2a)]^2}}.
 \end{alignat}
\end{subequations}

\noindent{\sl Transformation between local direction $y^{(i)}$ and coordinate direction $y^{\mu}$:}\\
\begin{equation}
 y^0 = y^{(0)}\Gamma + y^{(2)}\Delta\Gamma A,\quad y^1=y^{(1)}\sqrt{1+[r/(2a)]^2},\quad y^2=y^{(0)}\Gamma\zeta + y^{(2)}\Delta\Gamma B,\quad y^3=y^{(3)}.
\end{equation}
with the above abbreviations.

\subsection{Scaled cylindrical coordinates}
If we apply the simple transformation
\begin{equation}\label{eq:scaledcoordinates}
 T=\frac{t}{r_G}, \qquad R=\frac{r}{r_G}, \qquad \phi=\varphi,\qquad Z=\frac{z}{r_G},
\end{equation}
with $r_G=2a$, we find a formulation for the metric scaling with $r_G$, which is
\metricEq{
 ds^2 = r_G^2\left(-c^2dT^2 + \frac{dR^2}{1+R^2} + R^2(1-R^2)D\phi^2 +dZ^2 -2\sqrt{2}cR^2\,dT d\phi\right).
}{goedel_ds_grave}

% %% -------------------- Christoffel symbols --------------------
\SecChristoffel
\begin{subequations}
\begin{alignat}{3}
  \Gamma_{TR}^T & = \frac{2R}{1+R^2}, &\quad   \Gamma_{TR}^\phi & = -\frac{\sqrt{2}c}{R(1+R^2)},\\
  \Gamma_{T\phi}^R & = \sqrt{2}cR(1+R^2),&\quad     \Gamma_{RR}^R & = -\frac{R}{1+R^2},\\
  \Gamma_{R\phi}^T & = \frac{\sqrt{2}R^3}{c(1+R^2)}, &\quad     \Gamma_{R\phi}^\phi & = \frac{1}{R(1+R^2)},\\
  \Gamma_{\phi\phi}^R & =R(1+R^2)(2R^2-1).\phantom{,} &\quad&
\end{alignat}
\end{subequations}

% %% -------------------- Riemann tensor --------------------
\SecRiemann
\begin{subequations}
\begin{alignat}{3}
 R_{TRTR} &= \frac{2r_G^2c^2}{1+R^2}, &\quad   R_{TRR\phi} &=  -\frac{2\sqrt{2}r_G^2cR^2}{1+R^2},\\
 R_{T\phi T\phi} &= 2c^2r_G^2R^2(1+R^2) ,&\quad R_{R\phi R\phi} &= \frac{2 r_G^2 R^2(1+3R^2)}{1+R^2}.
\end{alignat}
\end{subequations}
% 
% %% -------------------- Ricci tensor --------------------
\SecRicci
\begin{equation}
 R_{TT}       = 4c^2, \quad   
 R_{T\phi}    = 4\sqrt{2}cR^2, \quad
 R_{\phi\phi} = 8R^4.
\end{equation}

\noindent {\bf Ricci and Kretschmann scalar}
\begin{equation}
  \mathcal{R} = -\frac{4}{r_G^2},\qquad \mathcal{K} = \frac{48}{r_G^4}.
\end{equation}

\noindent {\bf cosmological constant}:
\begin{equation}
 \Lambda = \frac{R}{2} 
\end{equation}

\SecKilling\\
The Killing vectors read
\begin{subequations}
\begin{alignat}{1}
 \underset{a}{\xi}^\mu =\left(\begin{array}{c}1\\0\\0\\0\end{array}\right),&\quad
 \underset{b}{\xi}^\mu = \frac{1}{\sqrt{1+R^2}}\left(\begin{array}{c}\frac{R}{\sqrt{2}c}\cos\varphi\\\frac{1}{2}(1+R^2)\sin\varphi\\\frac{1}{2R}(1+2R^2)\cos\varphi\\0 \end{array}\right),\quad
  \underset{c}{\xi}^\mu = \left(\begin{array}{c}0\\0\\1\\0\end{array}\right),\\
  \underset{d}{\xi}^\mu =\left(\begin{array}{c}0\\0\\0\\1\end{array}\right),&\quad
  \underset{e}{\xi}^\mu = \frac{1}{\sqrt{1+R^2}}\left(\begin{array}{c}\frac{R}{\sqrt{2}c}\sin\varphi\\-\frac{1}{2}(1+R^2)\cos\varphi\\\frac{1}{2R}(1+2R^2)\sin\varphi\\0 \end{array}\right).
\end{alignat}
\end{subequations}

% 
% %% -------------------- Local tetrad --------------------
\SecLocal\\
After the transformation to scaled cylindrical coordinates, the local tetrad reads
\begin{subequations}
\begin{alignat}{1}
 \mathbf{e}_{(0)} &=\frac{\Gamma}{r_G}\left(\partial_T+\zeta\partial_{\phi}\right),\quad \mathbf{e}_{(1)} =\frac{1}{r_G} \sqrt{1+R^2}\,\partial_R,\qquad \mathbf{e}_{(2)} =\frac{\Delta\Gamma}{r_G}\left(A\partial_T+B\partial_\phi\right),\quad \mathbf{e}_{(3)} = \frac{1}{r_G}\,\partial_Z,
\end{alignat}
\end{subequations}
where
\begin{subequations}
 \begin{alignat}{3}
  A &= R^2\left[-\sqrt{2}c+(1-R^2)\zeta\right],&\qquad B &= c^2+\sqrt{2}R^2 c\zeta,\\
  \Gamma &= \frac{1}{\sqrt{c^2+2\sqrt{2}R^2c\zeta- R^2(1-R^2)\zeta^2}},&\qquad\Delta&=\frac{1}{Rc\sqrt{1+R^2}}.
 \end{alignat}
\end{subequations}

\noindent{\sl Transformation between local direction $y^{(i)}$ and coordinate direction $y^{\mu}$:}\\
\begin{equation}
  y^0 = \frac{\Gamma}{r_G} y^{(0)} + \frac{\Delta \Gamma A}{r_G}y^{(2)},\qquad y^1 =\frac{1}{r_G}\sqrt{1+R^2}y^{(1)},\qquad y^2 = \frac{\Gamma \zeta}{r_G}y^{(0)} + \frac{\Delta\Gamma B}{r_G}y^{(2)},\qquad y^3 = \frac{1}{r_G}y^{(3)},
\end{equation}
and the back transformation is given by
\begin{subequations}\label{eq:coordtolocal}
\begin{alignat}{1}
  y^{(0)} &= \frac{r_G}{\Gamma}\frac{By^0-Ay^2}{B-\zeta A},\qquad y^{(1)} = \frac{r_G}{\sqrt{1+R^2}}y^1,\qquad  y^{(2)} = \frac{r_G}{\Delta\Gamma}\frac{y^2-\zeta y^0}{B-\zeta A},\qquad y^{(3)} = r_G y^3.
\end{alignat}
\end{subequations}

}{

}

%% ------------------------------------------------------------------------
%%    Halilsoy standing wave
%% ------------------------------------------------------------------------
\clearpage
\section{Halilsoy standing wave}
\setcounter{equation}{0}
\ifthenelse{\boolean{isARXIV}}{
% ******** Start of file halilsoyStandingWave.tex *********
%
%  Copyright (c) 2010 Thomas Mueller,
%                     Universitaet Stuttgart, VISUS
%

The standing wave metric by Halilsoy\cite{halilsoy1988} reads
\metricEq{
  ds^2 = V\left[e^{2K}\left(d\rho^2-dt^2\right)+\rho^2d\varphi^2\right] + \frac{1}{V}\left(dz+A\,d\varphi\right)^2,
}{halilsoy}
where 
\begin{subequations}
 \begin{align}
   V &= \cosh^2\alpha e^{-2CJ_0(\rho)\cos(t)} + \sinh^2\alpha e^{2CJ_0(\rho)\cos(t)},\\
   K &= \frac{C^2}{2}\left[\rho^2\left(J_0(\rho)^2+J_1(\rho)^2\right)-2\rho J_0(\rho)J_1(\rho)\cos^2t\right],\\
   A &= -2C\sinh(2\alpha)\rho J_1(\rho)\sin(t).
 \end{align}
\end{subequations}
with spherical Bessel functions $J_{1,2}$ and parameters $\alpha$ and $C$.

\SecLocal
\begin{equation}
  \mathbf{e}_{(0)} = \frac{e^{-K}}{\sqrt{V}}\partial_t, \qquad \mathbf{e}_{(1)} = \frac{e^{-K}}{\sqrt{V}}\partial_{\rho}, \qquad \mathbf{e}_{(2)} = \frac{1}{\rho\sqrt{V}}\partial_{\varphi}-\frac{A}{\rho\sqrt{V}}\partial_z, \qquad \mathbf{e}_{(3)} = \sqrt{V}\partial_z.
\end{equation}
dual tetrad:
\begin{equation}
  \boldsymbol{\theta}^{(0)} = \sqrt{V}e^K\,dt,\qquad \boldsymbol{\theta}^{(2)} = \sqrt{V}e^K\,d\rho,\qquad \boldsymbol{\theta}^{(2)} = \sqrt{V}\rho\,d\varphi,\qquad \boldsymbol{\theta}^{(3)}=\frac{1}{\sqrt{V}}\left(dz+A\,d\varphi\right).
\end{equation}
 
% 
% %% -------------------- Ricci rotation coefficients --------------------
% \SecRicRotCoef
% \begin{equation}
%   \gamma_{(\vartheta)(r)(\vartheta)} = \gamma_{(\varphi)(r)(\varphi)} = \frac{1}{r},\qquad \gamma_{(\varphi)(\vartheta)(\varphi)} = \frac{\cot\vartheta}{kr}.
% \end{equation}
% The contractions of the Ricci rotation coefficients read
% \begin{equation}
%   \gamma_{(r)} = \frac{2}{r},\qquad \gamma_{(\vartheta)} = \frac{\cot\vartheta}{kr}.
% \end{equation}
% 
% %% -------------------- Riemann tensor LT--------------------
% \SecRiemannLT
% \begin{equation}
%   R_{(\vartheta)(\varphi)(\vartheta)(\varphi)} = \frac{1-k^2}{k^2r^2}.
% \end{equation}
% 
% %% -------------------- Ricci tensor LT--------------------
% \SecRicciLT
% \begin{equation}
%  R_{(\vartheta)(\vartheta)} = R_{(\varphi)(\varphi)} = \frac{1-k^2}{k^2r^2}.
% \end{equation}
% 
% %% -------------------- Weyl tensor LT--------------------
% \SecWeylLT
% \begin{subequations}
%  \begin{align}
%   C_{(t)(r)(t)(r)} &= -C_{(\vartheta)(\varphi)(\vartheta)(\varphi)} = -\frac{1-k^2}{3k^2r^2},\\
%   C_{(t)(\vartheta)(t)(\vartheta)} &= C_{(t)(\varphi)(t)(\varphi)} = -C_{(r)(\vartheta)(r)(\vartheta)} = -C_{(r)(\varphi)(r)(\varphi)} = \frac{1-k^2}{6k^2r^2}.
%  \end{align}
% \end{subequations}

}{

}

%% ------------------------------------------------------------------------
%%      J a n i s - N e w m a n - W i n i c o u r
%% ------------------------------------------------------------------------
\clearpage
\section{Janis-Newman-Winicour}
\setcounter{equation}{0}
\ifthenelse{\boolean{isARXIV}}{
% ******** Start of file janewi.tex *********
%
%  Copyright (c) 2009 Thomas Mueller,
%                     Universitaet Stuttgart, VISUS
%

The Janis-Newman-Winicour\cite{janis1968} spacetime in spherical coordinates $(t,r,\vartheta,\varphi)$ is represented by the line element
\metricEq{
  ds^2 = -\alpha^{\gamma}c^2dt^2+\alpha^{-\gamma}dr^2 + r^2\alpha^{-\gamma+1}\left(d\vartheta^2+\sin^2\vartheta d\varphi^2\right),
}{janewi}
where $\alpha=1-r_s/(\gamma r)$. The Schwarzschild radius $r_s=2GM/c^2$ is defined by Newton's constant $G$, the speed of light $c$, and the mass parameter $M$. For $\gamma=1$, we obtain the Schwarzschild metric (\ref{eqM:schwarzschildSpherical}).

%% -------------------- Christoffel symbols --------------------
\SecChristoffel
\begin{subequations}
\begin{alignat}{5}
  \Gamma_{tt}^r &= \frac{r_sc^2}{2r^2}\alpha^{2\gamma-1}, &\qquad\Gamma_{tr}^t &= \frac{r_s}{2\gamma r^2\alpha}, &\qquad\Gamma_{rr}^r &= -\frac{r_s}{2\gamma r^2\alpha},\\
  \Gamma_{r\vartheta}^{\vartheta} &= \frac{2\gamma r-r_s(\gamma+1)}{2\gamma r^2\alpha}, &\qquad \Gamma_{r\varphi}^{\varphi} &= \frac{2\gamma r-r_s(\gamma+1)}{2\gamma r^2\alpha}, &\qquad \Gamma_{\vartheta\vartheta}^{r}&=-\frac{2\gamma r-r_s(\gamma+1)}{2\gamma},\\
  \Gamma_{\varphi\varphi}^{r}&=\Gamma_{\vartheta\vartheta}^{r}\sin^2\vartheta, &\qquad \Gamma_{\vartheta\varphi}^{\varphi} &= \cot\vartheta, &\qquad \Gamma_{\varphi\varphi}^{\vartheta}&=-\sin\vartheta\cos\vartheta.
\end{alignat}
\end{subequations}

%% -------------------- Riemann tensor --------------------
\SecRiemann
\begin{subequations}
\begin{alignat}{3}
  R_{trtr} &= -\frac{r_sc^2\left[2\gamma r-r_s(\gamma+1)\right]\alpha^{\gamma-2}}{2\gamma r^4}, &\: R_{t\vartheta t\vartheta} &= \frac{r_sc^2\left[2\gamma r -r_s(\gamma+1)\right]\alpha^{\gamma-1}}{4\gamma r^2},\\
  R_{t\varphi t\varphi} &= \frac{r_sc^2\left[2\gamma r -r_s(\gamma+1)\right]\alpha^{\gamma-1}\sin^2\vartheta}{4\gamma r^2}, & R_{r\vartheta r\vartheta} &= -\frac{r_s\left[2\gamma^2 r-r_s(\gamma+1)\right]}{4\gamma^2r^2\alpha^{\gamma-1}},\\
  R_{r\varphi r\varphi} &= -\frac{r_s\left[2\gamma^2 r-r_s(\gamma+1)\right]\sin^2\vartheta}{4\gamma^2r^2\alpha^{\gamma-1}}, & R_{\vartheta\varphi\vartheta\varphi} &= \frac{r_s\left[4\gamma^2 r-r_s(\gamma+1)^2\right]\sin^2\vartheta}{4\gamma^2\alpha^{\gamma}}.
\end{alignat}
\end{subequations}

%% -------------------- Weyl tensor --------------------
\SecWeyl
\begin{subequations}
\begin{alignat}{3}
  C_{trtr} &= -\frac{r_sc^2\alpha^{\gamma-2}\beta}{6\gamma^2 r^4}, &\qquad C_{t\vartheta t\vartheta} &= \frac{r_sc^2\alpha^{\gamma-1}\beta}{12\gamma^2 r^2},\\
  C_{t\varphi t\varphi} &= \frac{r_sc^2\alpha^{\gamma-1}\beta\sin^2\vartheta}{12\gamma^2 r^2}, & C_{r\vartheta r\vartheta} &= -\frac{r_s\beta}{12\gamma^2r^2\alpha^{\gamma-1}},\\
  C_{r\varphi r\varphi} &= -\frac{r_s\beta\sin^2\vartheta}{12\gamma^2r^2\alpha^{\gamma-1}}, & C_{\vartheta\varphi\vartheta\varphi} &= \frac{r_s\beta\sin^2\vartheta}{6\gamma^2\alpha^{\gamma}},
\end{alignat}
\end{subequations}
where $\beta=6\gamma^2 r-r_s(\gamma+1)(2\gamma+1)$.

%% -------------------- Ricci tensor --------------------
\SecRicci
\begin{equation}
  R_{rr} = \frac{r_s^2(1-\gamma^2)}{2\gamma^2r^4\alpha^2}.
\end{equation}

\noindent The {\sl Ricci scalar} reads
\begin{equation}
  \mathcal{R} = \frac{r_s^2(1-\gamma^2)\alpha^{\gamma-2}}{2\gamma^2r^4},
\end{equation}
whereas the {\sl Kretschmann scalar} is given by
\begin{equation}
 {\cal K} = \frac{r_s^2\alpha^{2\gamma-4}}{4\gamma^4 r^8}\left[7\gamma^2r_s^2(2+\gamma^2)+48\gamma^4r^2\alpha + 8\gamma r_s(2\gamma^2+1)(r_s-2\gamma r)+3r_s^2\right].
\end{equation}

%% -------------------- Local tetrad --------------------
\SecLocal
\begin{equation}
  \mathbf{e}_{(t)} = \frac{1}{c\alpha^{\gamma/2}}\partial_t, \qquad \mathbf{e}_{(r)} = \alpha^{\gamma/2}\partial_r,\qquad \mathbf{e}_{(\vartheta)} = \frac{\alpha^{(\gamma-1)/2}}{r}\partial_{\vartheta}, \qquad \mathbf{e}_{(\varphi)} = \frac{\alpha^{(\gamma-1)/2}}{r\sin\vartheta}\partial_{\varphi}.
\end{equation}
Dual tetrad:
\begin{equation}
 \boldsymbol{\theta}^{(t)}=c\alpha^{\gamma/2}dt,\qquad \boldsymbol{\theta}^{(r)} = \frac{dr}{\alpha^{\gamma/2}},\qquad \boldsymbol{\theta}^{(\vartheta)} = \frac{r}{\alpha^{(\gamma-1)/2}}d\vartheta,\qquad \boldsymbol{\theta}^{(\varphi)} = \frac{r\sin\vartheta}{\alpha^{(\gamma-1)/2}}d\varphi.
\end{equation}

%% -------------------- Ricci rotation coefficients --------------------
\SecRicRotCoef
\begin{subequations}
\begin{align}
  \gamma_{(r)(t)(t)} &= \frac{r_s}{2r^2}\alpha^{(\gamma-2)/2},\qquad \gamma_{(\vartheta)(r)(\vartheta)} = \gamma_{(\varphi)(r)(\varphi)} = \frac{2\gamma r-r_s(\gamma+1)}{2\gamma r^2}\alpha^{(\gamma-2)/2},\\
  \gamma_{(\varphi)(\vartheta)(\varphi)} &= \frac{\cot\vartheta}{r}\alpha^{(\gamma-1)/2}.
\end{align}
\end{subequations}
The contractions of the Ricci rotation coefficients read
\begin{equation}
  \gamma_{(r)} = \frac{4\gamma r-r_s(2+\gamma)}{2\gamma r^2}\alpha^{(\gamma-1)/2},\qquad \gamma_{(\vartheta)} = \frac{\cot\vartheta}{r}\alpha^{(\gamma-1)/2}.
\end{equation}

%% -------------------- Lie coefficients --------------------
\SecLieCoef
\begin{subequations}
\begin{align}
  c_{(t)(r)}^{(t)} &= \frac{r_s}{2r^2}\alpha^{(\gamma-2)/2},\qquad c_{(r)(\vartheta)}^{(\vartheta)} = c_{(r)(\varphi)}^{(\varphi)} = -\frac{2\gamma r-r_s(\gamma+1)}{2\gamma r^2}\alpha^{(\gamma-2)/2},\\
  c_{(\vartheta)(\varphi)}^{(\varphi)} &= -\frac{\cot\vartheta}{r}\alpha^{(\gamma-1)/2}.
\end{align}
\end{subequations}

%% -------------------- Euler Lagrange --------------------
\SecEulLag

The Euler-Lagrangian formalism, Sec.~\ref{subsec:EL}, for geodesics in the $\vartheta=\pi/2$ hyperplane yields the effective potential
\begin{equation}
  V_{\text{eff}} = \frac{1}{2}\alpha^{\gamma}\left(\frac{h^2\alpha^{\gamma-1}}{r^2}-\kappa c^2\right)
\end{equation}
with the constants of motion $h=r^2\alpha^{-\gamma+1}\dot{\varphi}$ and $k=\alpha^{\gamma}c^2\dot{t}$.
For null geodesics $(\kappa=0)$ and $\gamma>\frac{1}{2}$, there is an extremum at
\begin{equation}
  r = r_s\frac{1+2\gamma}{2\gamma}.
\end{equation}

%% -------------------- Embedding --------------------
\SecEmbedding

The embedding function $z=z(r)$ for $r\in[r_s(\gamma+1)^2/(4\gamma^2),\infty)$ follows from
\begin{equation}
 \frac{dz}{dr} = \sqrt{\frac{r_s\left[4r\gamma^2-r_s(1+\gamma)^2\right]}{4r^2\gamma^2\alpha^{\gamma+1}}}.
\end{equation}
However, the analytic solution
\begin{equation}
 z(r)=2\sqrt{r_sr}\,F_1\left(-\frac{1}{2};\frac{\gamma+1}{2},-\frac{1}{2};\frac{1}{2},\frac{r_s}{r\gamma},\frac{r_s(1+\gamma)^2}{4r\gamma^2}\right)-\frac{2\pi\gamma}{\gamma+1}\leftidx{_2}{F}{_1}\left(-\frac{1}{2},\frac{\gamma+1}{2};1;\frac{4\gamma}{(\gamma+1)^2}\right),
\end{equation}
depends on the Appell-$F_1$- and the Hypergeometric-$\leftidx{_2}{F}{_1}$-function.

}{

}

%% ------------------------------------------------------------------------
%%    K a s n e r
%% ------------------------------------------------------------------------
\clearpage
\section{Kasner}
\setcounter{equation}{0}
\ifthenelse{\boolean{isARXIV}}{
% ******** Start of file kasner.tex *********
%
%  Copyright (c) 2009 Thomas Mueller,
%                     Universitaet Stuttgart, VISUS
%

The Kasner spacetime in Cartesian coordinates $(t,x,y,z)$ is represented by the line element\cite{mtw,kasner1921} $(c=1)$
\metricEq{
  ds^2 = -dt^2+t^{2p_1}dx^2+t^{2p_2}dy^2+t^{2p_3}tz^2,
}{kasner}
where $p_1, p_2, p_3$ have to fulfill the two conditions
\begin{equation}
 p_1+p_2+p_3=1\qquad\text{and}\qquad p_1^2+p_2^2+p_3^2=1.
\end{equation}
These two conditions can also be represented by the Khalatnikov-Lifshitz parameter $u$ with
\begin{equation}
 p_1 = -\frac{u}{1+u+u^2},\qquad p_2 = \frac{1+u}{1+u+u^2},\qquad p_3 = \frac{u(1+u)}{1+u+u^2}.
\end{equation}

%% -------------------- Christoffel symbols --------------------
\SecChristoffel
\begin{subequations}
\begin{alignat}{5}
  \Gamma_{tx}^x &= \frac{p_1}{t}, &\qquad\Gamma_{ty}^y &= \frac{p_2}{t}, &\qquad\Gamma_{tz}^z &= \frac{p_3}{t},\\
  \Gamma_{xx}^t &= \frac{p_1t^{2p_1}}{t}, &\qquad \Gamma_{yy}^t &= \frac{p_2t^{2p_2}}{t}, &\qquad \Gamma_{zz}^t &= \frac{p_3t^{2p_3}}{t}.
\end{alignat}
\end{subequations}

Partial derivatives
\begin{subequations}
 \begin{alignat}{5}
    \Gamma_{tx,t}^x &= -\frac{p_1}{t^2}, &\qquad \Gamma_{ty,t}^t &= -\frac{p_2}{t^2}, &\qquad \Gamma_{tz,t}^z &= -\frac{p_3}{t^2},\\    
    \Gamma_{xx,t}^t &= p_1(2p_1-1)t^{2p_1-2}, & \Gamma_{yy,t}^t &= p_2(2p_2-1)t^{2p_2-2}, & \Gamma_{zz,t}^t &= p_3(2p_3-1)t^{2p_3-2}.
 \end{alignat}
\end{subequations}

%% -------------------- Riemann tensor --------------------
\SecRiemann
\begin{subequations}
\begin{alignat}{5}
  R_{txtx} &= \frac{p_1(1-p_1)t^{2p_1}}{t^2}, &\qquad R_{tyty} &= \frac{p_2(1-p_2)t^{2p_2}}{t^2}, &\qquad  R_{tztz} &= \frac{p_3(1-p_3)t^{2p_3}}{t^2},\\
  R_{xyxy} &= \frac{p_1p_2t^{2p_1}t^{2p_2}}{t^2}, & R_{xzxz} &= \frac{p_1p_3t^{2p_1}t^{2p_3}}{t^2}, &. R_{yzyz} &= \frac{p_2p_3t^{2p_2}t^{2p_3}}{t^2}.
\end{alignat}
\end{subequations}

The Ricci tensor as well as the Ricci scalar vanish identically. The Kretschmann scalar reads
\begin{subequations}
\begin{align}
 \mathcal{K} &= \frac{4}{t^4}\left(p_1^2-2p_1^3+p_1^4+p_2^2-2p_2^3+p_2^4+p_1^2p_3^2+p_3^2-2p_3^3+p_3^4+p_1^2p_2^2+p_2^2p_3^2\right)\\
  &= \frac{16u^2(1+u)^2}{t^4(1+u+u^2)^3}.
\end{align}
\end{subequations}

%% -------------------- Local tetrad --------------------
\SecLocal
\begin{equation}
  \mathbf{e}_{(t)} = \partial_t, \qquad \mathbf{e}_{(x)} = t^{-p_1}\partial_x,\qquad \mathbf{e}_{(y)} = t^{-p_2}\partial_y,\qquad \mathbf{e}_{(z)} = t^{-p_3}\partial_z.
\end{equation}
Dual tetrad:
\begin{equation}
  \boldsymbol{\theta}^{(t)} = dt,\qquad \boldsymbol{\theta}^{(x)} = t^{p_1}dx,\qquad \boldsymbol{\theta}^{(y)} = t^{p_2}dy,\qquad \boldsymbol{\theta}^{(z)} = t^{p_3}dz.
\end{equation}

%% -------------------- Ricci rotation coefficients --------------------
\SecRicRotCoef
\begin{equation}
  \boldsymbol{\gamma}_{(t)(r)(r)} = \frac{p_1}{t},\qquad \boldsymbol{\gamma}_{(t)(\vartheta)(\vartheta)} = \frac{p_2}{t},\qquad \boldsymbol{\gamma}_{(t)(\varphi)(\varphi)} = \frac{p_3}{t}.
\end{equation}
The contractions of the Ricci rotation coefficients read
\begin{equation}
  \boldsymbol{\gamma}_{(t)} = -\frac{1}{t}.
\end{equation}

%% -------------------- Riemann tensor LT--------------------
\SecRiemannLT
\begin{subequations}
 \begin{alignat}{5}
  R_{(t)(x)(y)(x)} &= \frac{p_1(1-p_1)}{t^2}, &\qquad R_{(t)(y)(t)(y)} &= \frac{p_2(1-p_2)}{t^2}, &\qquad R_{(t)(z)(t)(z)} &= \frac{p_3(1-p_3)}{t^2},\\
  R_{(x)(y)(x)(y)} &= \frac{p_1p_2}{t^2}, & R_{(x)(z)(x)(z)} &= \frac{p_1p_3}{t^2}, & R_{(y)(z)(y)(z)} &= \frac{p_2p_3}{t^2}.
 \end{alignat}
\end{subequations}

}{

}

%% ------------------------------------------------------------------------
%%      K e r r
%% ------------------------------------------------------------------------
\clearpage
\section{Kerr}
\setcounter{equation}{0}
\ifthenelse{\boolean{isARXIV}}{
% ******** Start of file kerr.tex *********
%
%  Copyright (c) 2009 Thomas Mueller,
%                     Universitaet Stuttgart, VISUS
%

The Kerr spacetime, found by Roy Kerr in 1963\cite{kerr1963}, describes a rotating black hole.
\subsection{Boyer-Lindquist coordinates}
The Kerr metric in Boyer-Lindquist coordinates
% \begin{subequations}
% \begin{align}
%   \label{eq:kerrBLbardeen}
%   ds^2 &= -\left(1-\frac{r_sr}{\Sigma}\right)c^2dt^2-\frac{2r_sar\sin^2\vartheta}{\Sigma}c\,dt\,d\varphi + \frac{\Sigma}{\Delta}dr^2 + \Sigma d\vartheta^2\\
%       &\quad + \left(r^2+a^2+\frac{r_sa^2r\sin^2\vartheta}{\Sigma}\right)\sin^2\vartheta d\varphi^2,
% \end{align}
% \end{subequations}
\metricEqAlign{
  ds^2 &= \dst -\left(1-\frac{r_sr}{\Sigma}\right)c^2dt^2-\frac{2r_sar\sin^2\vartheta}{\Sigma}c\,dt\,d\varphi + \frac{\Sigma}{\Delta}dr^2 + \Sigma d\vartheta^2\\[0.8em]
      &\quad \dst + \left(r^2+a^2+\frac{r_sa^2r\sin^2\vartheta}{\Sigma}\right)\sin^2\vartheta d\varphi^2,
}{kerrBLbardeen}
with $\Sigma=r^2+a^2\cos^2\vartheta$, $\Delta=r^2-r_sr+a^2$, and $r_s=2GM/c^2$, is taken from Bardeen\cite{bardeen1972}. $M$ is the mass and $a$ is the angular momentum per unit mass of the black hole. The contravariant form of the metric reads
\begin{equation}
  \label{eq:kerrBLbardeenCov}
  \partial_s^2 = -\frac{A}{c^2\Sigma\Delta}\partial_t^2 - \frac{2r_sar}{c\Sigma\Delta}\partial_t\partial_{\varphi}+\frac{\Delta}{\Sigma}\partial_r^2+\frac{1}{\Sigma}\partial_{\vartheta}^2 + \frac{\Delta-a^2\sin^2\vartheta}{\Sigma\Delta\sin^2\vartheta}\partial_{\varphi}^2,
\end{equation}
where $A=\left(r^2+a^2\right)^2-a^2\Delta\sin^2\vartheta=\left(r^2+a^2\right)\Sigma+r_sa^2r\sin^2\vartheta$. 

The event horizon $r_{+}$ is defined by the outer root of $\Delta$,
\begin{equation}
  r_{+} = \frac{r_s}{2}+\sqrt{\frac{r_s^2}{4}-a^2},
\end{equation}
whereas the outer boundary $r_0$ of the ergosphere follows from the outer root of $\Sigma-r_sr$,
\begin{equation}
  r_0 = \frac{r_s}{2}+\sqrt{\frac{r_s^2}{4}-a^2\cos^2\vartheta},
\end{equation}

\begin{SCfigure}[1][htb]
\ifthenelse{\boolean{isARXIV}}{
  \includegraphics[scale=0.9]{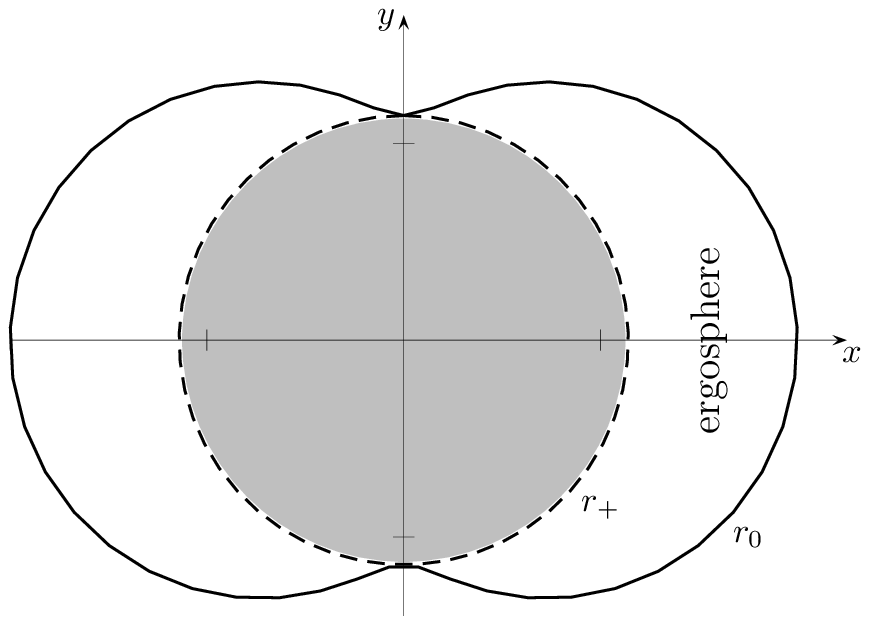}
}{
  \includegraphics[scale=0.9]{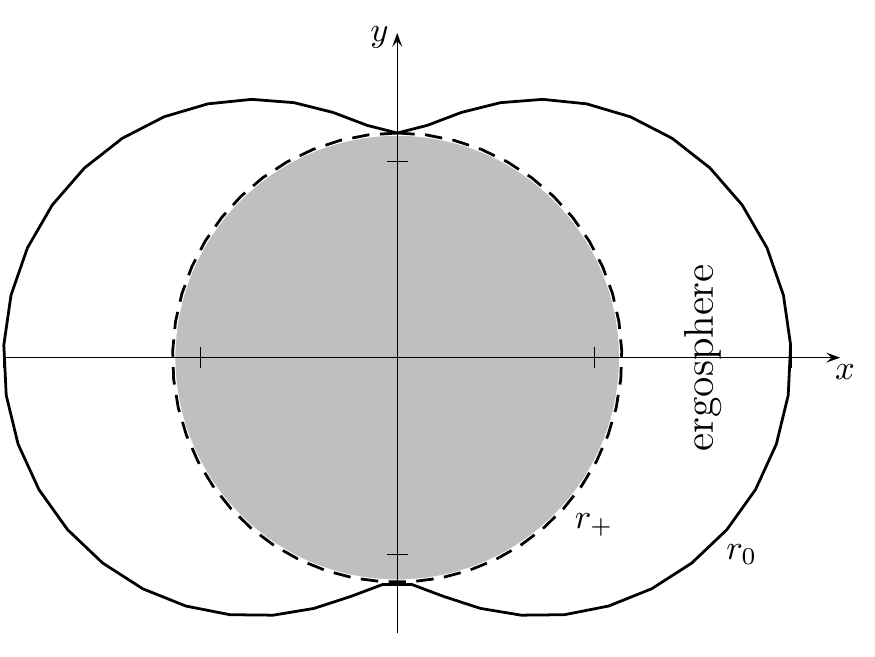}
}
  \caption{Ergosphere and horizon (dashed circle) for $a=0.99\frac{r_s}{2}$.}
\end{SCfigure}

%% -------------------- Christoffel symbols --------------------
\SecChristoffel
\begin{subequations}
\begin{alignat}{3}
  \Gamma_{tt}^r &= \frac{c^2r_s\Delta (r^2-a^2\cos^2\vartheta)}{2\Sigma^3}, &\qquad\Gamma_{tt}^{\vartheta} &= -\frac{c^2r_sa^2r\sin\vartheta\cos\vartheta}{\Sigma^3},\\
  \Gamma_{tr}^t &= \frac{r_s(r^2+a^2)(r^2-a^2\cos^2\vartheta)}{2\Sigma^2\Delta}, & \Gamma_{tr}^{\varphi} &= \frac{cr_sa(r^2-a^2\cos^2\vartheta)}{2\Sigma^2\Delta},\\
  \Gamma_{t\vartheta}^t &= -\frac{r_sa^2r\sin\vartheta\cos\vartheta}{\Sigma^2}, & \Gamma_{t\vartheta}^{\varphi} &= -\frac{cr_sar\cot\vartheta}{\Sigma^2},\\
  \Gamma_{t\varphi}^r &= -\frac{c\Delta r_sa\sin^2\vartheta(r^2-a^2\cos^2\vartheta)}{2\Sigma^3}, & \Gamma_{t\varphi}^{\vartheta} &= \frac{cr_sar(r^2+a^2)\sin\vartheta\cos\vartheta}{\Sigma^3},\\
  \Gamma_{rr}^r &= \frac{2ra^2\sin^2\vartheta-r_s(r^2-a^2\cos^2\vartheta)}{2\Sigma\Delta}, & \Gamma_{rr}^{\vartheta} &= \frac{a^2\sin\vartheta\cos\vartheta}{\Sigma\Delta},\\
  \Gamma_{r\vartheta}^r &= -\frac{a^2\sin\vartheta\cos\vartheta}{\Sigma}, & \Gamma_{r\vartheta}^{\vartheta} &= \frac{r}{\Sigma},\\
  \Gamma_{\vartheta\vartheta}^r &= -\frac{r\Delta}{\Sigma}, & \Gamma_{\vartheta\vartheta}^{\vartheta} &= -\frac{a^2\sin\vartheta\cos\vartheta}{\Sigma},\\
  \Gamma_{\vartheta\varphi}^{\varphi} &= \frac{\cot\vartheta}{\Sigma^2}\left[\Sigma^2+r_sa^2r\sin^2\vartheta\right], & \Gamma_{\vartheta\varphi}^t &= \frac{r_sa^3r\sin^3\vartheta\cos\vartheta}{c\Sigma^2},
\end{alignat}
\begin{align}  
  \Gamma_{r\varphi}^t &= \frac{r_sa\sin^2\vartheta\left[a^2\cos^2\vartheta(a^2-r^2)-r^2(a^2+3r^2)\right]}{2c\Sigma^2\Delta},\\
  \Gamma_{r\varphi}^{\varphi} &= \frac{2r\Sigma^2+r_s\left[a^4\sin^2\vartheta\cos^2\vartheta-r^2(\Sigma+r^2+a^2)\right]}{2\Sigma^2\Delta},\\
  \Gamma_{\varphi\varphi}^r &= \frac{\Delta\sin^2\vartheta}{2\Sigma^3}\left[-2r\Sigma^2+r_sa^2\sin^2\vartheta(r^2-a^2\cos^2\vartheta)\right],\\ 
  \Gamma_{\varphi\varphi}^{\vartheta} &= -\frac{\sin\vartheta\cos\vartheta}{\Sigma^3}\left[A\Sigma+\left(r^2+a^2\right)r_sa^2r\sin^2\vartheta\right],
\end{align}
\end{subequations}

\noindent {\bf General local tetrad:}
\begin{subequations}
\begin{alignat}{3}
  \mathbf{e}_{(0)} &= \Gamma\left(\partial_t + \zeta\partial_{\varphi}\right), &\qquad \mathbf{e}_{(1)} &= \sqrt{\frac{\Delta}{\Sigma}}\partial_r,\\
  \mathbf{e}_{(2)} &= \frac{1}{\sqrt{\Sigma}}\partial_{\vartheta}, & \mathbf{e}_{(3)} &= \frac{\Gamma}{c}\left(\mp\frac{g_{t\varphi}+\zeta g_{\varphi\varphi}}{\sqrt{\Delta}\,\sin\vartheta}\partial_t \pm\frac{g_{tt}+\zeta g_{t\varphi}}{\sqrt{\Delta}\,\sin\vartheta}\partial_{\varphi}\right),
\end{alignat}
\end{subequations}
where $-\Gamma^{-2} = g_{tt}+2\zeta g_{t\varphi}+\zeta^2 g_{\varphi\varphi}$,
\begin{equation}
  \Gamma^{-2} = \left(1-\frac{r_sr}{\Sigma}\right) + \frac{2r_sar\sin^2\vartheta}{\Sigma}\frac{\zeta}{c}-\left(r^2+a^2+\frac{r_sa^2r\sin^2\vartheta}{\Sigma}\right)\frac{\zeta^2}{c^2}\sin^2\vartheta\
\end{equation}

\noindent {\bf Non-rotating local tetrad $(\zeta=\omega)$:}
\begin{equation}
  \mathbf{e}_{(0)} = \sqrt{\frac{A}{\Sigma\Delta}}\left(\frac{1}{c}\partial_t + \omega\partial_{\varphi}\right), \quad \mathbf{e}_{(1)} = \sqrt{\frac{\Delta}{\Sigma}}\partial_r, \quad \mathbf{e}_{(2)} = \frac{1}{\sqrt{\Sigma}}\partial_{\vartheta}, \quad  \mathbf{e}_{(3)} = \sqrt{\frac{\Sigma}{A}}\frac{1}{\sin\vartheta}\partial_{\varphi},
\end{equation}
where $\omega = -g_{t\varphi}/g_{\varphi\varphi}= r_sar/A$.\\[0.5em]
Dual tetrad:
\begin{equation}
 \Cdlt{2} = \sqrt{\frac{\Sigma\Delta}{A}}c\,dt, \quad \Cdlt{1} = \sqrt{\frac{\Sigma}{\Delta}}dr,\quad \Cdlt{2} = \sqrt{\Sigma}d\vartheta,\quad \Cdlt{3} = \sqrt{\frac{A}{\Sigma}}\sin\vartheta\left(d\varphi-\omega\,d\varphi\right).
\end{equation}
\newpage

The relation between the constants of motion $E$, $L$, $Q$, and $\mu$ (defined in Bardeen\cite{bardeen1972}) and the initial direction $\boldsymbol{\upsilon}$, compare Sec. (\ref{subsec:initDir}), with respect to the LNRF reads $(c=1)$
\begin{subequations}
\begin{alignat}{3}
   \upsilon^{(0)} &= \sqrt{\frac{A}{\Sigma\Delta}}\,E - \frac{r_sra}{\sqrt{A\Sigma\Delta}}L, &\qquad \upsilon^{(1)} &= \sqrt{\frac{\Delta}{\Sigma}}p_r,\\
   \upsilon^{(2)} &= \frac{1}{\sqrt{\Sigma}}\sqrt{Q-\cos^2\vartheta\left[a^2\left(\mu^2-E^2\right)+\frac{L^2}{\sin^2\vartheta}\right]}, & \upsilon^{(3)} &= \sqrt{\frac{\Sigma}{A}}\frac{L}{\sin\vartheta}.
\end{alignat}
\end{subequations}

\noindent {\bf Static local tetrad $(\zeta=0)$:}
\begin{subequations}
\begin{align}
  \mathbf{e}_{(0)} &= \frac{1}{c\sqrt{1-r_sr/\Sigma}}\partial_t, \qquad \mathbf{e}_{(1)} = \sqrt{\frac{\Delta}{\Sigma}}\partial_r, \qquad \mathbf{e}_{(2)} = \frac{1}{\sqrt{\Sigma}}\partial_{\vartheta},\\
  \mathbf{e}_{(3)} &= \pm\frac{r_sar\sin\vartheta}{c\sqrt{1-r_sr/\Sigma}\sqrt{\Delta}\Sigma}\partial_t\mp\frac{\sqrt{1-r_sr/\Sigma}}{\sqrt{\Delta}\sin\vartheta}\partial_{\varphi}.
\end{align}
\end{subequations}

\noindent {\bf Photon orbits:}\\
The direct(-) and retrograd(+) photon orbits have radius
\begin{equation}
  r_{\text{po}} = r_s\left[1+\cos\left(\frac{2}{3}\arccos\frac{\mp 2a}{r_s}\right)\right].
\end{equation}

\noindent {\bf Marginally stable timelike circular orbits}\\
are defined via
\begin{equation}
 r_{\text{ms}} = \frac{r_s}{2}\left(3+Z_2\mp\sqrt{(3-Z_1)(2+Z_1+2Z_2)}\right),
\end{equation}
where 
\begin{subequations}
 \begin{align}
  Z_1 &= 1+\left(1-\frac{4a^2}{r_s^2}\right)^{1/3}\left[\left(1+\frac{2a}{r_s}\right)^{1/3}+\left(1-\frac{2a}{r_s}\right)^{1/3}\right],\\
  Z_2 &= \sqrt{\frac{12a^2}{r_s^2}+Z_1^2}.
 \end{align}
\end{subequations}

%% -------------------- Euler Lagrange --------------------
\SecEulLag

The Euler-Lagrangian formalism, Sec.~\ref{subsec:EL}, for geodesics in the $\vartheta=\pi/2$ hyperplane yields
\begin{equation}
  \frac{1}{2}\dot{r}^2 + V_{\text{eff}} = 0
\end{equation}
with the effective potential
\begin{equation}
  V_{\text{eff}}=\frac{1}{2r^3}\left\{h^2(r-r_s)+2\frac{ahk}{c}r_s-\frac{k^2}{c^2}\left[r^3+a^2(r+r_s)\right]\right\}-\frac{\kappa c^2\Delta}{r^2}
\end{equation}
and the constants of motion 
\begin{equation}
  k = \left(1-\frac{r_s}{r}\right)c^2\dot{t}+\frac{cr_sa}{r}\dot{\varphi},\qquad  h=\left(r^2+a^2+\frac{r_sa^2}{r}\right)\dot{\varphi}-\frac{cr_sa}{r}\dot{t}.
\end{equation}

%% -------------------- Further reading --------------------
\FurtherReading

Boyer and Lindquist\cite{boyer1967}, Wilkins\cite{wilkins1972}, Brill\cite{brill1966}.

}{

}

%% ------------------------------------------------------------------------
%%    K o t t l e r
%% ------------------------------------------------------------------------
\clearpage
\section{Kottler spacetime}
\setcounter{equation}{0}
\ifthenelse{\boolean{isARXIV}}{
% ******** Start of file kottler.tex *********
%
%  Copyright (c) 2009 Thomas Mueller,
%                     Universitaet Stuttgart, VISUS
%

The Kottler spacetime is represented in spherical coordinates $(t,r,\vartheta,\varphi)$ by the line element\cite{perlick2004}
\metricEq{
  ds^2 = -\left(1-\frac{r_s}{r}-\frac{\Lambda r^2}{3}\right)c^2dt^2+\frac{1}{1-r_s/r-\Lambda r^2/3}dr^2 + r^2d\Omega^2,
}{kottler}
where $r_s=2GM/c^2$ is the Schwarzschild radius, $G$ is Newton's constant, $c$ is the speed of light, $M$ is the mass of the black hole, and $\Lambda$ is the cosmological constant. If $\Lambda>0$ the metric is also known as Schwarzschild-deSitter metric, whereas if $\Lambda<0$ it is called Schwarzschild-anti-deSitter.

For the following, we define the two abbreviations
\begin{equation}
  \alpha = 1-\frac{r_s}{r} - \frac{\Lambda r^2}{3}\qquad\text{and}\qquad \beta = \frac{r_s}{r}-\frac{2\Lambda}{3}r^2.
\end{equation}
The critical points of the Kottler metric follow from the roots of the cubic equation $\alpha=0$. These can be found by means of the parameters $p=-1/\Lambda$ and $q=3r_s/(2\Lambda)$. If $\Lambda<0$, we have only one real root 
\begin{equation}
 r_1=\frac{2}{\sqrt{-\Lambda}}\sinh\left[\frac{1}{3}\arsinh\left(\frac{3r_s}{2}\sqrt{-\Lambda}\right)\right].
\end{equation}
If $\Lambda>0$, we have to distinguish whether $D\equiv q^2+p^3=9r_s^2/(4\Lambda^2)-\Lambda^{-3}$ is positive or negative. If $D>0$, there is no real positive root. For $D<0$, the two real positive roots read
\begin{equation}
 r_{\pm}=\frac{2}{\sqrt{\Lambda}}\cos\left[\frac{\pi}{3}\pm\frac{1}{3}\arccos\left(\frac{3r_s}{2}\sqrt{\Lambda}\right)\right]
\end{equation}

%% -------------------- Christoffel symbols --------------------
\SecChristoffel
\begin{subequations}
\begin{alignat}{5}
  \Gamma_{tt}^r &= \frac{c^2\alpha\beta}{2r}, &\qquad\Gamma_{tr}^t &= \frac{\beta}{2r\alpha}, &\qquad\Gamma_{rr}^r &= -\frac{\beta}{2r\alpha},\\
  \Gamma_{r\vartheta}^{\vartheta} &= \frac{1}{r}, &\qquad \Gamma_{r\varphi}^{\varphi} &= \frac{1}{r}, &\qquad \Gamma_{\vartheta\vartheta}^{r}&=-\alpha r,\\
   \Gamma_{\vartheta\varphi}^{\varphi} &= \cot\vartheta, &\qquad \Gamma_{\varphi\varphi}^{r}&=-\alpha r\sin^2\vartheta,&\quad\Gamma_{\varphi\varphi}^{\vartheta}&=-\sin\vartheta\cos\vartheta.
\end{alignat}
\end{subequations}

%% -------------------- Riemann tensor --------------------
\SecRiemann
\begin{subequations}
\begin{alignat}{3}
  R_{trtr} &= -\frac{c^2\left(3r_s+\Lambda r^3\right)}{3r^3}, &\qquad R_{t\vartheta t\vartheta} &= \frac{1}{2}c^2\alpha\beta,\\
  R_{t\varphi t\varphi} &= \frac{1}{2}c^2\alpha\beta\sin^2\vartheta, & R_{r\vartheta r\vartheta} &= -\frac{\beta}{2\alpha},\\
  R_{r\varphi r\varphi} &= -\frac{\beta}{2\alpha}\sin^2\vartheta, &\quad R_{\vartheta\varphi\vartheta\varphi} &= r\left(r_s+\frac{\Lambda r^3}{3}\right)\sin^2\vartheta.
\end{alignat}
\end{subequations}

%% -------------------- Ricci tensor --------------------
\SecRicci
\begin{equation}
  R_{tt} = -c^2\alpha\Lambda,\qquad R_{rr}=\frac{\Lambda}{\alpha},\qquad R_{\vartheta\vartheta} = \Lambda r^2,\qquad R_{\varphi\varphi} = \Lambda r^2\sin^2\vartheta.
\end{equation}

\noindent The {\sl Ricci scalar} and the {\sl Kretschmann scalar} read
\begin{equation}
  \mathcal{R} = 4\Lambda,\qquad \mathcal{K} = 12\frac{r_s^2}{r^6} + \frac{8\Lambda^2}{3}.
\end{equation}

%% -------------------- Weyl tensor --------------------
\SecWeyl
\begin{subequations}
\begin{alignat}{5}
  C_{trtr} &= -\frac{c^2r_s}{r^3}, &\qquad C_{t\vartheta t\vartheta} &= \frac{c^2\alpha r_s}{2r}, &\qquad C_{t\varphi t\varphi} &= \frac{c^2\alpha r_s\sin^2\vartheta}{2r},\\
  C_{r\vartheta r\vartheta} &= -\frac{r_s}{2r\alpha}, &  C_{r\varphi r\varphi} &= -\frac{r_s\sin^2\vartheta}{2r\alpha}, & C_{\vartheta\varphi\vartheta\varphi} &= rr_s\sin^2\vartheta.
\end{alignat}
\end{subequations}

%% -------------------- Local tetrad --------------------
\SecLocal
\begin{equation}
  \mathbf{e}_{(t)} = \frac{1}{c\sqrt{\alpha}}\partial_t, \qquad \mathbf{e}_{(r)} = \sqrt{\alpha}\partial_r,\qquad \mathbf{e}_{(\vartheta)} = \frac{1}{r}\partial_{\vartheta}, \qquad \mathbf{e}_{(\varphi)} = \frac{1}{r\sin\vartheta}\partial_{\varphi}.
\end{equation}
Dual tetrad:
\begin{equation}
  \boldsymbol{\theta}^{(t)} = c\sqrt{\alpha}\,dt,\qquad \boldsymbol{\theta}^{(r)} = \frac{dr}{\sqrt{\alpha}},\qquad \boldsymbol{\theta}^{(\vartheta)} = r\,d\vartheta,\qquad \boldsymbol{\theta}^{(\varphi)} = r\sin\vartheta\,d\varphi.
\end{equation}

%% -------------------- Ricci rotation coefficients --------------------
\SecRicRotCoef
\begin{equation}
  \boldsymbol{\gamma}_{(r)(t)(t)} = \frac{r_s-\frac{2}{3}\Lambda r^3}{2r^2\sqrt{\alpha}},\quad \boldsymbol{\gamma}_{(\vartheta)(r)(\vartheta)} = \boldsymbol{\gamma}_{(\varphi)(r)(\varphi)} = \frac{\sqrt{\alpha}}{r},\quad \boldsymbol{\gamma}_{(\varphi)(\vartheta)(\varphi)} = \frac{\cot\vartheta}{r}.
\end{equation}
The contractions of the Ricci rotation coefficients read
\begin{equation}
  \boldsymbol{\gamma}_{(r)} = \frac{4r-3r_s-2\Lambda r^3}{2r^2\sqrt{\alpha}},\qquad \boldsymbol{\gamma}_{(\vartheta)} = \frac{\cot\vartheta}{r}.
\end{equation}

%% -------------------- Riemann tensor LT --------------------
\SecRiemannLT
\begin{subequations}
 \begin{align}
  R_{(t)(r)(t)(r)} &= -R_{(\vartheta)(\varphi)(\vartheta)(\varphi)} = -\frac{\Lambda r^3+3r_s}{3r^3},\\
  R_{(t)(\vartheta)(t)(\vartheta)} &= R_{(t)(\varphi)(t)(\varphi)} = -R_{(r)(\vartheta)(r)(\vartheta)} = -R_{(r)(\varphi)(r)(\varphi)} = \frac{3r_s-2\Lambda r^3}{6r^3}.
 \end{align}
\end{subequations}

%% -------------------- Weyl tensor LT --------------------
\SecWeylLT
\begin{subequations}
 \begin{align}
  C_{(t)(r)(t)(r)} &= -C_{(\vartheta)(\varphi)(\vartheta)(\varphi)} = -\frac{r_s}{r^3},\\
  C_{(t)(\vartheta)(t)(\vartheta)} &= C_{(t)(\varphi)(t)(\varphi)} = -C_{(r)(\vartheta)(r)(\vartheta)} = -C_{(r)(\varphi)(r)(\varphi)} = \frac{r_s}{2r^3}.
 \end{align}
\end{subequations}

%% -------------------- Embedding --------------------
\SecEmbedding

The embedding function follows from the numerical integration of
\begin{equation}
  \frac{dz}{dr} = \sqrt{\frac{r_s/r+\Lambda r^2/3}{1-r_s/r-\Lambda r^2/3}}.
\end{equation}

%% -------------------- Euler Lagrange --------------------
\SecEulLag

The Euler-Lagrangian formalism\cite{rindler} yields the effective potential
\begin{equation}
  V_{\text{eff}}=\frac{1}{2}\left(1-\frac{r_s}{r}-\frac{\Lambda r^2}{3}\right)\left(\frac{h^2}{r^2}-\kappa c^2\right)
\end{equation}
with the constants of motion $k=(1-r_s/r-\Lambda r^2/3)c^2\dot{t}$, $h=r^2\dot{\varphi}$, and $\kappa$ as in Eq.~(\ref{eq:constrEq}).\\
As in the Schwarzschild metric, the effective potential has only one extremum for null geodesics, the so called photon orbit at $r=\frac{3}{2}r_s$. For timelike geodesics, however, we have
\begin{equation}
 \frac{dV_{\text{eff}}}{dr} = \frac{h^2(-6r+9r_s)+c^2r^2(3r_s-2r^3\Lambda)}{3r^4}\stackrel{!}{=}0.
\end{equation}
This polynomial of fifth order might have up to five extrema.
\vspace*{0.5cm}

%% -------------------- Further Reading --------------------
\FurtherReading

Kottler\cite{kottler1918}, Weyl\cite{weyl1919}, Hackmann\cite{hackmann2008}, Cruz\cite{cruz2005}.

}{

}

%% ------------------------------------------------------------------------
%%      M o r r i s  -  T h o r n e
%% ------------------------------------------------------------------------
\clearpage
\section{Morris-Thorne}
\setcounter{equation}{0}
\ifthenelse{\boolean{isARXIV}}{
% ******** Start of file morristhornesimple.tex *********
%
%  Copyright (c) 2009 Thomas Mueller,
%                     Universitaet Stuttgart, VISUS
%

The most simple wormhole geometry is represented by the metric of Morris and Thorne\cite{morris1988},
\metricEq{
  ds^2 = -c^2dt^2 + dl^2 + (b_0^2+l^2)\left(d\vartheta^2+\sin^2\!\vartheta\,d\varphi^2\right),
}{morrisThorne}
where $b_0$ is the throat radius and $l$ is the proper radial coordinate; and $\left\{t\in\setR,l\in\setR,\vartheta\in(0,\pi),\varphi\in[0,2\pi)\right\}$.

%% -------------------- Christoffel symbols --------------------
\SecChristoffel
\begin{subequations}
  \begin{alignat}{3}
    \Gamma_{l\vartheta}^{\vartheta} &= \frac{l}{b_0^2+l^2}, &\qquad \Gamma_{l\varphi}^{\varphi} &= \frac{l}{b_0^2+l^2}, &\qquad \Gamma_{\vartheta\vartheta}^{l} &= -l,\\
   \Gamma_{\vartheta\varphi}^{\varphi} &= \cot\vartheta, &\qquad \Gamma_{\varphi\varphi}^{l} &= -l\sin^2\!\vartheta, & \Gamma_{\varphi\varphi}^{\vartheta} &= -\sin\vartheta\,\cos\vartheta.
  \end{alignat}
\end{subequations}

Partial derivatives
\begin{subequations}
 \begin{alignat}{5}
    \Gamma_{l\vartheta,l}^{\vartheta} &= -\frac{l^2-b_0^2}{(b_0^2+l^2)^2}, &\qquad \Gamma_{l\varphi,l}^{\varphi} &= -\frac{l^2-b_0^2}{(b_0^2+l^2)^2}, &\qquad \Gamma_{\vartheta\vartheta,l}^l &= -1,\\
    \Gamma_{\vartheta\varphi,\vartheta}^{\varphi} &= -\frac{1}{\sin^2\vartheta}, & \Gamma_{\varphi\varphi,l}^l &= -\sin^2\vartheta, & \Gamma_{\varphi\varphi,\vartheta}^l &= -l\sin(2\vartheta),\\
    \Gamma_{\varphi\varphi,\vartheta}^{\vartheta} &= -\cos(2\vartheta).
 \end{alignat}
\end{subequations}

%% -------------------- Riemann tensor --------------------
\SecRiemann
\begin{equation}
  R_{l\vartheta l\vartheta}=-\frac{b_0^2}{b_0^2+l^2},\quad R_{l\varphi l\varphi}=-\frac{b_0^2\sin^2\!\vartheta}{b_0^2+l^2},\quad R_{\vartheta\varphi\vartheta\varphi}=b_0^2\sin^2\!\vartheta.
\end{equation}
{\bf Ricci tensor, Ricci and Kretschmann scalar:}
\begin{equation}
  R_{ll} = -2\frac{b_0^2}{\left(b_0^2+l^2\right)^2},\qquad \mathcal{R}=-2\frac{b_0^2}{\left(b_0^2+l^2\right)^2}, \qquad \mathcal{K} = \frac{12b_0^4}{\left(b_0^2+l^2\right)^4}.
\end{equation}

%% -------------------- Weyl tensor --------------------
\SecWeyl
\begin{subequations}
 \begin{alignat}{5}
  C_{tltl} &= -\frac{2}{3}\frac{c^2b_0^2}{\left(b_0^2+l^2\right)^2}, &\qquad C_{t\vartheta t\vartheta} &= \frac{1}{3}\frac{c^2b_0^2}{b_0^2+l^2}, &\qquad C_{t\varphi t\varphi} &= \frac{1}{3}\frac{c^2b_0^2\sin^2\vartheta}{b_0^2+l^2},\\
  C_{l\vartheta l\vartheta} &= -\frac{1}{3}\frac{b_0^2}{b_0^2+l^2}, & C_{l\varphi l\varphi} &= -\frac{1}{3}\frac{b_0^2\sin^2\vartheta}{b_0^2+l^2}, & C_{\vartheta\varphi\vartheta\varphi} &= \frac{2}{3}b_0^2\sin^2\vartheta.
 \end{alignat}
\end{subequations}

%% -------------------- Local tetrad --------------------
\SecLocal
\begin{equation}
  \mathbf{e}_{(t)} = \frac{1}{c}\partial_t, \qquad \mathbf{e}_{(l)} = \partial_l, \qquad \mathbf{e}_{(\vartheta)} =\frac{1}{\sqrt{b_0^2+l^2}}\partial_{\vartheta}, \qquad \mathbf{e}_{(\varphi)} = \frac{1}{\sqrt{b_0^2+l^2}\,\sin\vartheta}\partial_{\varphi}.
\end{equation}
Dual tetrad
\begin{equation}
 \boldsymbol{\theta}^{(t)} = c\,dt,\qquad \boldsymbol{\theta}^{(l)} = dl,\qquad \boldsymbol{\theta}^{(\vartheta)}=\sqrt{b_0^2+l^2}\,d\vartheta,\qquad \boldsymbol{\theta}^{(\varphi)}=\sqrt{b_0^2+l^2}\sin\vartheta\,d\varphi.
\end{equation}

%% -------------------- Ricci rotation coefficients --------------------
\SecRicRotCoef
\begin{equation}
   \boldsymbol{\gamma}_{(\vartheta)(r)(\vartheta)} = \boldsymbol{\gamma}_{(\varphi)(r)(\varphi)} = \frac{l}{b_0^2+l^2},\qquad \boldsymbol{\gamma}_{(\varphi)(\vartheta)(\varphi)} = \frac{\cot\vartheta}{\sqrt{b_0^2+l^2}}.
\end{equation}
The contractions of the Ricci rotation coefficients read
\begin{equation}
  \boldsymbol{\gamma}_{(r)} = \frac{2l}{b_0^2+l^2},\qquad \boldsymbol{\gamma}_{(\vartheta)} = \frac{\cot\vartheta}{\sqrt{b_0^2+l^2}}.
\end{equation}

%% -------------------- Riemann tensor LT--------------------
\SecRiemannLT
\begin{equation}
 R_{(l)(\vartheta)(l)(\vartheta)} = R_{(l)(\varphi)(l)(\varphi)} = -R_{(\vartheta)(\varphi)(\vartheta)(\varphi)} = -\frac{b_0^2}{\left(b_0^2+l^2\right)^2}.
\end{equation}

%% -------------------- Ricci tensor LT--------------------
\SecRicciLT
\begin{equation}
 R_{(l)(l)} = -\frac{2b_0^2}{\left(b_0^2+l^2\right)^2}.
\end{equation}

%% -------------------- Weyl tensor LT--------------------
\SecWeylLT
\begin{subequations}
 \begin{align}
  C_{(t)(l)(t)(l)} &= -C_{(\vartheta)(\varphi)(\vartheta)(\varphi)} = -\frac{2b_0^2}{3\left(b_0^2+l^2\right)^2},\\
  C_{(t)(\vartheta)(t)(\vartheta)} &= C_{(t)(\varphi)(t)(\varphi)} = -C_{(l)(\vartheta)(l)(\vartheta)} = -C_{(l)(\varphi)(l)(\varphi)} = \frac{b_0^2}{3\left(b_0^2+l^2\right)^2}.
 \end{align}
\end{subequations}

%% -------------------- Embedding --------------------
\SecEmbedding

The embedding function reads
\begin{equation}
  z(r) = \pm b_0\ln\left[\frac{r}{b_0}+\sqrt{\left(\frac{r}{b_0}\right)^2-1}\right]
\end{equation}
with $r^2 = b_0^2+l^2$.
\vspace*{0.5cm}

%% --------------------Euler Lagrange --------------------
\SecEulLag

The Euler-Lagrangian formalism, Sec.~\ref{subsec:EL}, for geodesics in the $\vartheta=\pi/2$ hyperplane yields \begin{equation}
  \frac{1}{2}\dot{l}^2+V_{\text{eff}} = \frac{1}{2}\frac{k^2}{c^2},\qquad V_{\text{eff}} = \frac{1}{2}\left(\frac{h^2}{b_0^2+l^2}-\kappa c^2\right),
\end{equation}
with the constants of motion $k=c^2\dot{t}$ and $h=(b_0^2+l^2)\dot{\varphi}$. The shape of the effective potential $V_{\text{eff}}$ is independend of the geodesic type. The maximum of the effective potential is located at $l=0$.\\[0.5em]
A geodesic that starts at $l=l_i$ with direction $\mathbf{y}=\pm\Clt{t}+\cos\xi\Clt{l}+\sin\xi\Clt{\varphi}$ approaches the wormhole throat asymptotically for $\xi=\xi_{\text{crit}}$ with
\begin{equation}
  \xi_{\text{crit}} = \arcsin\frac{b_0}{\sqrt{b_0^2+l_i^2}}.
\end{equation}
This critical angle is independent of the type of the geodesic.
\vspace*{0.5cm}

%% -------------------- Further Reading --------------------
\FurtherReading

Ellis\cite{ellis1973}, Visser\cite{visser1995}, M{\"u}ller\cite{mueller2004,mueller2008prdA}

}{

}

%% ------------------------------------------------------------------------
%%      O p p e n h e i m e r  -  Snyder  collapse
%% ------------------------------------------------------------------------
\clearpage
\section{Oppenheimer-Snyder collapse}
\setcounter{equation}{0}
\ifthenelse{\boolean{isARXIV}}{
% ******** Start of file oppenheimer.tex *********
%
%  Copyright (c) 2009 Thomas Mueller,
%                     Universitaet Stuttgart, VISUS
%

% -------------------------------------------------------------------
%   outer metric
% -------------------------------------------------------------------
\subsection{Outer metric}
The metric of the outer spacetime, $R>R_b$, in comoving coordinates $(\tau,R,\vartheta,\varphi)$ with $(c=1)$ is given by
\metricEq{
   ds^2 =-d\tau^2+\frac{R}{\left(R^{3/2}-\frac{3}{2}\sqrt{r_s}\tau\right)^{2/3}}dR^2 + \left(R^{3/2}-\frac{3}{2}\sqrt{r_s}\tau\right)^{4/3}\left(d\vartheta^2+\sin^2\vartheta d\varphi^2\right).
}{oppiOut}

%% -------------------- Christoffel symbols --------------------
\SecChristoffel
\begin{subequations}
\begin{alignat}{3}
  \Gamma_{\tau R}^R &= \frac{1}{2}\frac{\sqrt{r_s}}{R^{3/2}-\frac{3}{2}\sqrt{r_s}\tau}, &\qquad \Gamma_{\tau\vartheta}^{\vartheta} &= -\frac{\sqrt{r_s}}{R^{3/2}-\frac{3}{2}\sqrt{r_s}\tau},\\
  \Gamma_{\tau\varphi}^{\varphi} &= -\frac{\sqrt{r_s}}{R^{3/2}-\frac{3}{2}\sqrt{r_s}\tau}, & \Gamma_{RR}^{\tau} &= \frac{R\sqrt{r_s}}{2\left(R^{3/2}-\frac{3}{2}\sqrt{r_s}\tau\right)^{5/3}},\\
  \Gamma_{RR}^R &= -\frac{3\sqrt{r_s}\tau}{4\left(R^{3/2}-\frac{3}{2}\sqrt{r_s}\tau\right)R}, & \Gamma_{R\vartheta}^{\vartheta} &= \frac{\sqrt{R}}{R^{3/2}-\frac{3}{2}\sqrt{r_s}\tau},\\
  \Gamma_{R\varphi}^{\varphi} &= \frac{\sqrt{R}}{R^{3/2}-\frac{3}{2}\sqrt{r_s}\tau}, & \Gamma_{\vartheta\vartheta}^{\tau} &= -\sqrt{r_s}\left(R^{3/2}-\frac{3}{2}\sqrt{r_s}\tau\right)^{1/3},\\
   \Gamma_{\vartheta\vartheta}^R &= -\frac{R^{3/2}-\frac{3}{2}\sqrt{r_s}\tau}{\sqrt{R}}, & \Gamma_{\vartheta\varphi}^{\varphi} &= \cot\vartheta,\\
   \Gamma_{\varphi\varphi}^{\tau} &= -\sqrt{r_s}\left(R^{3/2}-\frac{3}{2}\sqrt{r_s}\tau\right)^{1/3}\!\sin^2\!\vartheta, & \Gamma_{\varphi\varphi}^{\vartheta} &= -\sin\vartheta\,\cos\vartheta,\\
  \Gamma_{\varphi\varphi}^R &= -\frac{\left(R^{3/2}-\frac{3}{2}\sqrt{r_s}\tau\right)\sin^2\!\vartheta}{\sqrt{R}}.
\end{alignat}
\end{subequations}

%% -------------------- Riemann tensor --------------------
\SecRiemann
\begin{subequations}
\begin{alignat}{3}
  R_{\tau R\tau R} &=-\frac{Rr_s}{\left(R^{3/2}-\frac{3}{2}\sqrt{r_s}\,\tau\right)^{8/3}}, &\qquad R_{\tau\vartheta\tau\vartheta} &= \frac{1}{2}\frac{r_s}{\left(R^{3/2}-\frac{3}{2}\sqrt{r_s}\,\tau\right)^{2/3}},\\
  R_{\tau\varphi\tau\varphi} &= \frac{1}{2}\frac{r_s\sin^2\vartheta}{\left(R^{3/2}-\frac{3}{2}\sqrt{r_s}\,\tau\right)^{2/3}}, &  R_{R\vartheta R\vartheta} &= -\frac{1}{2}\frac{Rr_s}{\left(R^{3/2}-\frac{3}{2}\sqrt{r_s}\,\tau\right)^{4/3}},\\
  R_{R\varphi R\varphi} &= -\frac{1}{2}\frac{Rr_s\sin^2\vartheta}{\left(R^{3/2}-\frac{3}{2}\sqrt{r_s}\,\tau\right)^{4/3}}, & R_{\vartheta\varphi\vartheta\varphi} &= \left(R^{3/2}-\frac{3}{2}\sqrt{r_s}\,\tau\right)^{2/3}\,r_s\sin^2\vartheta.
\end{alignat}
\end{subequations}

The Ricci tensor and the Ricci scalar vanish identically.

%% -------------------- Kretschmann scalar --------------------
\SecKretsch
\begin{equation}
  \mathcal{K}=12\frac{r_s^2}{\left(R^{3/2}-\frac{3}{2}\sqrt{r_s}\,\tau\right)^4}.
\end{equation}

%% -------------------- Local tetrad --------------------
\SecLocal
\begin{subequations}
\begin{alignat}{3}
  \mathbf{e}_{(\tau)} &= \partial_{\tau}, &\qquad \mathbf{e}_{(R)} &= \frac{\left(R^{3/2}-\frac{3}{2}\sqrt{r_s}\tau\right)^{1/3}}{\sqrt{R}}\partial_R,\\
  \mathbf{e}_{(\vartheta)} &= \frac{1}{\left(R^{3/2}-\frac{3}{2}\sqrt{r_s}\tau\right)^{2/3}}\partial_{\vartheta}, & \mathbf{e}_{(\varphi)} &=\frac{1}{\left(R^{3/2}-\frac{3}{2}\sqrt{r_s}\tau\right)^{2/3}\sin\vartheta}\partial_{\varphi}.
\end{alignat}
\end{subequations}

%% -------------------- Ricci rotation coefficients --------------------
\SecRicRotCoef
\begin{subequations}
\begin{align}  
  \gamma_{(\tau)(R)(R)} &= -\frac{\sqrt{r_s}}{2R^{3/2}-3\sqrt{r_s}\tau},\quad \gamma_{(\tau)(\vartheta)(\vartheta)} = \gamma_{(\tau)(\varphi)(\varphi)} = \frac{2\sqrt{r_s}}{2R^{3/2}-3\sqrt{r_s}\tau},\\
  \gamma_{(R)(\varphi)(\varphi)} &= \gamma_{(R)(\vartheta)(\vartheta)} = -\left(R^{3/2}-\frac{3}{2}\sqrt{r_s}\tau\right)^{-2/3}.
\end{align}
\end{subequations}
The contractions of the Ricci rotation coefficients read
\begin{equation}
  \gamma_{(\tau)} = -\frac{3\sqrt{r_s}}{2R^{3/2}-3\sqrt{r_s}\tau},\quad \gamma_{(R)} = 2\left(R^{3/2}-\frac{3}{2}\sqrt{r_s}\tau\right)^{-2/3},\quad \gamma_{(\vartheta)} = \cot\vartheta\left(R^{3/2}-\frac{3}{2}\sqrt{r_s}\tau\right)^{-2/3}.
\end{equation}

%% -------------------- Riemann tensor LT--------------------
\SecRiemannLT
\begin{subequations}
 \begin{align}
  R_{(\tau)(R)(\tau)(R)} &= -R_{(\vartheta)(\varphi)(\vartheta)(\varphi)} = -\frac{4r_s}{\left(2R^{3/2}-3\sqrt{r_s}\tau\right)^2},\\
  R_{(\tau)(\vartheta)(\tau)(\vartheta)} &= R_{(\tau)(\varphi)(\tau)(\varphi)} = -R_{(R)(\vartheta)(R)(\vartheta)} = -R_{(R)(\varphi)(R)(\varphi)} = \frac{2r_s}{\left(2R^{3/2}-3\sqrt{r_s}\tau\right)^2}.
 \end{align}
\end{subequations}

The Ricci tensor with respect to the local tetrad vanishes identically.

% -------------------------------------------------------------------
%   inner metric
% -------------------------------------------------------------------
\subsection{Inner metric}
The metric of the inside, $R\leq R_b$, reads
\metricEq{
   ds^2 =-d\tau^2+\left(1-\frac{3}{2}\sqrt{r_s}R_b^{-3/2}\tau\right)^{4/3}\left[dR^2 + R^2\left(d\vartheta^2+\sin^2\vartheta d\varphi^2\right)\right].
}{oppiIn}
For the following components, we define
\begin{equation}
  A_{\text{Oin}} := 1-\frac{3}{2}\sqrt{r_s}R_b^{-3/2}\tau.
\end{equation}

%% -------------------- Christoffel symbols --------------------
\SecChristoffel
\begin{subequations}
\begin{alignat}{5}
  \Gamma_{\tau R}^R &= -\frac{\sqrt{r_s}R_b^{-3/2}}{A_{\text{Oin}}}, &\quad \Gamma_{\tau\vartheta}^{\vartheta} &= -\frac{\sqrt{r_s}R_b^{-3/2}}{A_{\text{Oin}}}, &\quad \Gamma_{\tau\varphi}^{\varphi} &= -\frac{\sqrt{r_s}R_b^{-3/2}}{A_{\text{Oin}}},\\
  \Gamma_{RR}^{\tau} &= -A_{\text{Oin}}^{1/3}\!\!\sqrt{r_s}R_b^{-3/2}, & \Gamma_{R\vartheta}^{\vartheta} &= \frac{1}{R}, & \Gamma_{R\varphi}^{\varphi} &= \frac{1}{R},\\
  \Gamma_{\vartheta\vartheta}^R &= -R, & \Gamma_{\vartheta\varphi}^{\varphi} &= \cot\vartheta, & \Gamma_{\vartheta\vartheta}^{\tau} &= -A_{\text{Oin}}^{1/3}\!\!\sqrt{r_s}R_b^{-3/2}R^2,\\   
  \Gamma_{\varphi\varphi}^R &= -R\sin^2\vartheta, & \Gamma_{\varphi\varphi}^{\vartheta} &= -\sin\vartheta\,\cos\vartheta, & \Gamma_{\varphi\varphi}^{\tau} &= -A_{\text{Oin}}^{1/3}\!\!\sqrt{r_s}R_b^{-3/2}R^2\sin^2\vartheta.
\end{alignat}
\end{subequations}

%% -------------------- Riemann tensor --------------------
\SecRiemann
\begin{subequations}
\begin{alignat}{5}
  R_{\tau R\tau R} &= \frac{1}{2}\frac{r_s}{R_b^3A_{\text{Oin}}^{2/3}}, &\qquad R_{\tau\vartheta\tau\vartheta} &= \frac{1}{2}\frac{r_sR^2}{R_b^3A_{\text{Oin}}^{2/3}}, &\qquad R_{\tau\varphi\tau\varphi} &= \frac{1}{2}\frac{r_sR^2\sin^2\vartheta}{R_b^3A_{\text{Oin}}^{2/3}},\\
  R_{R\varphi R\varphi} &= \frac{r_sR^2\sin^2\vartheta}{R_b^3}A_{\text{Oin}}^{2/3}, &  R_{R\vartheta R\vartheta} &= \frac{r_sR^2}{R_b^3}A_{\text{Oin}}^{2/3},  & R_{\vartheta\varphi\vartheta\varphi} &= \frac{r_sR^4\sin^2\vartheta}{R_b^3}A_{\text{Oin}}^{2/3}.
\end{alignat}
\end{subequations}

%% -------------------- Ricci tensor --------------------
\SecRicci
\begin{equation}
  R_{\tau\tau} = \frac{3}{2}\frac{r_s}{R_b^3A_{\text{Oin}}^2},\qquad  R_{RR} = \frac{3}{2}\frac{r_s}{R_b^3A_{\text{Oin}}^{2/3}},\quad R_{\vartheta\vartheta} = \frac{3}{2}\frac{r_sR^2}{R_b^3A_{\text{Oin}}^{2/3}},\quad R_{\varphi\varphi} = \frac{3}{2}\frac{r_sR^2\sin^2\vartheta}{R_b^3A_{\text{Oin}}^{2/3}}.
\end{equation}

\noindent The {\sl Ricci} and {\sl Kretschmann} scalars read:
\begin{equation}
  \mathcal{R} = \frac{3r_s}{R_b^3A_{\text{Oin}}^2},\qquad \mathcal{K} =  15\frac{r_s^2}{R_b^6A_{\text{Oin}}^4}.
\end{equation}

%% -------------------- Local tetrad --------------------
\SecLocal
\begin{equation}
  \mathbf{e}_{(\tau)} = \partial_{\tau}, \qquad \mathbf{e}_{(R)} = \frac{1}{A_{\text{Oin}}^{2/3}}\partial_R,\qquad \mathbf{e}_{(\vartheta)} = \frac{1}{RA_{\text{Oin}}^{2/3}}\partial_{\vartheta}, \qquad \mathbf{e}_{(\varphi)} = \frac{1}{A_{\text{Oin}}^{2/3}R\sin\vartheta}\partial_{\varphi}.
\end{equation}

%% -------------------- Ricci rotation coefficients --------------------
\SecRicRotCoef
\begin{subequations}
\begin{align}  
  \gamma_{(\tau)(R)(R)} &= \gamma_{(\tau)(\vartheta)(\vartheta)} = \gamma_{(\tau)(\varphi)(\varphi)} = \frac{\sqrt{r_s}R_b^{-3/2}}{A_{\text{Oin}}},\\ 
  \gamma_{(R)(\vartheta)(\vartheta)} &= \gamma_{(R)(\varphi)(\varphi)} = -\frac{1}{RA_{\text{Oin}}^{2/3}},\\ \gamma_{(\vartheta)(\varphi)(\varphi)} &= -\frac{\cot\vartheta}{RA_{\text{Oin}}^{2/3}}.
\end{align}
\end{subequations}
The contractions of the Ricci rotation coefficients read
\begin{equation}
  \gamma_{(\tau)} = -\frac{3\sqrt{r_s}R_b^{-3/2}}{A_{\text{Oin}}},\qquad \gamma_{(R)} = \frac{2}{RA_{\text{Oin}}^{2/3}},\qquad \gamma_{(\vartheta)} = \frac{\cot\vartheta}{RA_{\text{Oin}}^{2/3}}.
\end{equation}

%% -------------------- Riemann tensor LT--------------------
\SecRiemannLT
\begin{subequations}
 \begin{align}
  R_{(\tau)(R)(\tau)(R)} &= R_{(\tau)(\vartheta)(\tau)(\vartheta)} = R_{(\tau)(\varphi)(\tau)(\varphi)} = \frac{r_sR_b^{-3}}{2A_{\text{Oin}}^2},\\
  R_{(R)(\vartheta)(R)(\vartheta)} &= R_{(R)(\varphi)(R)(\varphi)} = R_{(\vartheta)(\varphi)(\vartheta)(\varphi)} = \frac{r_sR_b^{-3}}{A_{\text{Oin}}^2}.
 \end{align}
\end{subequations}

%% -------------------- Ricci tensor LT--------------------
\SecRicciLT
\begin{equation}
 R_{(\tau)(\tau)} = R_{(R)(R)} = R_{(\vartheta)(\vartheta)} = R_{(\varphi)(\varphi)} = \frac{3r_sR_b^{-3}}{2A_{\text{Oin}}^2}.
\end{equation}

%% -------------------- Further reading --------------------
\FurtherReading

Oppenheimer and Snyder\cite{oppenheimer1939b}.
}{

}

%% ------------------------------------------------------------------------
%%     Petrov-Type D
%% ------------------------------------------------------------------------
\clearpage
\section{Petrov-Type D -- Levi-Civita spacetimes}
\setcounter{equation}{0}
The Petrov type D static vacuum spacetimes AI-C are taken from Stephani et al.\cite{exact2003}, Sec. 18.6, with the coordinate and parameter ranges given in "Exact solutions of the gravitational field equations" by Ehlers and Kundt \cite{ehlers1962}.
% \input{metrics/PTD_AI}
% \input{metrics/PTD_AII}
% \input{metrics/PTD_AIII}
% \input{metrics/PTD_BI}
% \input{metrics/PTD_BII}
% \input{metrics/PTD_BIII}
% \input{metrics/PTD_C}

%% ------------------------------------------------------------------------
%%    Petrov Type D ES
%% ------------------------------------------------------------------------
%\newcommand{\ESCstr}{Petrov Type D Exact Solutions Case}
%\newcommand{\PTDRangeStart}{Coordinates and parameters are restricted to\begin{center} \begin{tabular}{l}}
%\newcommand{\PTDRangeEnd}{  \end{tabular} \end{center}}
\newcommand{\PTDRangeStart}{Coordinates and parameters are restricted to\begin{center} }
\newcommand{\PTDRangeEnd}{ \end{center}}

%% ------------------------------------------------------------------------
%%    Petrov Type D ES AI
%% ------------------------------------------------------------------------
%\setcounter{equation}{0}
\ifthenelse{\boolean{isARXIV}}{
% ******** Start of file PTD_AI.tex *********
%
%  Copyright (c) 2010 Felix Beslmeisl,
%                     Universitaet Stuttgart
%
\subsection{\ESCstr~AI}

%\subsection{Spherical Coordinates}
In spherical coordinates, $(t,r,\vartheta,\varphi)$, the metric is given by the line element
\metricEq{
  ds^2 = r^2\left( d\vartheta^2 + \sin^2 \vartheta d\varphi^2 \right) + \frac{r}{r-b}dr^2-\frac{r-b}{r}dt^2.
}{PTD_AI_Spherical}
This is the well known Schwarzschild solution if $b=r_s$, cf. Eq.~(\ref{eqM:schwarzschildSpherical}).
\PTDRangeStart
  $t\in\setR$, \qquad $0< \vartheta< \pi$, \qquad $\varphi \in [0,2\pi)$, \qquad $(0<b<r) \vee (b<0<r)$.
\PTDRangeEnd

% -------------------- Local tetrad --------------------
 \SecLocal
 \begin{equation}
   \mathbf{e}_{(t)} = \sqrt{{\frac {r}{r-b}}}\partial_t, \qquad 
   \mathbf{e}_{(r)} = \sqrt{{\frac {r-b}{r}}}\partial_r, \qquad
   \mathbf{e}_{(\vartheta)} = \frac{1}{r}\partial_{\vartheta}, \qquad 
   \mathbf{e}_{(\varphi)} = \frac {1}{r\sin{\vartheta}}\partial_{\varphi}.
 \end{equation}
 Dual tetrad:
 \begin{equation}
  \boldsymbol{\theta}^{(t)} = \sqrt{{\frac {r-b}{r}}}\,dt, \qquad 
  \boldsymbol{\theta}^{(r)} = \sqrt {{\frac {r}{r-b}}}\,dr, \qquad 
  \boldsymbol{\theta}^{(\vartheta)}= r\,d\vartheta, \qquad
  \boldsymbol{\theta}^{(\varphi)} = r\sin{\vartheta}\,d\varphi.
 \end{equation}
 
% -------------------- effective Potential -------------------- 
 \SecEffPoti \\
With the Hamilton-Jacobi formalism it is possible to obtain an effective potential fulfilling $\frac{1}{2}\dot{r}^2 + \frac{1}{2}V_{\text{eff}}(r)=\frac{1}{2}C_0^2$ with
\begin{equation}
  V_{\text{eff}}(r)=K\frac{r-b}{r^3}-\kappa\frac{r-b}{r}
\end{equation}
and the constants of motion
\begin{subequations}
\begin{eqnarray}
   C_0^2 &=& \dot{t}^2\left(\frac{r-b}{r}\right)^2,  \\
   K     &=& \dot{\vartheta}^2 r^4 + \dot{\varphi}^2 r^4 \sin^2{\vartheta}. %\\
   %-\kappa c^2 &=& -K\frac1{r^2}-\dot{r}^2 \frac{r}{r - b} +\dot{t}^2\frac{r - b}{r}.
\end{eqnarray}
\end{subequations}

}{

}

%% ------------------------------------------------------------------------
%%    Petrov Type D ES AII
%% ------------------------------------------------------------------------
%\clearpage
%\setcounter{equation}{0}
\ifthenelse{\boolean{isARXIV}}{
% ******** Start of file PTD_AII.tex *********
%
%  Copyright (c) 2010 Felix Beslmeisl,
%                     Universitaet Stuttgart
%
\subsection{\ESCstr~AII}

%\subsection{Cylindrical Coordinates}
In cylindrical coordinates, the metric is given by the line element
\metricEq{
  ds^2 = z^2\left( dr^2 + \sinh^2r\,d\varphi^2 \right) + \frac{z}{b-z}dz^2-\frac{b-z}{z}dt^2.
}{PTD_AII_Spherical}
\PTDRangeStart
  $t\in\setR$, \qquad $0< r$, \qquad $\varphi \in [0,2\pi)$, \qquad $0<z < b$.
\PTDRangeEnd

% -------------------- Local tetrad --------------------
 \SecLocal
 \begin{equation}
   \mathbf{e}_{(t)} = \sqrt{{\frac {z}{b-z}}}\partial_t, \qquad 
   \mathbf{e}_{(r)} = \frac{1}{z}\partial_r, \qquad
   \mathbf{e}_{(\varphi)} = \frac {1}{z\sinh{r}}\partial_{\varphi}, \qquad 
   \mathbf{e}_{(z)} = \sqrt{{\frac {b-z}{z}}}\partial_{z}.
 \end{equation}
 Dual tetrad:
 \begin{equation}
  \boldsymbol{\theta}^{(t)} = \sqrt{{\frac {b-z}{z}}}\,dt, \qquad 
  \boldsymbol{\theta}^{(r)} = z\,dr, \qquad 
  \boldsymbol{\theta}^{(\varphi)}= z\sinh{r}\,d\varphi, \qquad
  \boldsymbol{\theta}^{(z)} = \sqrt{{\frac {z}{b-z}}}\,dz.
 \end{equation}

}{

}

%% ------------------------------------------------------------------------
%%    Petrov Type D ES AIII
%% ------------------------------------------------------------------------
%\clearpage
%\setcounter{equation}{0}
\ifthenelse{\boolean{isARXIV}}{
% ******** Start of file PTD_AIII.tex *********
%
%  Copyright (c) 2010 Felix Beslmeisl,
%                     Universitaet Stuttgart
%
\subsection{\ESCstr~AIII}

%\subsection{Cylindrical Coordinates}
In cylindrical coordinates, the metric is given by the line element
\metricEq{
  ds^2 = z^2\left( dr^2 + r^2 d\varphi^2 \right) + z dz^2-\frac{1}{z}dt^2.
}{PTD_AIII_Spherical}
\PTDRangeStart
 $t\in\setR$, \qquad $0< r $, \qquad  $\varphi \in [0,2\pi)$, \qquad $0<z$.
\PTDRangeEnd

% -------------------- Local tetrad --------------------
 \SecLocal
 \begin{equation}
   \mathbf{e}_{(t)} = \sqrt{z}\partial_t, \qquad 
   \mathbf{e}_{(r)} = \frac{1}{z}\partial_r, \qquad
   \mathbf{e}_{(\varphi)} = \frac{1}{zr}\partial_{\varphi}, \qquad 
   \mathbf{e}_{(z)} = \frac{1}{\sqrt{z}}\partial_{z}.
 \end{equation}
 Dual tetrad:
 \begin{equation}
  \boldsymbol{\theta}^{(t)} = \frac{1}{\sqrt{z}}dt, \qquad 
  \boldsymbol{\theta}^{(r)} = z\,dr, \qquad 
  \boldsymbol{\theta}^{(\varphi)}= zr\,d\varphi, \qquad
  \boldsymbol{\theta}^{(z)} = \sqrt{z}\,dz.
 \end{equation}

}{

}

%% ------------------------------------------------------------------------
%%    Petrov Type D ES BI
%% ------------------------------------------------------------------------
%\clearpage
%\setcounter{equation}{0}
\ifthenelse{\boolean{isARXIV}}{
% ******** Start of file PTD_BI.tex *********
%
%  Copyright (c) 2010 Felix Beslmeisl,
%                     Universitaet Stuttgart
%
\subsection{\ESCstr~BI}

%\subsection{Spherical Coordinates}
In spherical coordinates, the metric is given by the line element
\metricEq{
  ds^2 = r^2\left( d\vartheta^2 - \sin^2 \vartheta dt^2 \right) + \frac{r}{r-b}dr^2+\frac{r-b}{r}d\varphi^2.
}{PTD_BI_Spherical}
\PTDRangeStart
 $t\in\setR$, \qquad $0< \vartheta< \pi$, \qquad $\varphi\in [0,2\pi)$, \qquad $(0<b<r) \vee (b<0<r)$.
\PTDRangeEnd

% -------------------- Local tetrad --------------------
 \SecLocal
 \begin{equation}
   \mathbf{e}_{(t)} = \frac {1}{r\sin{\vartheta}}\partial_t, \qquad 
   \mathbf{e}_{(r)} = \sqrt{{\frac {r-b}{r}}}\partial_r, \qquad
   \mathbf{e}_{(\vartheta)} = \frac{1}{r}\partial_{\vartheta}, \qquad 
   \mathbf{e}_{(\varphi)} = \sqrt{{\frac {r}{r-b}}}\partial_{\varphi}.
 \end{equation}
 Dual tetrad:
 \begin{equation}
  \boldsymbol{\theta}^{(t)} = r\sin{\vartheta}\,dt, \qquad 
  \boldsymbol{\theta}^{(r)} = \sqrt {{\frac {r}{r-b}}}\,dr, \qquad 
  \boldsymbol{\theta}^{(\vartheta)}= r\,d\vartheta, \qquad
  \boldsymbol{\theta}^{(\varphi)} = \sqrt{{\frac {r-b}{r}}}\,d\varphi.
 \end{equation}
 
 % -------------------- effective Potential -------------------- 
 \SecEffPoti\\
 With the Hamilton-Jacobi formalism, an effective potential for the radial coordinate can be calculated fulfilling $\frac{1}{2}\dot{r}^2 + \frac{1}{2}V_{\text{eff}}(r)=\frac{1}{2}C_0^2$ with
\begin{equation}
  V_{\text{eff}}(r)=K\frac{r-b}{r^3}-\kappa\frac{r-b}{r}
\end{equation}
and the constants of motion
\begin{subequations}
\begin{eqnarray}
   C_0^2 &=& \dot{\varphi}^2 \left(\frac{r-b}{r}\right)^2,  \\
   K     &=& \dot{\vartheta}^2 r^4 - \dot{t}^2 r^4 \sin^2{\vartheta}. %\\
   %-\kappa c^2 &=& -K\frac1{r^2}-\dot{r}^2 \frac{r}{r - b} -\dot{\varphi}^2\frac{r - b}{r},
\end{eqnarray}
\end{subequations}
Note that the metric is not spherically symmetric. Particles or light rays fall into one of the poles if they are not moving in the $\vartheta=\frac{\pi}{2}$ plane.

}{

}

%% ------------------------------------------------------------------------
%%    Petrov Type D ES BII
%% ------------------------------------------------------------------------
%\clearpage
%\setcounter{equation}{0}
\ifthenelse{\boolean{isARXIV}}{
% ******** Start of file PTD_BII.tex *********
%
%  Copyright (c) 2010 Felix Beslmeisl,
%                     Universitaet Stuttgart
%
\subsection{\ESCstr~BII}

%\subsection{Cylindrical Coordinates}
In cylindrical coordinates, the metric is given by the line element
\metricEq{
  ds^2 = z^2\left( dr^2 - \sinh^2r\,dt^2 \right) + \frac{z}{b-z}dz^2+\frac{b-z}{z}d\varphi^2.
}{PTD_BII_Spherical}
\PTDRangeStart
  $t\in\setR$, \qquad $\varphi\in [0,2\pi)$, \qquad $0<z<b$, \qquad $0<r$.
\PTDRangeEnd

% -------------------- Local tetrad --------------------
 \SecLocal
 \begin{equation}
   \mathbf{e}_{(t)} = \frac {1}{z\sinh{r}}\partial_t, \qquad 
   \mathbf{e}_{(r)} = \frac{1}{z}\partial_r, \qquad
   \mathbf{e}_{(\varphi)} = \sqrt{{\frac {z}{b-z}}}\partial_{\varphi}, \qquad 
   \mathbf{e}_{(z)} = \sqrt{{\frac {b-z}{z}}}\partial_{z}.
 \end{equation}
 Dual tetrad:
 \begin{equation}
  \boldsymbol{\theta}^{(t)} = z\sinh{r}\,dt, \qquad 
  \boldsymbol{\theta}^{(r)} = z\,dr, \qquad 
  \boldsymbol{\theta}^{(\varphi)}= \sqrt{{\frac {b-z}{z}}}\,d\varphi, \qquad
  \boldsymbol{\theta}^{(z)} = \sqrt{{\frac {z}{b-z}}}\,dz.
 \end{equation}

}{

}

%% ------------------------------------------------------------------------
%%    Petrov Type D ES BIII
%% ------------------------------------------------------------------------
%\clearpage
%\setcounter{equation}{0}
\ifthenelse{\boolean{isARXIV}}{
% ******** Start of file PTD_BIII.tex *********
%
%  Copyright (c) 2010 Felix Beslmeisl,
%                     Universitaet Stuttgart
%
\subsection{\ESCstr~BIII}

%\subsection{Cylindrical Coordinates}
In cylindrical coordinates, the metric is given by the line element
\metricEq{
  ds^2 = z^2\left( dr^2 - r^2 dt^2 \right) + z dz^2+\frac{1}{z}d\varphi^2.
}{PTD_BIII_Spherical}
\PTDRangeStart
  $t\in\setR$, \qquad $\varphi\in [0,2\pi)$, \qquad $0<z$, \qquad $0<r$.
\PTDRangeEnd

% -------------------- Local tetrad --------------------
 \SecLocal
 \begin{equation}
   \mathbf{e}_{(t)} = \frac{1}{zr}\partial_t, \qquad 
   \mathbf{e}_{(r)} = \frac{1}{z}\partial_r, \qquad
   \mathbf{e}_{(\varphi)} = \sqrt{z}\,\partial_{\varphi}, \qquad 
   \mathbf{e}_{(z)} = \frac{1}{\sqrt{z}}\partial_{z}.
 \end{equation}
 Dual tetrad:
 \begin{equation}
  \boldsymbol{\theta}^{(t)} = zr\,dt, \qquad 
  \boldsymbol{\theta}^{(r)} = z\,dr, \qquad 
  \boldsymbol{\theta}^{(\varphi)}= \frac{1}{\sqrt{z}}\,d\varphi, \qquad
  \boldsymbol{\theta}^{(z)} = \sqrt{z}\,dz.
 \end{equation}

}{

}

%% ------------------------------------------------------------------------
%%    Petrov Type D ES C
%% ------------------------------------------------------------------------
%\clearpage
%\setcounter{equation}{0}
\ifthenelse{\boolean{isARXIV}}{
% ******** Start of file PTD_C.tex *********
%
%  Copyright (c) 2010 Felix Beslmeisl,
%                     Universitaet Stuttgart
%
\subsection{\ESCstr~C}

%\subsection{Standard Coordinates}
The metric is given by the line element
\metricEq{
  ds^2 =\frac{1}{(x+y)^2}\left( \frac1{f(x)} dx^2 + f(x) d\varphi^2 - \frac1{f(-y)}dy^2 +f(-y)dt^2 \right)
}{PTD_C}
with $f(u) := \pm (u^3 +au +b)$.
\PTDRangeStart
   $0<x+y$,\qquad $f(-y)>0$,\qquad $0>f(x)$.
\PTDRangeEnd

% -------------------- Local tetrad --------------------
 \SecLocal
 \begin{subequations}
 \begin{alignat}{3}
  \mathbf{e}_{(t)} &= (x+y)\frac{1}{\sqrt{-y^3-ay+b}}\,\partial_t,  &\qquad
  \mathbf{e}_{(x)} &= (x+y)\sqrt{x^3+ax+b}\,\partial_x, \\
  \mathbf{e}_{(y)} &= (x+y)\sqrt{-y^3-ay+b}\,\partial_y, &\qquad
  \mathbf{e}_{(\varphi)} &= (x+y)\frac{1}{\sqrt{x^3+ax+b}}\,\partial_{\varphi},
%   &\mathbf{e}_{(t)} = \left (1+\alpha\,r\cos(\theta)\right )\left (-r\right )\sqrt {{\frac {r}{\left (-1+{
%\alpha}^{2}{r}^{2}\right )\left (-r+2\,m\right )}}}\partial_t, \\&    
%    \mathbf{e}_{(r)} = \left (1+\alpha\,r\cos(\theta)\right ){\sqrt {{\frac {\left 
%(-1+{\alpha}^{2}{r}^{2}\right )\left (-r+2\,m\right )}{r}}}}
%\partial_r, \\&
%   \mathbf{e}_{(\vartheta)} = {\left (1+\alpha\,r\cos(\theta)\right )}{\frac {\sqrt {1+2\,\alpha\,m
%\cos(\theta)}}{r}}\partial_{\vartheta}, \\&  
%   \mathbf{e}_{(\varphi)} = {\left (1+\alpha\,r\cos(\theta)\right )}{\frac {1}{r\sin(\theta)\sqrt {1+2\,%\alpha\,m\cos(\theta)}
%}}\partial_{\varphi}.
 \end{alignat}
 \end{subequations}
 Dual tetrad:
 \begin{subequations}
  \begin{alignat}{3}
    \boldsymbol{\theta}^{(t)} &= \frac1{x+y}\sqrt{-y^3-ay+b}\,dt, &\qquad
   \boldsymbol{\theta}^{(x)} &=  \frac1{x+y}\frac{1}{\sqrt{x^3+ax+b}}\,dx, \\
   \boldsymbol{\theta}^{(y)} &=   \frac1{x+y}\frac{1}{\sqrt{-y^3-ay+b}}\,dy, &\qquad
   \boldsymbol{\theta}^{(\varphi)} &=  \frac1{x+y}\sqrt{x^3+ax+b} \,d\varphi,
 \end{alignat}
 \end{subequations}
 
A coordinate change can eliminate the linear term in the polynom f generating a quadratic term instead. This brings the line element to the form 
\metricEq{
ds^2 = \frac{1}{A(x+y)^2}\left[ \frac1{f(x)} dx^2 + f(x) dp^2 - \frac1{f(-y)}dy^2 +f(-y)dq^2 \right]
}{PTD_Pravda_C}
with $f(u) := \pm (-2mAu^3 -u^2 + 1)$ given in \cite{pravda2000}.

Furthermore, coordinates can be adapted to the boost-rotation symmetry with the line element in \cite{pravda2000} from in \cite{bonnor1982}

\metricEq{
ds^2 = \frac{1}{z^2-t^2}\left[e^{\rho}r^2(z\,dt-t\,dz)^2 -e^{\lambda}(z\,dz-t\,dt)^2 \right] - e^{\lambda}\,dr^2-r^2e^{-\rho}\,d\varphi^2
}{PTD_Pravda_C_C}
with $$e^{\rho}=\frac{R_3+R+Z_3-r^2}{4\alpha^2\left(R_1+R+Z_1-r^2\right)},$$
                  $$e^{\lambda}=\frac{2\alpha^2\left[R(R+R_1+Z_1)-Z_1r^2\right] \left[R_1R_3+(R+Z_1)(R+Z_3)-(Z_1+Z_3)r^2 \right]}{R_iR_3\left[ R(R +R_3+Z_3) - Z_3r^2\right]},$$
                  $$R=\frac12\left(z^2-t^2+r^2\right),$$
                  $$R_i=\sqrt{(R + Z_i)^2 - 2Z_i r^2},$$
                  $$Z_i=z_i-z_2,$$
                  $$\alpha^2=\frac14 \frac{m^2}{A^6(z_2-z_1)^2(z3-z1)^2},$$
                  $$q=\frac1{4\alpha^2},$$
                  and $z_3<z_1<z_2$ the roots of $2A^4z^3-A^2z^2+m^2.$

% -------------------- Local tetrad --------------------
 \SecLocal \\ \\
  Case $z^2-t^2>0$:
 \begin{subequations}
 \begin{alignat}{3}
 \mathbf{e}_{(t)} &= \frac{1}{\sqrt{z^2-t^2}}\left( qze^{-\rho/2}\,\partial_t + te^{-\lambda/2}\,\partial_z,\right), &\qquad 
  \mathbf{e}_{(r)} &= e^{-\lambda/2}\,\partial_r,\\
  \mathbf{e}_{(z)} &= \frac{1}{\sqrt{z^2-t^2}}\left( qte^{-\rho/2}\,\partial_t + ze^{-\lambda/2}\,\partial_z,\right), &\qquad \mathbf{e}_{(\varphi)} &= re^{\rho/2}\,\partial_{\varphi}.  
 \end{alignat}
 \end{subequations}
   Case $z^2-t^2<0$:
 \begin{subequations}
 \begin{alignat}{3}
  \mathbf{e}_{(t)} &= \frac{1}{\sqrt{t^2-z^2}}\left( qte^{-\rho/2}\,\partial_t + ze^{-\lambda/2}\,\partial_z,\right),  &\qquad 
   \mathbf{e}_{(r)} &= e^{-\lambda/2}\,\partial_r,\\
   \mathbf{e}_{(z)} &= \frac{1}{\sqrt{t^2-z^2}}\left( qze^{-\rho/2}\,\partial_t + te^{-\lambda/2}\,\partial_z,\right), &\qquad \mathbf{e}_{(\varphi)} &= re^{\rho/2}\,\partial_{\varphi}.
 \end{alignat}
 \end{subequations}
 
 Dual tetrad: \\ \\
 Case $z^2-t^2>0$:
 \begin{subequations}
  \begin{alignat}{3}
    \boldsymbol{\theta}^{(t)} &= \sqrt{\frac{e^{\rho}}{z^2-t^2}}\frac{1}{q}\left(z\,dt + t\,dz\right), &\qquad 
     \boldsymbol{\theta}^{(r)} &=  e^{\lambda}\,dr,\\
     \boldsymbol{\theta}^{(z)} &= \sqrt{\frac{e^{\lambda}}{z^2-t^2}}\left(t\,dt +z\,dz\right), &\qquad \boldsymbol{\theta}^{(\varphi)} &= \frac1{re^{\rho}} \,d\varphi.     
  \end{alignat}
 \end{subequations}
 
 Case $z^2-t^2>0$:
 \begin{subequations}
  \begin{alignat}{3}
     \boldsymbol{\theta}^{(t)} &= \sqrt{\frac{e^{\lambda}}{t^2-z^2}}\left(t\,dt +z\,dz\right), &\qquad 
     \boldsymbol{\theta}^{(r)} &= e^{\lambda}\,dr,\\
     \boldsymbol{\theta}^{(z)} &= \sqrt{\frac{e^{\rho}}{t^2-z^2}}\frac{1}{q}\left(z\,dt + t\,dz\right), &\qquad 
      \boldsymbol{\theta}^{(\varphi)} &= \frac1{re^{\rho}} \,d\varphi.
  \end{alignat}
 \end{subequations}
 
}{

}

%% ------------------------------------------------------------------------
%%      P l a n e -  W a v e 
%% ------------------------------------------------------------------------
\clearpage
\section{Plane gravitational wave}
\setcounter{equation}{0}
\ifthenelse{\boolean{isARXIV}}{
% ******** Start of file planeWave.tex *********
%
%  Copyright (c) 2009 Thomas Mueller, Frank Grave,
%                     Universitaet Stuttgart, VISUS
%

W. Rindler described in \cite{rindler} an exact plane gravitational wave which is bounded between two planes. The metric
of the so called 'sandwich wave' with $u:=t-x$ reads 
\metricEq{
 ds^{2} = -dt^{2} + dx^{2} + p^{2}\left(u\right) dy^{2} + q^{2}\left(u\right) dz^{2}.
}{planeWave}
The functions $p\left(u\right)$ and $q\left(u\right)$ are given by 
\begin{equation}
 p\left(u\right):=\begin{cases}
                    p_0 = \mathrm{const.}&        u<-a\\
                    1-u&                          0<u\\
                    L\left(u\right)\mathrm{e}^{m\left(u\right)}&  \text{else}
                  \end{cases}
  \qquad\text{and}\qquad
 q\left(u\right):=\begin{cases}
                    q_0 = \mathrm{const.}&        u<-a\\
                    1-u&                          0<u\\
                    L\left(u\right)\mathrm{e}^{-m\left(u\right)}&  \text{else}\\
                  \end{cases},
\end{equation}
where a is the longitudinal extension of the wave. The functions $L\left(u\right)$ and $m\left(u\right)$ are 
\begin{equation}
 L\left(u\right) = 1-u+\frac{u^{3}}{a^{2}}+\frac{u^{4}}{2a^{3}}, \qquad 
 m\left(u\right) = \pm2\sqrt{3}\int\sqrt{\frac{u^{2}+au}{2a^{3}u-2au^{3}-u^{4}-2a^{3}}} \, du.
\end{equation}

%% -------------------- Christoffel symbols --------------------
\SecChristoffel
\begin{equation}
 \Cchris{t y}{y} = -\Cchris{x y}{y} = \frac{1}{p}\pdiff{p}{u}, \qquad      
 \Cchris{z z}{t} =  \Cchris{z z}{x} = q\pdiff{q}{u}, \qquad           
 \Cchris{t z}{z} = -\Cchris{x z}{z} = \frac{1}{q}\pdiff{q}{u}, \qquad  
 \Cchris{y y}{t} =  \Cchris{y y}{x} = p\pdiff{p}{u}.
\end{equation}

%% -------------------- Riemann tensor --------------------
\SecRiemann
\begin{equation}
  R_{tyty} = R_{xyxy} = -R_{tyxy} = -p\p2diff{p}{u},\qquad R_{tztz} = R_{xzxz} = -R_{tzxz} = -q\p2diff{q}{u}.
\end{equation}

%% -------------------- Local tetrad --------------------
\SecLocal
\begin{equation}
  \Clt{t} = \partial_{t}, \quad \Clt{x} = \partial_{x}, \quad 
  \Clt{y} = \frac{1}{p}\partial_{y}, \quad \Clt{z} = \frac{1}{q}\partial_{z}.
\end{equation}
Dual tetrad:
\begin{equation}
  \Cdlt{t} =  dt, \quad \Cdlt{x} =  dx, \quad 
  \Cdlt{y} =  pdy, \quad \Cdlt{z} =  qdz.
\end{equation}

}{

}

%% ------------------------------------------------------------------------
%%      R e i s s n e r  -  N o r d s t r o m
%% ------------------------------------------------------------------------
\clearpage
\section{Reissner-Nordstr{\o}m}
\setcounter{equation}{0}
\ifthenelse{\boolean{isARXIV}}{
% ******** Start of file reissnernordstrom.tex *********
%
%  Copyright (c) 2009 Thomas Mueller,
%                     Universitaet Stuttgart, VISUS
%

The Reissner-Nordstr{\o}m black hole in spherical coordinates $\left\{t\in\setR,r\in\setR^{+},\vartheta\in(0,\pi),\varphi\in[0,2\pi)\right\}$ is defined by the metric\cite{mtw}
\metricEq{
  ds^2 = -A_{\text{RN}}c^2dt^2 + A_{\text{RN}}^{-1}dr^2 + r^2\left(d\vartheta^2+\sin^2\vartheta\,d\varphi^2\right),
}{reissner}
where
\begin{equation}
  A_{\text{RN}} = 1-\frac{r_s}{r}+\frac{\rho Q^2}{r^2}
\end{equation}
with $r_s=2GM/c^2$, the charge $Q$, and $\rho=G/(\epsilon_0c^4)\approx 9.33\cdot 10^{-34}$. As in the Schwarzschild case, there is a true curvature singularity at $r=0$. However, for $Q^2<r_s^2/(4\rho)$ there are also two critical points at 
\begin{equation}
  r=\frac{r_s}{2}\pm\frac{r_s}{2}\sqrt{1-\frac{4\rho Q^2}{r_s^2}}.
\end{equation}

%% -------------------- Christoffel symbols --------------------
\SecChristoffel
\begin{subequations}
\begin{alignat}{5}
  \Gamma_{tt}^r &= \frac{A_{\text{RN}}c^2(r_sr-2\rho Q^2)}{2r^3}, &\qquad \Gamma_{tr}^t &= \frac{r_sr-2\rho Q^2}{2r^3A_{\text{RN}}}, &\qquad \Gamma_{rr}^r &= -\frac{r_sr-2\rho Q^2}{2r^3A_{\text{RN}}},\\
  \Gamma_{r\vartheta}^{\vartheta} &= \frac{1}{r}, & \Gamma_{r\varphi}^{\varphi} &= \frac{1}{r}, & \Gamma_{\vartheta\vartheta}^r &= -rA_{\text{RN}},\\
  \Gamma_{\vartheta\varphi}^{\varphi} &= \cot\vartheta, & \Gamma_{\varphi\varphi}^r &= -rA_{\text{RN}}\sin^2\vartheta, & \Gamma_{\varphi\varphi}^{\vartheta} &= -\sin\vartheta\cos\vartheta.
\end{alignat}
\end{subequations}

%% -------------------- Riemann tensor --------------------
\SecRiemann
\begin{subequations}
\begin{alignat}{3}
  R_{trtr} &= -\frac{c^2(r_sr-3\rho Q^2)}{r^4}, &\qquad R_{t\vartheta t\vartheta} &= \frac{A_{\text{RN}}c^2(rs_r-2\rho Q^2)}{2r^2},\\
   R_{t\varphi t\varphi} &= \frac{A_{\text{RN}}c^2(r_sr-2\rho Q^2)\sin^2\vartheta}{2r^2}, & R_{r\vartheta r\vartheta} &= -\frac{r_sr-2\rho Q^2}{2r^2A_{\text{RN}}},\\
  R_{r\varphi r\varphi} &= -\frac{(r_sr-2\rho Q^2)\sin^2\vartheta}{2r^2A_{\text{RN}}}, & R_{\vartheta\varphi\vartheta\varphi} &= (r_sr-\rho Q^2)\sin^2\vartheta.
\end{alignat}
\end{subequations}

%% -------------------- Ricci tensor --------------------
\SecRicci
\begin{equation}
  R_{tt} = \frac{c^2\rho Q^2A_{\text{RN}}}{r^4},\qquad R_{rr} = -\frac{\rho Q^2}{r^4A_{\text{RN}}}, \qquad R_{\vartheta\vartheta} = \frac{\rho Q^2}{r^2},\qquad R_{\varphi\varphi} = \frac{\rho Q^2\sin^2\vartheta}{r^2}.
\end{equation}

\noindent While the Ricci scalar vanishes identically, the Kretschmann scalar reads
\begin{equation}
  \mathcal{K} = 4\frac{3r_s^2r^2-12r_sr\rho Q^2 + 14\rho^2Q^4}{r^8}.
\end{equation}

%% -------------------- Weyl tensor --------------------
\SecWeyl
\begin{subequations}
\begin{alignat}{3}
  C_{trtr} &= -\frac{c^2(r_sr-2\rho Q^2)}{r^4}, &\qquad C_{t\vartheta t\vartheta} &= -\frac{A_{\text{RN}}c^2(rs_r-2\rho Q^2)}{2r^2},\\
   C_{t\varphi t\varphi} &= \frac{A_{\text{RN}}c^2(r_sr-2\rho Q^2)\sin^2\vartheta}{2r^2}, & C_{r\vartheta r\vartheta} &= -\frac{r_sr-2\rho Q^2}{2r^2A_{\text{RN}}},\\
  C_{r\varphi r\varphi} &= -\frac{(r_sr-2\rho Q^2)\sin^2\vartheta}{2r^2A_{\text{RN}}}, & C_{\vartheta\varphi\vartheta\varphi} &= (r_sr-2\rho Q^2)\sin^2\vartheta.
\end{alignat}
\end{subequations}

%% -------------------- Local tetrad --------------------
\SecLocal
\begin{equation}
  \mathbf{e}_{(t)} = \frac{1}{c\sqrt{A_{\text{RN}}}}\partial_t, \qquad \mathbf{e}_{(r)} = \sqrt{A_{\text{RN}}}\partial_r,\qquad \mathbf{e}_{(\vartheta)} = \frac{1}{r}\partial_{\vartheta}, \qquad \mathbf{e}_{(\varphi)} = \frac{1}{r\sin\vartheta}\partial_{\varphi}.
\end{equation}
Dual tetrad:
\begin{equation}
 \boldsymbol{\theta}^{(t)} = c\sqrt{A_{\text{RN}}}\,dt,\qquad \boldsymbol{\theta}^{(r)} = \frac{dr}{\sqrt{A_{\text{RN}}}},\qquad \boldsymbol{\theta}^{(\vartheta)}=r\,d\vartheta,\qquad \boldsymbol{\theta}^{(\varphi)} = r\sin\vartheta\,d\varphi.
\end{equation}

%% -------------------- Ricci rotation coefficients --------------------
\SecRicRotCoef
\begin{equation}
  \boldsymbol{\gamma}_{(r)(t)(t)} = \frac{rr_s-2\rho Q^2}{2r^3\sqrt{A_{\text{RN}}}},\quad \boldsymbol{\gamma}_{(\vartheta)(r)(\vartheta)} = \boldsymbol{\gamma}_{(\varphi)(r)(\varphi)} = \frac{\sqrt{A_{\text{RN}}}}{r},\quad \boldsymbol{\gamma}_{(\varphi)(\vartheta)(\varphi)} = \frac{\cot\vartheta}{r}.
\end{equation}
The contractions of the Ricci rotation coefficients read
\begin{equation}
  \boldsymbol{\gamma}_{(r)} = \frac{4r^2-3rr_s+2\rho Q^2}{2r^3\sqrt{A_{\text{RN}}}},\qquad \boldsymbol{\gamma}_{(\vartheta)} = \frac{\cot\vartheta}{r}.
\end{equation}

%% -------------------- Riemann tensor LT--------------------
\SecRiemannLT
\begin{subequations}
 \begin{align}
  R_{(t)(r)(t)(r)} &= -\frac{r_sr-3\rho Q^2}{r^4},\qquad R_{(\vartheta)(\varphi)(\vartheta)(\varphi)} = \frac{r_sr-\rho Q^2}{r^4},\\
  R_{(t)(\vartheta)(t)(\vartheta)} &= R_{(t)(\varphi)(t)(\varphi)} = -R_{(r)(\vartheta)(r)(\vartheta)} = -R_{(r)(\varphi)(r)(\varphi)} = \frac{r_sr-2\rho Q^2}{2r^4}.
 \end{align}
\end{subequations}

%% -------------------- Ricci tensor LT--------------------
\SecRicciLT
\begin{equation}
  R_{(t)(t)} = -R_{(r)(r)} = R_{(\vartheta)(\vartheta)} = R_{(\varphi)(\varphi)} = \frac{\rho Q^2}{r^4}.
\end{equation}

%% -------------------- Weyl tensor LT--------------------
\SecWeylLT
\begin{subequations}
 \begin{align}
  C_{(t)(r)(t)(r)} &= -C_{(\vartheta)(\varphi)(\vartheta)(\varphi)} = -\frac{r_sr-2\rho Q^2}{r^4},\\
  C_{(t)(\vartheta)(t)(\vartheta)} &= C_{(t)(\varphi)(t)(\varphi)} = -C_{(r)(\vartheta)(r)(\vartheta)} = -C_{(r)(\varphi)(r)(\varphi)} = \frac{r_sr-2\rho Q^2}{2r^4}.
 \end{align}
\end{subequations}

%% -------------------- Embedding --------------------
\SecEmbedding

The embedding function follows from the numerical integration of
\begin{equation}
  \frac{dz}{dr} = \sqrt{\frac{1}{1-r_s/r+\rho Q^2/r^2}-1}.
\end{equation}
\vspace*{0.5cm}

%% -------------------- Euler Lagrange --------------------
\SecEulLag

The Euler-Lagrangian formalism, Sec.~\ref{subsec:EL}, for geodesics in the $\vartheta=\pi/2$ hyperplane yields
\begin{equation}
  \frac{1}{2}\dot{r}^2+V_{\text{eff}} = \frac{1}{2}\frac{k^2}{c^2},\qquad V_{\text{eff}} = \frac{1}{2}\left(1-\frac{r_s}{r}+\frac{\rho Q^2}{r^2}\right)\left(\frac{h^2}{r^2}-\kappa c^2\right)
\end{equation}
with constants of motion $k=A_{\text{RN}}c^2\dot{t}$ and $h=r^2\dot{\varphi}$. For null geodesics, $\kappa=0$, there are two extremal points
\begin{equation}
  r_{\pm} = \frac{3}{4}r_s\left(1\pm \sqrt{1-\frac{32\rho Q^2}{9r_s^2}}\right),
\end{equation}
where $r_{+}$ is a maximum and $r_{-}$ a minimum.

%% -------------------- Further reading --------------------
\FurtherReading

Eiroa\cite{eiroa2002}

}{

}

%% ------------------------------------------------------------------------
%%    d e S i t t e r
%% ------------------------------------------------------------------------
\clearpage
\section{de Sitter spacetime}
\setcounter{equation}{0}
\ifthenelse{\boolean{isARXIV}}{
% ******** Start of file deSitter.tex *********
%
%  Copyright (c) 2009 Thomas Mueller, Frank Grave,
%                     Universitaet Stuttgart, VISUS
%

The de Sitter spacetime with $\Lambda>0$ is a solution of the Einstein field equations with constant curvature. A detailed discussion can be found for example in Hawking and Ellis\cite{hawking1999}. Here, we use the coordinate transformations given by Bi\v{c}\'ak\cite{bicak2001}.

%% ------------------------------------------------------
%%    Standard coordinates
%% ------------------------------------------------------
\subsection{Standard coordinates}
The de Sitter metric in standard coordinates $\left\{\tau\in\setR,\chi\in [-\pi,\pi],\vartheta\in(0,\pi),\varphi\in [0,2\pi)\right\}$ reads
\metricEq{
  ds^2 = -d\tau^2 + \alpha^2\cosh^2\frac{\tau}{\alpha}\left[d\chi^2+\sin^2\chi\left(d\vartheta^2 + \sin^2\vartheta\,d\varphi^2\right)\right],
}{deSitterStandard}
where $\alpha^2=3/\Lambda$.

%% -------------------- Christoffel symbols --------------------
\SecChristoffel
\begin{subequations}
\begin{alignat}{5}
  \Cchris{\tau\chi}{\chi} &= \frac{1}{\alpha}\tanh\frac{\tau}{\alpha}, &\,\Cchris{\tau\vartheta}{\vartheta} &= \frac{1}{\alpha}\tanh\frac{\tau}{\alpha}, &\,\Cchris{\tau\varphi}{\varphi} &= \frac{1}{\alpha}\tanh\frac{\tau}{\alpha}, \\
  \Cchris{\chi\chi}{\tau} &= \alpha\sinh\frac{\tau}{\alpha}\cosh\frac{\tau}{\alpha}, &\,\Cchris{\chi\vartheta}{\vartheta} &= \cot\chi, &\, \Cchris{\chi\varphi}{\varphi} &= \cot\chi, \\
  \Cchris{\vartheta\vartheta}{\tau} &= \alpha\sin^2\chi\sinh\frac{\tau}{\alpha}\cosh\frac{\tau}{\alpha}, & \Cchris{\vartheta\vartheta}{\chi} &= -\sin\chi\cos\chi, & \Cchris{\vartheta\varphi}{\varphi} &= \cot\vartheta,\\
  \Cchris{\varphi\varphi}{\tau} &= \alpha\sin^2\chi\sin^2\vartheta\sinh\frac{\tau}{\alpha}\cosh\frac{\tau}{\alpha}, & \Cchris{\varphi\varphi}{\chi} &= -\sin^2\vartheta\sin\chi\cos\chi, & \Cchris{\varphi\varphi}{\vartheta} &= -\sin\vartheta\cos\vartheta.
\end{alignat}
\end{subequations}

%% -------------------- Riemann tensor --------------------
\SecRiemann
\begin{subequations}
\begin{alignat}{3}
  R_{\tau\chi\tau\chi} &= -\cosh^2\frac{\tau}{\alpha}, &\qquad R_{\tau\vartheta\tau\vartheta} &= -\cosh^2\frac{\tau}{\alpha}\sin^2\chi,\\ 
  R_{\tau\varphi\tau\varphi} &= -\cosh^2\frac{\tau}{\alpha}\sin^2\chi\sin^2\vartheta, & R_{\chi\vartheta\chi\vartheta} &= \alpha^2\left(1+\sinh^2\frac{\tau}{\alpha}\right)^2\sin^2\chi,\\
  R_{\chi\varphi\chi\varphi} &= \alpha^2\left(1+\sinh^2\frac{\tau}{\alpha}\right)^2\sin^2\chi\sin^2\vartheta, & R_{\vartheta\varphi\vartheta\varphi} &= \alpha^2\left(1+\sinh^2\frac{\tau}{\alpha}\right)^2\sin^4\chi\sin^2\vartheta.
\end{alignat}
\end{subequations}

%% -------------------- Ricci tensor --------------------
\SecRicci
\begin{equation}
  R_{\tau\tau} = -\frac{3}{\alpha^2},\quad R_{\chi\chi} = 3\cosh^2\frac{\tau}{\alpha},\quad R_{\vartheta\vartheta} = 3\cosh^2\frac{\tau}{\alpha}\sin^2\chi,\quad R_{\varphi\varphi} = 3\cosh^2\frac{\tau}{\alpha}\sin^2\chi\sin^2\vartheta.
\end{equation}

\noindent {\bf Ricci and Kretschmann scalars:}
\begin{equation}
 \mathcal{R} = \frac{12}{\alpha^2},\qquad \mathcal{K} = \frac{24}{\alpha^4}.
\end{equation}

%% -------------------- Local tetrad --------------------
\SecLocal
\begin{equation}
  \Clt{\tau} = \partial_{\tau}, \quad \Clt{\chi} = \frac{1}{\alpha\cosh\frac{\tau}{\alpha}}\partial_{\chi},\quad \Clt{\vartheta} = \frac{1}{\alpha\cosh\frac{\tau}{\alpha}\sin\chi}\partial_{\vartheta}, \quad \Clt{\varphi} = \frac{1}{\alpha\cosh\frac{\tau}{\alpha}\sin\chi\sin\vartheta}\partial_{\varphi}.
\end{equation}
Dual tetrad:
\begin{equation}
 \Cdlt{\tau} = d\tau,\quad \Cdlt{\chi} = \alpha\cosh\frac{\tau}{\alpha}d\chi,\quad \Cdlt{\vartheta}=\alpha\cosh\frac{\tau}{\alpha}\sin\chi\,d\vartheta,\quad \Cdlt{\varphi} = \alpha\cosh\frac{\tau}{\alpha}\sin\chi\sin\vartheta\,d\varphi.
\end{equation}

%% ------------------------------------------------------
%%    Conformally Einstein coordinates
%% ------------------------------------------------------
\subsection{Conformally Einstein coordinates}
In conformally Einstein coordinates $\left\{\eta\in [0,\pi],\chi\in [-\pi,\pi],\vartheta\in [0,\pi],\varphi\in [0,2\pi)\right\}$, the de Sitter metric reads
\metricEq{
  ds^2 = \frac{\alpha^2}{\sin^2\eta}\left[-d\eta^2+d\chi^2+\sin^2\chi\,\left(d\vartheta^2 + \sin^2\vartheta\,d\varphi^2\right)\right].
}{deSitterConfEinstein}
It follows from the standard form (\ref{eqM:deSitterStandard}) by the transformation
\begin{equation}
  \eta = 2\arctan\left(e^{\tau/\alpha}\right).
\end{equation}

%% ------------------------------------------------------
%%    Conformally flat coordinates
%% ------------------------------------------------------
\subsection{Conformally flat coordinates}
Conformally flat coordinates $\left\{T\in\setR,r\in\setR,\vartheta\in (0,\pi),\varphi\in [0,2\pi)\right\}$ follow from conformally Einstein coordinates by means of the transformations
\begin{equation}
 T = \frac{\alpha\sin\eta}{\cos\chi+\cos\eta},\quad r = \frac{\alpha\sin\chi}{\cos\chi+\cos\eta},\quad\text{or}\quad \eta = \arctan\frac{2T\alpha}{\alpha^2-T^2+r^2},\quad \chi = \arctan\frac{2r\alpha}{\alpha^2+T^2-r^2}.
\end{equation}
For the transformation $(T,R)\rightarrow (\eta,\chi)$, we have to take care of the coordinate domains. In that case, if $\kappa^2-T^2+r^2<0$, we have to map $\eta\rightarrow \eta+\pi$. On the other hand, if $\kappa^2+T^2-r^2<0$, we have to consider the sign of $r$. If $r>0$, then $\chi\rightarrow \chi+\pi$, otherwise $\chi\rightarrow \chi-\pi$.

The resulting metric reads
\metricEq{
  ds^2 = \frac{\alpha^2}{T^2}\left[-dT^2+dr^2+r^2\left(d\vartheta^2 + \sin^2\vartheta\,d\varphi^2\right)\right].
}{deSitterConfFlat}
Note that we identify points $(r<0,\vartheta,\varphi)$ with $(r>0,\pi-\vartheta,\varphi-\pi)$.

%% -------------------- Christoffel symbols --------------------
\SecChristoffel
\begin{subequations}
  \begin{align}
     \Cchris{TT}{T} &= \Cchris{Tr}{r} = \Cchris{T\vartheta}{\vartheta} = \Cchris{T\varphi}{\varphi} = \Cchris{rr}{T} = -\frac{1}{T}, \quad \Cchris{r\vartheta}{\vartheta} = \Cchris{r\varphi}{\varphi} = \frac{1}{r}, \quad \Cchris{\vartheta\vartheta}{T} = -\frac{r^2}{T},\quad \Cchris{\vartheta\vartheta}{r} = -r,\\
     \Cchris{\vartheta\varphi}{\varphi} &= \cot\vartheta, \quad \Cchris{\varphi\varphi}{T} = -\frac{r^2\sin^2\vartheta}{T},\quad \Cchris{\varphi\varphi}{r} = -r\sin^2\vartheta,\quad \Cchris{\varphi\varphi}{\vartheta} = -\sin\vartheta\cos\vartheta.
  \end{align}
\end{subequations}

%% -------------------- Riemann tensor --------------------
\SecRiemann
\begin{subequations}
 \begin{alignat}{5}
   R_{TrTr} &= -\frac{\alpha^2}{T^4}, &\qquad R_{T\vartheta T\vartheta} &= -\frac{\alpha^2r^2}{T^4}, &\qquad R_{T\varphi T\varphi} &= -\frac{\alpha^2r^2\sin^2\vartheta}{T^4},\\
  R_{r\vartheta r\vartheta} &= \frac{\alpha^2r^2}{T^4}, & R_{r\varphi r\varphi} &= \frac{\alpha^2r^2\sin^2\vartheta}{T^4}, & R_{\vartheta\varphi\vartheta\varphi} &= \frac{\alpha^2r^4\sin^2\vartheta}{T^4}.
 \end{alignat}
\end{subequations}

%% -------------------- Ricci tensor --------------------
\SecRicci
\begin{equation}
  R_{TT} = -\frac{3}{T^2}, \qquad R_{rr} = \frac{3}{T^2}, \qquad R_{\vartheta\vartheta} = \frac{3r^2}{T^2},\qquad R_{\varphi\varphi} = \frac{3r^2\sin^2\vartheta}{T^2}.
\end{equation}

\noindent The {\sl Ricci and Kretschmann scalar read:}
\begin{equation}
  \mathcal{R} = \frac{12}{\alpha^2},\qquad \mathcal{K}=\frac{24}{\alpha^4}.
\end{equation}

%% -------------------- Local tetrad --------------------
\SecLocal
\begin{equation}
  \Clt{T} = \frac{T}{\alpha}\partial_T,\qquad \Clt{r} = \frac{T}{\alpha}\partial_r,\qquad \Clt{\vartheta} = \frac{T}{\alpha r}\partial_{\vartheta},\qquad \Clt{\varphi} = \frac{T}{\alpha r\sin\vartheta}\partial_{\varphi}.
\end{equation}

%% ------------------------------------------------------
%%    Static coordinates
%% ------------------------------------------------------
\subsection{Static coordinates}
The de Sitter metric in static spherical coordinates  $\left\{t\in\setR,r\in\setR^{+},\vartheta\in (0,\pi),\varphi\in [0,2\pi)\right\}$ reads
\metricEq{
  ds^2 = -\left(1 - \frac{\Lambda}{3}r^2\right)c^2 dt^2 + \left(1 - \frac{\Lambda}{3}r^2\right)^{-1}dr^2 + r^2\left(d\vartheta^2 + \sin^2\vartheta\,d\varphi^2\right).
}{deSitterStatic}
%If $\Lambda>0$, the metric is called \textit{de Sitter}, while for $\Lambda<0$ it is called \textit{anti-de Sitter}.
It follows from the conformally Einstein form (\ref{eqM:deSitterConfEinstein}) by the transformations
\begin{equation}
  t=\frac{\alpha}{2}\ln\frac{\cos\chi-\cos\eta}{\cos\chi+\cos\eta},\quad r=\alpha\frac{\sin\chi}{\sin\eta}.
\end{equation}

%% -------------------- Christoffel symbols --------------------
\SecChristoffel
\begin{subequations}
\begin{alignat}{5}
\Gamma_{tt}^{r} &= \frac{(\Lambda r^2 - 3)}{9}c^2\Lambda r, &\qquad\Gamma_{tr}^{t} &= \dfrac{\Lambda r}{\Lambda r^2 - 3}, &\qquad\Gamma_{rr}^{r} &= \dfrac{\Lambda r}{3-\Lambda r^2}, \\
\Gamma_{r\vartheta}^{\vartheta} &= \dfrac{1}{r}, &\qquad\Gamma_{r\phi}^{\phi} &= \dfrac{1}{r}, &\qquad\Gamma_{\vartheta\vartheta}^{r} &= \dfrac{(\Lambda r^2 - 3)r}{3}, \\
\Gamma_{\vartheta\phi}^{\phi} &= \cot(\vartheta), &\qquad\Gamma_{\phi\phi}^{r} &= \dfrac{\Lambda r^2 - 3}{3}r\sin^2(\vartheta), &\qquad\Gamma_{\phi\phi}^{\vartheta} &= -\sin(\vartheta)\cos(\vartheta).
\end{alignat}
\end{subequations}

%% -------------------- Riemann tensor --------------------
\SecRiemann
\begin{subequations}
\begin{align}
  R_{trtr} &= -\frac{\Lambda}{3}c^2, &\quad R_{t\vartheta t\vartheta} &= -\frac{3-\Lambda r^2}{9}c^2\Lambda r^2, &\quad R_{t\varphi t\varphi} &= -\frac{3-\Lambda r^2}{9}c^2\Lambda r^2\sin(\vartheta)^2,\\
  R_{r\vartheta r\vartheta} &= \frac{\Lambda r^2}{-\Lambda r^2 + 3}, &\quad R_{r\varphi r\varphi} &= \frac{\Lambda r^2\sin(\theta)^2}{-\Lambda r^2 + 3}, & R_{\vartheta\varphi\vartheta\varphi} &= \frac{r^4\sin^2(\theta)\Lambda}{3} .
\end{align}
\end{subequations}

%% -------------------- Ricci tensor --------------------
\SecRicci
\begin{equation}
  R_{tt} = \frac{\Lambda r^2 - 3}{3}c^2\Lambda,\qquad R_{rr} = \frac{3\Lambda}{3-\Lambda r^2}, \qquad R_{\vartheta\vartheta} = \Lambda r^2,\qquad R_{\varphi\varphi} = r^2\sin^2(\vartheta)\Lambda.
\end{equation}

\noindent The {\sl Ricci scalar and Kretschmann scalar read:}
\begin{equation}
  \mathcal{R} = 4\Lambda,\qquad \mathcal{K}=\frac{8}{3}\Lambda^2.
\end{equation}

%% -------------------- Local tetrad --------------------
\SecLocal
\begin{equation}
  \mathbf{e}_{(t)} = \sqrt{\frac{3}{3-\Lambda r^2}}\frac{\partial_t}{c}, \qquad \mathbf{e}_{(r)} = \sqrt{1-\frac{\Lambda r^2}{3}}\partial_{r}, \qquad \mathbf{e}_{(\vartheta)} = \frac{1}{r}\partial_{\vartheta}, \qquad \mathbf{e}_{(\varphi)} = \frac{1}{r\sin(\vartheta)}\partial_{\varphi}.
\end{equation}

%% -------------------- Ricci rotation coefficients --------------------
\SecRicRotCoef
\begin{equation}
  \gamma_{(t)(r)(t)} = -\frac{\Lambda r}{\sqrt{9-3\Lambda r^2}},\qquad \gamma_{(\vartheta)(r)(\vartheta)} = \gamma_{(\varphi)(r)(\varphi)} = \frac{\sqrt{9-3\Lambda r^2}}{3r}, \qquad \gamma_{(\varphi)(\vartheta)(\varphi)} = \frac{\cot\vartheta}{r}.
\end{equation}
The contractions of the Ricci rotation coefficients read
\begin{equation}
  \gamma_{(r)} = \frac{\sqrt{9 - 3\Lambda r^2}(\Lambda r^2 - 2)}{(\Lambda r^2 - 3)r},\qquad \gamma_{(\vartheta)} = \frac{\cot\vartheta}{r}.
\end{equation}

%% -------------------- Riemann tensor LT--------------------
\SecRiemannLT
\begin{equation} 
  -R_{(t)(r)(t)(r)} = -R_{(t)(\vartheta)(t)(\vartheta)} = -R_{(t)(\varphi)(t)(\varphi)} = R_{(r)(\vartheta)(r)(\vartheta)} = R_{(r)(\varphi)(r)(\varphi)} = R_{(\vartheta)(\varphi)(\vartheta)(\varphi)} = \frac{1}{3}\Lambda .
\end{equation}

%% -------------------- Ricci tensor LT--------------------
\SecRicciLT
\begin{equation} 
  -R_{(t)(t)} = R_{(r)(r)} = R_{(\vartheta)(\vartheta)} = R_{(\varphi)(\varphi)} = \Lambda .
\end{equation}

%% --------------------------------------
%%  lemaitre-robinson form
%% --------------------------------------
\subsection{Lema\^{\i}tre-Robertson form}
The de Sitter universe in the Lema\^{\i}tre-Robertson form reads
\metricEq{
  ds^2 = -c^2dt^2 + e^{2Ht}\left[dr^2 + r^2\left(d\vartheta^2 + \sin^2\vartheta\,d\varphi^2\right)\right],
}{deSitterRobLemaitre}
with Hubble's Parameter $H=\sqrt{\frac{\Lambda c^2}{3}}=\frac{c}{\alpha}$, which is assumed here to be time-independent. \\
%The transformation between the two representations of the metric is given by
%\begin{subequations}
%\begin{alignat}{3}
%\tilde{r} &= \frac{r}{\sqrt{1-H^2r^2}}e^{-Ht}, &\qquad \tilde{t} &= t + \frac{1}{2H}\ln\left(1 - H^2r^2\right), \\
%d\tilde{r} &= \frac{He^{-Ht}}{\sqrt{1-H^2r^2}}\left(\frac{dr}{1-H^2r^2} - rdt\right), &\qquad d\tilde{t} &= dt - \frac{Hr}{1 - H^2r^2}dr
%\end{alignat}
%\end{subequations}
%with \hspace{1mm}$\tilde{ }$\hspace{1mm} marking the coordinates in the Lema\^{\i}tre-Robinson form.
This a special case of the first and second form of the Friedman-Robertson-Walker metric defined in Eqs. (\ref{eqM:FRW1}) and (\ref{eqM:FRW2}) with $R(t) = e^{Ht}$ 
and $k=0$. \\
%% -------------------- Christoffel symbols --------------------
\SecChristoffel
\begin{subequations}
\begin{alignat}{5}
\Gamma_{tr}^r &= H, &\qquad \Gamma_{t\vartheta}^\vartheta &= H, &\qquad \Gamma_{t\varphi}^\varphi &= H, \\
\Gamma_{rr}^t &= \frac{e^{2Ht}H}{c^2}, &\qquad \Gamma_{r\vartheta}^\vartheta &= \frac{1}{r}, &\qquad \Gamma_{r\varphi}^\varphi &= \frac{1}{r}, \\
\Gamma_{\vartheta\vartheta}^t &= \frac{e^{2Ht}r^2 H}{c^2}, &\qquad \Gamma_{\vartheta\vartheta}^r &= -r, &\qquad \Gamma_{\vartheta\varphi}^\varphi &= \cot(\vartheta), \\
\Gamma_{\varphi\varphi}^t &= \frac{e^{2Ht}r^2\sin^2(\theta)H}{c^2}, &\qquad \Gamma_{\varphi\varphi}^r &= -r\sin(\vartheta)^2, &\qquad \Gamma_{\varphi\varphi}^\vartheta &= -\sin(\vartheta)\cos(\vartheta).
\end{alignat}
\end{subequations}

%% -------------------- Riemann tensor --------------------
\SecRiemann
\begin{subequations}
\begin{alignat}{3}
  R_{trtr} &= -e^{2Ht}H^2, &\qquad R_{t\vartheta t\vartheta} &= -e^{2Ht}r^2H^2, \\ 
  R_{t\varphi t\varphi} &= -e^{2Ht}r^2\sin^2(\vartheta)H^2, &\qquad R_{r\vartheta r\vartheta} &= \frac{e^{4Ht}r^2H^2}{c^2}, \\
  R_{r\varphi r\varphi} &= \frac{e^{4Ht}r^2\sin^2(\vartheta)H^2}{c^2}, &\qquad R_{\vartheta\varphi\vartheta\varphi} &= \frac{e^{4Ht}r^4\sin^2(\vartheta)H^2}{c^2} .
\end{alignat}
\end{subequations}

%% -------------------- Ricci tensor --------------------
\SecRicci
\begin{equation}
  R_{tt} = -3H^2,\qquad R_{rr} = 3\frac{e^{2Ht}H^2}{c^2}, \qquad R_{\vartheta\vartheta} = 3\frac{e^{2Ht}r^2H^2}{c^2},\qquad R_{\varphi\varphi} = 3\frac{e^{2Ht}r^2\sin^2(\vartheta)H^2}{c^2}.
\end{equation}

\noindent {\bf Ricci and Kretschmann scalars:}
\begin{equation}
  \mathcal{R} = \frac{12H^2}{c^2},\qquad \mathcal{K}=\frac{24H^4}{c^4}.
\end{equation}

%% -------------------- Local tetrad --------------------
\SecLocal
\begin{equation}
  \mathbf{e}_{(t)} = \frac{1}{c}\partial_{t}, \qquad \mathbf{e}_{(r)} = e^{-Ht}\partial_{r}, \qquad \mathbf{e}_{(\vartheta)} = \frac{e^{-Ht}}{r}\partial_{\vartheta}, \qquad \mathbf{e}_{(\varphi)} = \frac{e^{-Ht}}{r\sin{\vartheta}}\partial_{\varphi}.
\end{equation}

%% -------------------- Ricci rotation coefficients --------------------
\SecRicRotCoef
\begin{subequations}
\begin{align}
  \gamma_{(r)(t)(r)} &= \gamma_{(\vartheta)(t)(\vartheta)} = \gamma_{(\varphi)(t)(\varphi)} = \frac{H}{c} \\
  \gamma_{(\vartheta)(r)(\vartheta)} &= \gamma_{(\varphi)(r)(\varphi)} = \frac{1}{e^{Ht}r}, \quad\gamma_{(\varphi)(\vartheta)(\varphi)} = \frac{\cot(\theta)}{e^{Ht}r} .
\end{align}
\end{subequations}
The contractions of the Ricci rotation coefficients read
\begin{equation}
  \gamma_{(t)} = 3\frac{H}{c}, \qquad\gamma_{(r)} = \frac{2}{e^{Ht}r}, \gamma_{(\vartheta)} = \frac{\cot(\theta)}{e^{Ht}r}.
\end{equation}

%% -------------------- Riemann tensor LT--------------------
\SecRiemannLT
\begin{subequations}
\begin{align}
   R_{(t)(r)(t)(r)} &= R_{(t)(\vartheta)(t)(\vartheta)} = R_{(t)(\varphi)(t)(\varphi)} = -\frac{H^2}{c^2} \\
   R_{(r)(\vartheta)(r)(\vartheta)} &= R_{(r)(\varphi)(r)(\varphi)} = R_{(\vartheta)(\varphi)(\vartheta)(\varphi)} = \frac{H^2}{c^2}.
\end{align}
\end{subequations}

%% -------------------- Ricci tensor LT--------------------
\SecRicciLT
\begin{equation} 
  -R_{(t)(t)} = R_{(r)(r)} = R_{(\vartheta)(\vartheta)} = R_{(\varphi)(\varphi)} = 3\frac{H^2}{c^2}.
\end{equation}

%% ------------------------------------------------------
%%    Cartesian coordinates
%% ------------------------------------------------------
\subsection{Cartesian coordinates}
The de Sitter universe in Lema\^{\i}tre-Robertson form can also be expressed in Cartesian coordinates:
\metricEq{
  ds^2 = -c^2dt^2 + e^{2Ht}\left[dx^2 + dy^2 + dz^2\right].
}{deSitterRobLemaitreCart}

%% -------------------- Christoffel symbols --------------------
\SecChristoffel
\begin{subequations}
\begin{alignat}{5}
\Gamma_{tx}^x &= H, &\qquad \Gamma_{ty}^y &= H, &\qquad \Gamma_{tz}^z &= H, \\
\Gamma_{xx}^t &= \frac{e^{2Ht}H}{c^2}, &\qquad \Gamma_{yy}^t &= \frac{e^{2Ht}H}{c^2}, &\qquad \Gamma_{zz}^t &= \frac{e^{2Ht}H}{c^2}. \\
\end{alignat}
\end{subequations}

Partial derivatives
\begin{equation}
  \Gamma_{xx,t}^t = \Gamma_{yy,t}^t = \Gamma_{zz,t}^t = \frac{2H^2e^{2Ht}}{c^2}.
\end{equation}

%% -------------------- Riemann tensor --------------------
\SecRiemann
\begin{equation}
  R_{txtx} = R_{txtx} = R_{tztz} = -e^{2Ht}H^2, \qquad R_{xyxy} = R_{xzxz} = R_{yzyz} = \frac{e^{4Ht}H^2}{c^2}.
\end{equation}

%% -------------------- Ricci tensor --------------------
\SecRicci
\begin{equation}
  R_{tt} = -3H^2,\qquad R_{xx} = R_{yy} = R_{zz} = 3\frac{e^{2Ht}H^2}{c^2}.
\end{equation}

\noindent The Ricci and Kretschmann scalar read:
\begin{equation}
  \mathcal{R} = 12\frac{H^2}{c^2},\qquad \mathcal{K}=24\frac{H^4}{c^4}.
\end{equation}

%% -------------------- Local tetrad --------------------
\SecLocal
\begin{equation}
  \mathbf{e}_{(t)} = \frac{1}{c}\partial_{t}, \qquad \mathbf{e}_{(x)} = e^{-Ht}\partial_{x}, \qquad \mathbf{e}_{(y)} = e^{-Ht}\partial_{y}, \qquad \mathbf{e}_{(z)} = e^{-Ht}\partial_{z}.
\end{equation}

%% -------------------- Ricci rotation coefficients --------------------
\SecRicRotCoef
\begin{equation}
  \gamma_{(x)(t)(x)} = \gamma_{(y)(t)(y)} = \gamma_{(z)(t)(z)} = \frac{H}{c}.
\end{equation}
The only non-vanishing contraction of the Ricci rotation coefficients read
\begin{equation}
  \gamma_{(t)} = 3\frac{H}{c}.
\end{equation}

%% -------------------- Riemann tensor LT--------------------
\SecRiemannLT
\begin{subequations}
\begin{align}
   R_{(t)(x)(t)(x)} &= R_{(t)(y)(t)(y)} = R_{(t)(z)(t)(z)} = -\frac{H^2}{c^2}, \\
   R_{(x)(y)(x)(y)} &= R_{(x)(z)(x)(z)} = R_{(y)(z)(y)(z)} = \frac{H^2}{c^2}.
\end{align}
\end{subequations}

%% -------------------- Ricci tensor LT--------------------
\SecRicciLT
\begin{equation} 
  -R_{(t)(t)} = R_{(x)(x)} = R_{(y)(y)} = R_{(z)(z)} = 3\frac{H^2}{c^2}.
\end{equation}

%% ------------------------------------------------------
%%
%% ------------------------------------------------------
%\subsection{Embedded in five-dimensional Euclidean space}
%Using the transformation 
%\begin{equation}
%\begin{aligned}
% \alpha &= r\sin(\vartheta)\cos(\varphi), &\qquad\beta &= r\sin(\vartheta)\sin(\varphi), &\qquad \gamma &= r\cos(\vartheta) \\ 
% \delta &= R\sqrt{1-\frac{r^2}{R^2}}\cosh(\nicefrac{t}{R}), &\qquad\varepsilon &= R\sqrt{1-\frac{r^2}{R^2}}\sinh(\nicefrac{t}{R}).
%\end{aligned}
%\end{equation}
%or
%\begin{equation}
%  \delta + \varepsilon = Re^{\nicefrac{t}{R}}\sqrt{1-\frac{r^2}{R^2}}, \qquad \delta - \varepsilon = Re^{-\nicefrac{t}{R}}\sqrt{1-\frac{r^2}{R^2}}.
%\end{equation}
%respectively, the de-Sitter spacetime can be changed into the form
%\begin{equation}
% ds^2 = -d\alpha^2 - d\beta^2 - d\gamma^2 - d\delta^2 + d\varepsilon^2
%\end{equation}
%now transforming according to
%\begin{equation}
% z_{1} = i\alpha, \qquad z_{2} = i\beta, \qquad z_{3} = i\gamma, \qquad z_{4} = i\delta, \qquad z_{5} = \varepsilon
%\end{equation}
%leads finally to
%\begin{equation}
% ds^2 = dz_{1}^2 + dz_{2}^2 + dz_{3}^2 + dz_{4}^2 + dz_{5}^2
%\end{equation}
%with the constraint
%\begin{equation}
% z_{1}^2 + z_{2}^2 + z_{3}^2 + z_{4}^2 + z_{5}^2 = -R^2
%\end{equation}
%which describes the de-Sitter spacetime as a 4-dimensional sphere embedded in 5-dimensional Euclidean space.

%% -------------------- Further reading --------------------
\FurtherReading

Tolman\cite[sec. 142]{tolman}, Bi\v{c}\'ak\cite{bicak2001}

}{

}

%% ------------------------------------------------------------------------
%%    S t r a i g h  s p i n n i n g  s t r i n g
%% ------------------------------------------------------------------------
\clearpage
\section{Straight spinning string}
\setcounter{equation}{0}
\ifthenelse{\boolean{isARXIV}}{
% ******** Start of file spinstring.tex *********
%
%  Copyright (c) 2009 Thomas Mueller,
%                     Universitaet Stuttgart, VISUS
%

The metric of a straight spinning string in cylindrical coordinates $(t,\rho,\varphi,z)$ reads
\metricEq{
  ds^2 = -\left(c\,dt - a\,d\varphi\right)^2 + d\rho^2 + k^2\rho^2d\varphi^2+dz^2,
}{spinstring}
where $a\in\setR$ and $k>0$ are two parameters, see Perlick\cite{perlick2004}.

%% -------------------- Metric tensor --------------------
\SecMetric
\begin{equation}
 g_{tt} = -c^2,\qquad g_{t\varphi} = ac,\qquad g_{\rho\rho} = g_{zz} = 1, \qquad g_{\varphi\varphi} = k^2\rho^2-a^2.
\end{equation}

%% -------------------- Christoffel symbols --------------------
\SecChristoffel
\begin{equation}
   \Gamma_{\rho\varphi}^t = \frac{a}{c\rho},\qquad \Gamma_{\rho\varphi}^{\varphi} = \frac{1}{\rho},\qquad \Gamma_{\varphi\varphi}^{\rho} = -k^2\rho.
\end{equation}

Partial derivatives
\begin{equation}
  \Gamma_{\rho\varphi,\rho}^t = -\frac{\alpha}{c\rho^2},\qquad \Gamma_{\rho\varphi,\rho}^{\varphi} = -\frac{1}{\rho^2},\qquad \Gamma_{\varphi\varphi,\rho}^{\rho} = -k^2.
\end{equation}

The Riemann-, Ricci-, and Weyl-tensors as well as the Ricci- and Kretschmann-scalar vanish identically.

%% -------------------- Local tetrad --------------------
\SecStatLocal
\begin{equation}
  \mathbf{e}_{(0)} = \frac{1}{c}\partial_t, \qquad \mathbf{e}_{(1)} = \partial_{\rho}, \qquad \mathbf{e}_{(2)} =\frac{1}{k\rho}\left(\frac{a}{c}\partial_t+\partial_{\varphi}\right), \qquad \mathbf{e}_{(3)} = \partial_z.
\end{equation}
Dual tetrad:
\begin{equation}
 \boldsymbol{\theta}^{(0)} = c\,dt-a\,d\varphi,\qquad \boldsymbol{\theta}^{(1)} = d\rho,\qquad \boldsymbol{\theta}^{(2)}=k\rho\,d\varphi,\qquad \boldsymbol{\theta}^{(3)}=dz.
\end{equation}
Ricci rotation coefficients and their contractions read
\begin{equation}
   \gamma_{(2)(1)(2)} = \frac{1}{\rho},\qquad \gamma_{(0)}=\gamma_{(2)}=\gamma_{(3)} = 0,\qquad \gamma_{(1)}=\frac{1}{\rho}.
\end{equation}

\SecComLocal
\begin{subequations}
  \begin{alignat}{3}
  \mathbf{e}_{(0)} &= \frac{\sqrt{k^2\rho^2-a^2}}{k\rho }\left(\frac{1}{c}\partial_t-\frac{a}{k^2\rho^2-a^2}\partial_{\varphi}\right), &\qquad \mathbf{e}_{(1)} &=\partial_{\rho},\\
   \mathbf{e}_{(2)} &=\frac{1}{\sqrt{k^2\rho^2-a^2}}\partial_{\varphi}, & \mathbf{e}_{(3)} &= \partial_z.
  \end{alignat}
\end{subequations}
Dual tetrad:
\begin{equation}
 \boldsymbol{\theta}^{(0)} = \frac{k\rho}{\sqrt{k^2\rho^2-a^2}}c\,dt,\quad \boldsymbol{\theta}^{(1)} = d\rho,\quad \boldsymbol{\theta}^{(2)}=\frac{ac\,dt}{\sqrt{k^2\rho^2-a^2}}+\sqrt{k^2\rho^2-a^2}d\varphi,\quad \boldsymbol{\theta}^{(3)}=dz.
\end{equation}
Ricci rotation coefficients and their contractions read
\begin{subequations}
\begin{align}
  \gamma_{(0)(1)(0)}&=\frac{a^2}{\rho\left(k^2\rho^2-a^2\right)},\quad \gamma_{(2)(1)(0)}=\gamma_{(0)(2)(1)} =\gamma_{(0)(1)(2)}=\frac{ak}{k^2\rho^2-a^2},\\
  \gamma_{(2)(1)(2)} &= \frac{k^2\rho}{k^2\rho^2-a^2},\\
  \gamma_{(1)} &= \frac{1}{\rho}.
\end{align}
\end{subequations}
\newpage 

%% -------------------- Euler Lagrange --------------------
\SecEulLag

The Euler-Lagrangian formalism, Sec.~\ref{subsec:EL}, for geodesics in the $\vartheta=\pi/2$ hyperplane yields
\begin{equation}
 \dot{\rho}^2+\frac{1}{k^2\rho^2}\left(h_2-\frac{ah_1}{c}\right)^2 - \kappa c^2 = \frac{h_1^2}{c^2},
\end{equation}
with the constants of motion $h_1 = c(c\dot{t}-a\dot{\varphi})$ and $h_2=a(c\dot{t}-a\dot{\varphi})+k^2\rho^2\dot{\varphi}$.\\[0.5em]
The point of closest approach $\rho_{\text{pca}}$ for a null geodesic that starts at $\rho=\rho_i$ with $\mathbf{y}=\pm\Clt{0}+\cos\xi\Clt{1}+\sin\xi\Clt{2}$ with respect to the static tetrad is given by $\rho=\rho_i\sin\xi$. Hence, the $\rho_{\text{pca}}$ is independent of $a$ and $k$. The same is also true for timelike geodesics.

}{

}

%% ------------------------------------------------------------------------
%%    Sultana Dyer spacetime
%% ------------------------------------------------------------------------
\clearpage
\section{Sultana-Dyer spacetime}
\setcounter{equation}{0}
\ifthenelse{\boolean{isARXIV}}{
% ******** Start of file sultanaDyer.tex *********
%
%  Copyright (c) 2010 Thomas Mueller,
%                     Universitaet Stuttgart, VISUS
%

The Sultana-Dyer metric represents a black hole in the Einstein-de Sitter universe. In spherical coordinates $(t,r,\vartheta,\varphi)$, the metric reads\cite{sultana2005} $(G=c=1)$
\metricEq{
  ds^2 = t^4\left[\left(1-\frac{2M}{r}\right)dt^2-\frac{4M}{r}dt\,dr - \left(1+\frac{2M}{r}\right)dr^2-r^2\,d\Omega^2\right],
}{SultanaDyer}
where $M$ is the mass of the black hole and $\Omega^2=d\vartheta^2+\sin^2\vartheta d\varphi^2$ is the spherical surface element. Note that here, the signature of the metric is $\sign(\mathbf{g})=-2$.

%% -------------------- Christoffel symbols --------------------
\SecChristoffel
\begin{subequations}
 \begin{alignat}{5}
  \Cchris{tt}{t} &= \frac{2\left(r^3+4M^2r+M^2t\right)}{tr^3}, &\quad \Cchris{tt}{r} &= \frac{M(r-2M)(4r+t)}{tr^3}, &\quad \Cchris{tr}{t} &= \frac{M(r+2M)(4r+t)}{tr^3},\\
  \Cchris{tr}{r} &= \frac{2\left(r^3-4M^2r-M^2t\right)}{tr^3}, & \Cchris{t\vartheta}{\vartheta} &= \frac{2}{t}, &    \Cchris{t\varphi}{\varphi} &= \frac{2}{t},\\
  \Cchris{r\vartheta}{\vartheta} &= \frac{1}{r}, & \Cchris{r\varphi}{\varphi} &= \frac{1}{r}, & \Cchris{\vartheta\vartheta}{t} &= \frac{2\left(r^2+2Mr-Mt\right)}{t},\\
  \Cchris{\vartheta\vartheta}{r} &= -\frac{4Mr+tr-2Mt}{t}, & \Cchris{\vartheta\varphi}{\varphi} &= \cot\vartheta, & \Cchris{\varphi\varphi}{\vartheta} &= -\sin\vartheta\cos\vartheta,
 \end{alignat}
 \begin{alignat}{3}
  \Cchris{rr}{t} &= \frac{2\left(r^3+4Mr^2+4M^2r+M^2t+Mtr\right)}{tr^3}, &\quad \Cchris{rr}{r} &= -\frac{M\left(4r^2+8Mr+2Mt+tr\right)}{tr^3},\\
  \Cchris{\varphi\varphi}{t} &= \frac{2\left(r^2+2Mr-Mt\right)\sin^2\vartheta}{t}, & \Cchris{\varphi\varphi}{r} &= -\frac{\left(4Mr+tr-2Mt\right)\sin^2\vartheta}{t}.
 \end{alignat}
\end{subequations}

%% -------------------- Riemann tensor --------------------
\SecRiemann
\begin{subequations}
 \begin{align}
   R_{trtr} &= \frac{2t^2\left(-2Mr^2-r^3+Mt^2+2Mtr\right)}{r^3},\\ 
   R_{r\vartheta t\vartheta} &= -\frac{t^2\left(2r^4+16M^2r^2+4Mtr^2-4M^2r^2t+Mt^2r-2M^2t^2\right)}{r^2},\\
   R_{t\vartheta r\vartheta} &= -\frac{2Mt^2(4r+t)(r^2+2Mr-Mt)}{r^2},\\
   R_{r\varphi t\varphi} &= -\frac{t^2\sin^2\vartheta\left(2r^4+16M^2r^2+4Mtr^2-4M^2r^2t+Mt^2r-2M^2t^2\right)}{r^2},\\
   R_{t\varphi r\varphi} &= -\frac{2Mt^2\sin^2\vartheta(4r+t)(r^2+2Mr-Mt)}{r^2},\\
   R_{r\vartheta r\vartheta} &= -\frac{t^2\left(4r^4+16Mr^4-4M^2tr+16M^2r^2-2M^2t^2-Mt^2r\right)}{r^2},\\
   R_{r\varphi r\varphi} &= -\frac{t^2\sin^2\vartheta\left(4r^4+16Mr^4-4M^2tr+16M^2r^2-2M^2t^2-Mt^2r\right)}{r^2},\\
   R_{\vartheta\varphi\vartheta\varphi} &= -2t^2r\sin^2\vartheta\left(2r^3+4Mr^2-4Mtr+mt^2\right).
 \end{align}
\end{subequations}

%% -------------------- ricci tensor --------------------
\SecRicci
\begin{subequations}
 \begin{alignat}{5}
   R_{tt} &= \frac{2\left(3r^2+12M^2+2Mt\right)}{t^2r^2}, &\quad R_{tr} &= \frac{4M\left(3r+t+6M\right)}{t^2r^2},\\
   R_{rr} &= \frac{2\left(3r^2+12Mr+2Mt+12M^2\right)}{t^2r^2}, & R_{\vartheta\vartheta} &= \frac{6\left(r^2+2Mr-2Mt\right)}{t^2},\\
   R_{\varphi\varphi} &= \frac{6\left(r^2+2Mr-2Mt\right)\sin^2\vartheta}{t^2}.
 \end{alignat}
\end{subequations}

\noindent {\bf Ricci and Kretschmann scalars:}
\begin{subequations}
 \begin{align}
  R &= -\frac{12\left(r^2+2Mr-2Mt\right)}{t^6r^2},\\
  \mathcal{K} &= \frac{48\left(M^2t^4+20M^2r^4+20Mr^5+8M^2r^2t^2-4Mr^4t-16M^2r^3t+5r^6\right)}{t^12r^6}.
 \end{align}
\end{subequations}

%% -------------------- comoving local tetrad --------------------
\SecComLocal
\begin{equation}
  \Clt{0} = \frac{\sqrt{1+2M/r}}{t^2}\partial_t - \frac{2M/r}{t^2\sqrt{1+2M/r}}\partial_r,\quad \Clt{1} = \frac{1}{t^2\sqrt{1+2M/r}}\partial_r, \quad \Clt{2}=\frac{1}{t^2r}\partial_{\vartheta}, \quad \Clt{3} = \frac{1}{t^2r\sin\vartheta}\partial_{\varphi}.
\end{equation}

%% -------------------- static local tetrad --------------------
\SecStatLocal
\begin{equation}
 \Clt{0} = \frac{1}{t^2\sqrt{1-2M/r}}\partial_t,\quad \Clt{1} = \frac{2M/r}{t^2\sqrt{1-2M/r}}\partial_t+\frac{\sqrt{1-2M/r}}{t^2}\partial_r,\quad \Clt{2} = \frac{1}{t^2r}\partial_{\vartheta}, \quad \Clt{3} = \frac{1}{t^2r\sin\vartheta}\partial_{\varphi}.
\end{equation}

%% -------------------- Further reading --------------------
\FurtherReading

Sultana and Dyer\cite{sultana2005}.

}{

}

%% ------------------------------------------------------------------------
%%    T a u b N U T
%% ------------------------------------------------------------------------
\clearpage
\section{TaubNUT}
\setcounter{equation}{0}
\ifthenelse{\boolean{isARXIV}}{
% ******** Start of file taubNUT.tex *********
%
%  Copyright (c) 2009 Thomas Mueller,
%                     Universitaet Stuttgart, VISUS
%

The TaubNUT metric in Boyer-Lindquist like spherical coordinates $(t,r,\vartheta,\varphi)$ reads\cite{bini2002} $(G=c=1)$
\metricEq{
  ds^2 = -\frac{\Delta}{\Sigma}\left(dt+2\ell\cos\vartheta\,d\varphi\right)^2+\Sigma\left(\frac{dr^2}{\Delta}+d\vartheta^2+\sin^2\vartheta\,d\varphi^2\right),
}{TaubNUT}
where $\Sigma=r^2+\ell^2$ and $\Delta=r^2-2Mr-\ell^2$. Here, $M$ is the mass of the black hole and $\ell$ the magnetic monopol strength.

%% -------------------- Christoffel symbols --------------------
\SecChristoffel
\begin{subequations}
\begin{alignat}{5}
  \Gamma_{tt}^r &= \frac{\Delta\rho}{\Sigma^3}, &\quad\Gamma_{tr}^t &= \frac{\rho}{\Delta\Sigma}, &\quad\Gamma_{t\vartheta}^t &= -2\ell^2\cos\vartheta\frac{\Delta}{\Sigma^2},\\
  \Gamma_{t\vartheta}^{\varphi} &= \frac{\ell\Delta}{\Sigma^2\sin\vartheta}, &\quad \Gamma_{t\varphi}^r &= \frac{2\ell\rho\Delta\cos\vartheta}{\Sigma^3}, & \Gamma_{t\varphi}^{\vartheta} &= -\frac{\ell\Delta\sin\vartheta}{\Sigma^2},\\
  \Gamma_{rr}^r &= -\frac{\rho}{\Sigma\Delta}, & \Gamma_{r\vartheta}^{\vartheta} &= \frac{r}{\Sigma}, & \Gamma_{r\varphi}^{\varphi} &= \frac{r}{\Sigma},\qquad \Gamma_{\vartheta\vartheta}^r = -\frac{r\Delta}{\Sigma},
\end{alignat}
\begin{align}
  \Gamma_{r\varphi}^t &= \frac{-2\ell(r^3-3Mr^2-3r\ell^2+M\ell^2)\cos\vartheta}{\Sigma\Delta},\\
  \Gamma_{\vartheta\varphi}^t &= -\frac{\ell\left[\cos^2\vartheta\left(6r^2\ell^2-8\ell^2Mr-3\ell^4+r^4\right)+\Sigma^2\right]}{\Sigma^2\sin\vartheta},\\
  \Gamma_{\varphi\varphi}^r &= \frac{\Delta}{\Sigma^3}\left[\cos^2\vartheta\left(9r\ell^4+4\ell^2Mr^2-4\ell^4M+r^5+2r^3\ell^2\right)-r\Sigma^2\right],\\
  \Gamma_{\vartheta\varphi}^{\varphi} &= \frac{\left(4r^2\ell^2-4Mr\ell^2-\ell^4+r^4\right)\cot\vartheta}{\Sigma^2},\\
  \Gamma_{\varphi\varphi}^{\vartheta} &= -\frac{\left(6r^2\ell^2-8Mr\ell^2-3\ell^4+r^4\right)\sin\vartheta\cos\vartheta}{\Sigma^2},
\end{align}
\end{subequations}
where $\rho=2r\ell^2+Mr^2-M\ell^2$.

%% -------------------- static local tetrad --------------------
\SecStatLocal
\begin{equation}
  \mathbf{e}_{(0)} = \sqrt{\frac{\Sigma}{\Delta}}\partial_t, \quad \mathbf{e}_{(1)} = \sqrt{\frac{\Delta}{\Sigma}}\partial_r,\quad \mathbf{e}_{(2)} = \frac{1}{\sqrt{\Sigma}}\partial_{\vartheta}, \quad \mathbf{e}_{(3)} = -\frac{2\ell\cot\vartheta}{\sqrt{\Sigma}}\partial_t + \frac{1}{\sqrt{\Sigma}\sin\vartheta}\partial_{\varphi}.
\end{equation}
Dual tetrad:
\begin{equation}
 \boldsymbol{\theta}^{(0)}=\sqrt{\frac{\Delta}{\Sigma}}\left(dt+2\ell\cos\vartheta\,d\varphi\right),\quad \boldsymbol{\theta}^{(1)} = \sqrt{\frac{\Sigma}{\Delta}}dr,\quad \boldsymbol{\theta}^{(2)} = \sqrt{\Sigma}d\vartheta,\quad \boldsymbol{\theta}^{(3)} = \sqrt{\Sigma}\sin\vartheta\,d\varphi.
\end{equation}
% 
% %% -------------------- local comoving tetrad --------------------
% \SecComLocal
% \begin{equation}
%  \Clt{0} = \frac{\sqrt{Q/\Sigma}}{\sin\vartheta}\partial_t+\frac{2\ell\cot\vartheta}{\sqrt{Q\Sigma}}\partial_{\varphi},\quad \Clt{1} = \sqrt{\frac{\Delta}{\Sigma}}\partial_r,\quad \Clt{2} = \frac{1}{\sqrt{\Sigma}}\partial_{\vartheta},\quad \Clt{3} = \sqrt{\frac{\Sigma}{\Delta Q}}\partial_{\varphi},
% \end{equation}
% where $Q=(\Sigma^2/\Delta)\sin^2\vartheta - 4\ell^2\cos^2\vartheta$.\\[0.5em]
% Dual tetrad:
% \begin{equation}
% \Cdlt{0} = \sin\vartheta\sqrt{\frac{\Sigma}{Q}}dt,\quad \Cdlt{1} = \sqrt{\frac{\Sigma}{\Delta}}dr,\quad \Cdlt{2} = \sqrt{\Sigma}d\vartheta,\quad \Cdlt{3} = -2\ell\cos\vartheta\sqrt{\frac{\Delta}{\Sigma}}dt+\sqrt{\frac{\Delta Q}{\Sigma}}d\varphi.
% \end{equation}
% \newpage

%% -------------------- Euler Lagrange --------------------
\SecEulLag

The Euler-Lagrangian formalism, Sec.~\ref{subsec:EL}, for geodesics in the $\vartheta=\pi/2$ hyperplane yields
\begin{equation}
  \frac{1}{2}\dot{r}^2+V_{\text{eff}} = \frac{1}{2}\frac{k^2}{c^2},\qquad V_{\text{eff}}=\frac{1}{2}\frac{\Delta}{\Sigma}\left(\frac{h^2}{\Sigma}-\kappa\right)
\end{equation}
with the constants of motion $k=(\Delta/\Sigma)\dot{t}$ and $h=\Sigma\dot{\varphi}$. For null geodesics, we obtain a photon orbit at $r=r_{\text{po}}$ with
\begin{equation}
  r_{\text{po}} = M + 2\sqrt{M^2+\ell^2}\cos\left(\frac{1}{3}\arccos\frac{M}{\sqrt{M^2+\ell^2}}\right)
\end{equation}

%% -------------------- Further reading --------------------
\FurtherReading

Bini et al.\cite{bini2003b}.

}{

}

% -----------------------------------------------------------------
%                           thebibliography
% -----------------------------------------------------------------
\newpage %neue Seite, damit Link auf Seitenanfang zeigt
\phantomsection %notwendig, damit Link nicht unterhalb der Überschrift zeigt
\addcontentsline{toc}{chapter}{Bibliography}
\bibliographystyle{alphaurl}
\bibliography{lit_cos}

%% ---------------------------- end document ---------------------------
\end{document}